%% file: ms.tex
\documentclass{article}

\usepackage[english]{babel} % default is babel's ngerman
\usepackage[T1]{fontenc}
\usepackage[utf8x]{inputenc}

\usepackage[a4paper, total={5.7in, 8.8in}]{geometry}
\usepackage{url}

\usepackage{algorithm}
\usepackage{algorithmic}
\usepackage{amsmath}
\usepackage{subcaption}
\usepackage[dvipsnames]{xcolor}
\usepackage{pgf}
\usepackage{capt-of}
\usepackage{cite}

\usepackage[pdftex]{hyperref} 
\hypersetup{colorlinks=true}
\hypersetup{linkcolor=black}
\hypersetup{citecolor=black}

\usepackage{tikz}
\usepackage{pgfplots}
\pgfplotsset{compat=1.5.1}
\usetikzlibrary{calc}
\usetikzlibrary{arrows}

\begin{document}

\title{A comparative study of fluid-particle coupling methods for fully resolved lattice Boltzmann simulations}

\author{C. Rettinger\footnote{Correspondence to: Chair for System Simulation, Friedrich-Alexander-Universit\"{a}t Erlangen-N\"{u}rnberg, Cauerstr. 11, 91058 Erlangen, Germany. E-mail: christoph.rettinger@fau.de}, U. R\"{u}de\\Chair for System Simulation,\\ Friedrich-Alexander-Universität Erlangen-Nürnberg,\\ Cauerstraße 11, 91058 Erlangen, Germany}

\date{}

\maketitle

\begin{abstract}
		The direct numerical simulation of particulate systems offers a unique approach to study the dynamics of fluid-solid suspensions by fully resolving the submerged particles and without introducing empirical models.
		For the lattice Boltzmann method, different variants exist to incorporate the fluid-particle interaction into the simulation.
		This paper provides a detailed and systematic comparison of two different methods, namely the momentum exchange method and the partially saturated cells method by Noble and Torczynski.
		Three subvariants of each method are used in the benchmark scenario of a single heavy sphere settling in ambient fluid to study their characteristics and accuracy for particle Reynolds numbers from $185$ up to $365$.
		The sphere must be resolved with at least $24$ computational cells per diameter to achieve velocity errors below $5\%$.
		The momentum exchange method is found to be more accurate in predicting the streamwise velocity component whereas the partially saturated cells method is more accurate in the spanwise components.
		The study reveals that the resolution should be chosen with	respect to the coupling dynamics, and not only based on the flow properties, to avoid large errors in the fluid-particle interaction.
		
		Keywords: lattice Boltzmann method; particulate flow; direct numerical simulation; fluid-structure interaction; momentum exchange method; partially saturated cells method
\end{abstract}

\input{Introduction}

\input{NumericalMethod}

\input{BenchmarkDragForce}

\input{BenchmarkMSHS}

\input{Conclusion}

%\section*{References}
\bibliographystyle{plain}
\bibliography{Library} 

\include{Tables}	
\end{document}

%% file: Introduction.tex
\section{Introduction}

To study particulate flows, direct numerical simulations (DNS) have become a viable and important tool.
Resolving the flow structures and the submerged particles offers the unique possibility to trace the motion of single particles, to evaluate the hydrodynamic forces acting on each individual particle, and to investigate the flow field in detail.
As an alternative to conventional computational fluid dynamics (CFD) methods that solve the Navier-Stokes equations, the lattice Boltzmann method (LBM) has been applied successfully for DNS studies of such particulate systems.
Examples include coupled simulations of several thousand \cite{pan_fluidization_2002,third_comparison_2015} up to millions of spherical particles \cite{xiong_large-scale_2012,gotz_direct_2010} as they typically appear in fluidized beds at a laboratory scale.
Also non-spherical objects can be handled by this method which allows to study flows with blood cells \cite{janoschek_rotational_2011}, the motion of elongated particles \cite{mao_motion_2014, bartuschat_two_2016} and ice floes \cite{mierke2015gpu}.
Extensions exist that incorporate electrostatic forces between the particles \cite{bartuschat_parallel_2015,kuron2016moving}, enable deformable objects \cite{wu_simulating_2010,kruger_efficient_2011} or model self-propelled swimmers \cite{pickl_all_2012}.

A crucial component of such simulations is the accurate and efficient description of the interaction between the solid and the fluid phase which requires appropriate coupling methods.
Avoiding the costly remeshing due to changes in the geometry and topology that are caused by the moving particles is a key aspect here.
Therefore, classical CFD methods often make use of the immersed boundary method to satisfy the no-slip boundary condition on the particle surfaces and to compute the hydrodynamic forces acting on them \cite{peskin_numerical_1977, uhlmann_immersed_2005, tang_methodology_2014}.
The standard LBM employs uniform Cartesian meshes with cubic cells and thus it is natural here that coupling methods are utilized which do not alter the mesh.
This results in an excellent parallel performance and scalability of the LBM method \cite{godenschwager_framework_2013,hasert_complex_2014} and with suitable data structures also the fluid-solid coupling can be implemented with high parallel efficiency \cite{gotz_direct_2010}.

Several different fluid-solid coupling approaches have been proposed for the LBM: 
the momentum exchange method \cite{ladd_numerical_1994,aidun_direct_1998}, the partially saturated cells method \cite{noble_lattice-boltzmann_1998, holdych_lattice_2003}, and methods that rely on Lagrangian marker points like the immersed boundary method \cite{feng_immersed_2004,niu_momentum_2006, wu_particulate_2009} or the external boundary force method \cite{wu_simulating_2010}.
Beyond this general classification of the methods, a large number of subvariants can be found in literature.

It is apparent that a rigorous systematic comparison of these methods for systems with moving solid particles is needed to assist the choice for a suitable method.
Challenges here are to define appropriate benchmark scenarios for which reliable reference data exist and to cover the flow regimes at larger Reynolds numbers as they are most relevant for typical engineering applications.
One standard benchmark is to evaluate the drag force acting on an infinite periodic array of spheres in Stokes flow, as e.g. in \cite{peng_comparative_2008,bogner_drag_2015,khirevich_coarse-_2015,fattahi_lattice_2016}. 
This benchmark can make use of an analytic solution in the form of a series expansion \cite{hasimoto_periodic_1959, sangani_slow_1982} and is thus particularly suitable to validate implementations and investigate certain characteristics of the fluid-solid coupling methods as will be demonstrated later on in this article.
However, the significance of this benchmark is limited when the goal is to simulate scenarios with moving particles at larger Reynolds numbers since the motion of the particles will affect the accuracy and stability of the coupling mechanism. 
A suitable benchmark has to take into account all possible sources of numerical errors that might appear in a coupled simulation.
Besides the fluid simulation itself, errors can be introduced by the particle simulation and the coupling from the fluid to the solid phase and vice versa.

Some studies exist that aim at directly comparing different fluid-solid coupling methods for LBM but they are mostly restricted to two-dimensional setups:
In \cite{peng_comparative_2008}, the momentum exchange method with an interpolated bounce back scheme and the immersed boundary method are compared with respect to their accuracy and efficiency for a laminar flow over a stationary cylinder \cite{schafer_benchmark_1996}. 
The settling of circular and elliptical particles is simulated in \cite{wang_evaluation_2013} to establish a comparison between three different LBM collision operators in combination with the momentum exchange method.
Various coupling methods are compared in \cite{chen_comparative_2014} for two-dimensional objects with a prescribed motion and with the focus on the accuracy of acoustic properties. 
In a recent study, sedimentation of a single circular particle is simulated with different coupling methods \cite{wu_comparative_2017}. 

The benchmark proposed by Uhlmann and Dušek \cite{uhlmann_motion_2014} features a single sphere settling in an ambient fluid in a three-dimensional domain.
High accuracy results obtained with spectral elements are available for flow regimes with particle Reynolds numbers ranging from $185$ up to $365$.
This test case can be set up independent of the used CFD method and it is well suited to compare different methods.
Its applicability to LBM has already been demonstrated in a preparatory study by \cite{rettinger_simulations_2016}. 
In particular, it can be used to evaluate which resolution is required to achieve sufficiently accurate results in these flow regimes.
In this article, we therefore use this benchmark to establish a systematic comparison between two commonly used coupling approaches with the LBM, namely the momentum exchange method and the partially saturated cells method.
For each method, three distinct subvariants are applied and validated to illustrate their respective features and strengths.
For the momentum exchange method, the simple bounce back and two interpolated versions are compared, whereas different solid collision operators are used for the partially saturated cells method.
Methods that employ Lagrangian markers are not included in this study as they are typically more suitable for deformable objects in contrast to the rigid particles here.
However, future work should also investigate those methods to establish a complete overview over all available coupling methods.

The remainder of this paper is structured as follows: 
First the numerical methods are introduced in Sec.~\ref{chap:NumericalMethod}.
This includes a brief summary of the lattice Boltzmann method in Sec.~\ref{sec:LBM} as basis for the momentum exchange method and its variants presented in Sec.~\ref{sec:MEM}.
The partially saturated cells method with its variants is introduced in Sec.~\ref{sec:PSM}.
A short study in Sec.~\ref{sec:DragForce} compares the obtained force on a fixed sphere in Stokes flow for the resulting six approaches.
Section~\ref{sec:description} elaborates the setup of the benchmark test case from Uhlmann and Dušek. 
The six coupling approaches are then evaluated for four different flow regimes in Secs.~\ref{sec:regime1}--\ref{sec:regime4}.
The results are discussed in Sec.~\ref{sec:discussion}. 
Finally, the most important findings are summarized in Sec.~\ref{chap:conclusion}.

%% file: NumericalMethod.tex
\section{Numerical method}
\label{chap:NumericalMethod}

\subsection{Lattice Boltzmann method}
\label{sec:LBM}
For the simulation of hydrodynamics, the lattice Boltzmann approach \cite{chen_lattice_1998} with the \textit{D3Q19} lattice model \cite{qian_lattice_1992} is utilized.
Having its origin in statistical mechanics, the evolution of particle distribution functions (PDFs) on a Cartesian lattice is computed.
Each of these PDFs $f_q$, with $q \in \{0,\dots,18\}$, is associated with a lattice velocity $\boldsymbol{c}_q$.
In its most general form, the lattice Boltzmann equation is then given by the collision step
\begin{equation}
	\tilde{f}_q(\boldsymbol{x},t) = f_q(\boldsymbol{x},t) + \mathcal{C}_q(\boldsymbol{x},t) + \mathcal{F}_q(\boldsymbol{x},t),\label{eq:LBM_Collide}
\end{equation}
specified by the collision operator $\mathcal{C}_q$ and the external forcing operator $\mathcal{F}_q$.
In the succeeding stream step, the post collision values $\tilde{f}_q$ are distributed to the corresponding neighbor lattice cells via
\begin{equation}
	f_q( \boldsymbol{x} + \boldsymbol{c}_q\,\Delta t ,t+\Delta t) = \tilde{f}_q(\boldsymbol{x},t). \label{eq:LBM_Stream}
\end{equation}

The most commonly applied collision model is the BGK model \cite{bhatnagar_model_1954} that uses a single relaxation parameter to linearly relax the PDFs towards their equilibrium values $f_q^{\text{eq}}$. 
Those can be computed as 
\begin{equation}
	f_q^{\text{eq}}(\rho_f,\boldsymbol{U}) = w_q \left( \rho_f + \rho_0 \left(\frac{\boldsymbol{c}_q \cdot \boldsymbol{U}}{c_s^2} + \frac{(\boldsymbol{c}_q \cdot \boldsymbol{U})^2}{2c_s^4} - \frac{\boldsymbol{U} \cdot \boldsymbol{U}}{2c_s^2}\right) \right) \label{eq:LBM_EQ}
\end{equation}
for incompressible flows \cite{he1997lattice}.
The fluid density $\rho_f = \rho_0 + \delta \rho_f$, with the mean density $\rho_0$ and the fluctuation $\delta \rho_f$, and the velocity $\boldsymbol{U}$ are cell local quantities and calculated via moments of the PDFs:
\begin{equation}
\rho_f(\boldsymbol{x},t) = \sum_q f_q(\boldsymbol{x},t),\quad \boldsymbol{U}(\boldsymbol{x},t) = \frac{1}{\rho_0}\sum_q f_q(\boldsymbol{x},t) \boldsymbol{c}_q.
\end{equation}
The lattice weights $w_q$ are as given e.g. in \cite{qian_lattice_1992} and $c_s$ is the lattice speed of sound.
The collision operator for the BGK model is then 
\begin{equation}
\mathcal{C}_q^{\text{BGK}}(\boldsymbol{x},t) = \tfrac{\Delta t}{\tau}\left(f_q^{\text{eq}}(\rho_f, \boldsymbol{U} ) - f_q(\boldsymbol{x},t) \right). \label{eq:LBM_SRT}
\end{equation}
It features the relaxation time $\tau \in (\frac{1}{2}, \infty)$ which determines the kinematic fluid viscosity $\nu$ via
\begin{equation}
\nu = (\tau - \tfrac{\Delta t}{2})c_s^2 \label{eq:Viscosity}.
\end{equation}
The forcing operator in Eq.~\eqref{eq:LBM_Collide} is used to incorporate external forces and can be written as 
\begin{equation}
\mathcal{F}_q(\boldsymbol{x},t) = \Delta t w_q \left[\frac{\boldsymbol{c}_q-\boldsymbol{U}}{c_s^2} + \frac{\boldsymbol{c}_q \cdot \boldsymbol{U}}{c_s^4} \boldsymbol{c}_q\right] \cdot \boldsymbol{f}^{\text{ext}}, \label{eq:Forcing}
\end{equation}
with a constant force density $\boldsymbol{f}^{\text{ext}}$ \cite{ladd_lattice-boltzmann_2001}.
The cell local macroscopic velocity $\boldsymbol{u}$ is then obtained via
\begin{equation}
\boldsymbol{u}(\boldsymbol{x},t) =  \boldsymbol{U}(\boldsymbol{x},t) + \tfrac{\Delta t}{2\rho_0} \boldsymbol{f}^{\text{ext}}, \label{eq:MacVel}
\end{equation} 
and thus differs from the velocity $\boldsymbol{U}$, which is used to calculate $f_q^{\text{eq}}$ and $\mathcal{F}_q$ in Eqs.~\eqref{eq:LBM_EQ} and \eqref{eq:Forcing}, respectively, by a shift depending on $\boldsymbol{f}^{\text{ext}}$ \cite{guo_discrete_2002}.

However, it is well-known that the BGK collision operator has shortcomings with respect to stability and accuracy \cite{luo_numerics_2011}. 
In particular, the location of solid boundaries exhibits an undesired $\tau$ dependency.
To overcome this shortcoming, the two-relaxation-time collision operator from \cite{ginzburg_two-relaxation-time_2008} can be applied, which is given by
\begin{equation}
\mathcal{C}_q^{\text{TRT}}(\boldsymbol{x},t) = \lambda_+(f_q^+(\boldsymbol{x},t) - f_q^{\text{eq},+}(\rho_f, \boldsymbol{U})) + \lambda_-(f_q^-(\boldsymbol{x},t) - f_q^{\text{eq},-}(\rho_f, \boldsymbol{U})). \label{eq:MEM_TRT}
\end{equation}
Here, the PDFs and their equilibrium values are split into their symmetric and anti-symmetric parts,
\begin{equation}
f_q^+ = \tfrac{1}{2}(f_q + f_{\bar{q}}),\quad f_q^- = \tfrac{1}{2}(f_q - f_{\bar{q}}),
\end{equation}
where $\bar{q}$ denotes the opposite direction of $q$, such that $\boldsymbol{c}_{\bar{q}} = - \boldsymbol{c}_q$. 
This model features two independent relaxation parameters, $\lambda_+,\lambda_- \in(-2,0)$, where the first one is again related to the kinematic viscosity and $\lambda_+= -\tfrac{\Delta t}{\tau}$ holds. 
The boundary location can be rendered effectively independent of the viscosity by choosing $\lambda_-$ to satisfy 
\begin{equation}
\Lambda_\pm := \left( \tfrac{1}{2} + \tfrac{1}{\lambda_+}\right) \left(\tfrac{1}{2} + \tfrac{1}{\lambda_-}\right),
\end{equation}
where $\Lambda_\pm$ is a constant value \cite{ginzburg_two-relaxation-time_2008,khirevich_coarse-_2015}.% and typically taken as $\tfrac{3}{16}$ .

In the LBM context, it is common to use lattice units which results in the lattice spacing $\Delta x = 1$, the time step size $\Delta t = 1$, $\rho_0 = 1$ and $c_s= \tfrac{1}{\sqrt{3}}$. 
To allow for simulations of fluid-solid flows with fully resolved rigid objects, the LBM has to be extended by a suitable coupling method. 
Two possibilities are briefly explained in the following sections, namely the momentum exchange method in Sec.~\ref{sec:MEM} and the partially saturated cells method in Sec.~\ref{sec:PSM}.
All features are implemented in the LBM framework \textsc{waLBerla}\footnote{\url{http://walberla.net/}} \cite{godenschwager_framework_2013} that is used for the simulations in Secs.~\ref{sec:DragForce} and \ref{chap:benchmark}.

\subsection{Momentum exchange method}
\label{sec:MEM}
The momentum exchange method (MEM) for fluid-solid interactions was first presented by Ladd \cite{ladd_numerical_1994} and then extended by Aidun et al. \cite{aidun_direct_1998}. 
The main idea is to explicitly map the particle into the fluid domain by marking containing cells as \textit{solid} and to apply suitable no-slip boundary conditions along its surface. 
Using the link-based momentum exchange idea, the hydrodynamic force acting on the particle can be obtained. 
Next, the different parts of this method are presented.

\subsubsection{No-slip boundary condition}
On the particle's surface, the macroscopic flow velocity must match the particle's velocity. 
In the momentum exchange method, this is realized by imposing no-slip boundary conditions between the fluid and solid cells. 
The simplest variant of this boundary condition is the bounce back (BB) scheme \cite{ladd_numerical_1994}:
\begin{equation}
	f_{\bar{q}}(\boldsymbol{x},t+\Delta t) = \tilde{f}_q(\boldsymbol{x},t) - 2 \tfrac{w_q}{c_s^2}\rho_0 \boldsymbol{v}(\boldsymbol{x}_b,t) \cdot \boldsymbol{c}_q, \label{eq:MEM_BB}
\end{equation}
where $\boldsymbol{v}$ is the particle's velocity calculated at the boundary location $\boldsymbol{x}_b$ midway between fluid and solid cell center.
This velocity is obtained by
\begin{equation}
\boldsymbol{v}(\boldsymbol{x},t) = \boldsymbol{U}_p(t) + \boldsymbol{\Omega}_p(t) \times (\boldsymbol{x}-\boldsymbol{X}_p(t)), \label{eq:bodyVelocity}
\end{equation}
with the particle's translational velocity $\boldsymbol{U}_p$, rotational velocity $\boldsymbol{\Omega}_p$, and center of gravity $\boldsymbol{X}_p$.
The BB scheme results in a staircase approximation of the particle shape.
However, as the actual shape and thus the exact surface position of the particle is known, more advanced boundary conditions can be applied. 
They typically use the information about the actual boundary location by introducing a variable $\delta_q$ which is the ratio between the distance from the fluid cell center to the boundary and the distance from the fluid cell center to the solid cell center.
The boundary location is thus given as $\boldsymbol{x}_b = \boldsymbol{x} + \boldsymbol{c}_q\delta_q$. 
For example, the central linear interpolation (CLI) scheme from \cite{ginzburg_two-relaxation-time_2008} interpolates the PDFs along the direction of $\boldsymbol{c}_q$ to realize the no-slip boundary condition:
\begin{equation}
	f_{\bar{q}}(\boldsymbol{x},t+\Delta t) = \tilde{f}_q(\boldsymbol{x},t) + \kappa_0 \tilde{f}_q(\boldsymbol{x}-\boldsymbol{c}_q\,\Delta t,t) - \kappa_0 \tilde{f}_{\bar{q}}(\boldsymbol{x},t)-\alpha \tfrac{w_q}{c_s^2}\rho_0 \boldsymbol{v}(\boldsymbol{x}_b,t) \cdot \boldsymbol{c}_q, \label{eq:MEM_CLI}
\end{equation}
with the coefficients $\kappa_0 = (1-2\delta_q)/(1+2\delta_q)$ and $\alpha = 4/(1+2\delta_q)$.
It can be observed that this scheme is a generalization of the BB scheme which is recovered by setting $\delta_q=\tfrac{1}{2}$.

Another variant of the boundary condition is the multi-reflection (MR) scheme which applies a quadratic interpolation strategy for the PDFs \cite{ginzburg_two-relaxation-time_2008}:
\begin{align}
f_{\bar{q}}(\boldsymbol{x},t+\Delta t) =\ &\kappa_1 \tilde{f}_q(\boldsymbol{x},t) + \kappa_0 \tilde{f}_q(\boldsymbol{x}-\boldsymbol{c}_q\,\Delta t,t) + \kappa_{-1} \tilde{f}_q(\boldsymbol{x}-2\boldsymbol{c}_q\,\Delta t,t) + \nonumber \\ 
&\bar{\kappa}_{-1} \tilde{f}_{\bar{q}}(\boldsymbol{x},t) + \bar{\kappa}_{-2} \tilde{f}_{\bar{q}}(\boldsymbol{x}-\boldsymbol{c}_q\,\Delta t,t) - \alpha \tfrac{w_q}{c_s^2} \rho_0 \boldsymbol{v}(\boldsymbol{x}_b,t) \cdot \boldsymbol{c}_q + f_q^{\text{pc}}, \label{eq:MEM_MR}
\end{align}
where the expressions for $\kappa$, $\alpha$, and the correction term $f_q^{\text{pc}}$ can be found in Tabs. 3 and 4 from \cite{ginzburg_two-relaxation-time_2008}.

In contrast to other existing interpolation schemes, CLI and MR yield viscosity independent results in combination with the TRT collision operator \cite{pan_evaluation_2006,fattahi_lattice_2016}, as will be demonstrated in Sec.~\ref{sec:DragForce}.

\subsubsection{Fluid-solid interaction force}
\label{sec:MEM_force}
Following \cite{ladd_numerical_1994} and the proposed improvement by \cite{wen_galilean_2014}, the force $\boldsymbol{F}^{q_{\text{f-s}}}$ acting on a solid object at position $\boldsymbol{x}_b$ over a fluid-solid link $q_{\text{f-s}}$ can be calculated with the help of the momentum exchange idea:
\begin{equation}
\boldsymbol{F}^{q_{\text{f-s}}}(\boldsymbol{x}_b,t) = \tfrac{(\Delta x)^3}{\Delta t} \left[ (\boldsymbol{c}_{q_{\text{f-s}}} - \boldsymbol{v}(\boldsymbol{x}_b,t)) \tilde{f}_{q_{\text{f-s}}}(\boldsymbol{x},t) - (\boldsymbol{c}_{\bar{q}_{\text{f-s}}} - \boldsymbol{v}(\boldsymbol{x}_b,t)) f_{\bar{q}_{\text{f-s}}}(\boldsymbol{x},t+\Delta t) \right]. \label{eq:MEM_Force}
\end{equation}
This force is thus given as the difference between the momentum towards the boundary, i.e.\ multiplying the lattice velocity with the post-collision value of the PDF, and the momentum away from the boundary, calculated by multiplying the inverse lattice velocity with the PDF after the boundary handling from Eqs.~\eqref{eq:MEM_BB}, \eqref{eq:MEM_CLI} or \eqref{eq:MEM_MR}.
To ensure Galilean invariance for moving solid boundaries, it has been shown that the boundary velocity $\boldsymbol{v}(\boldsymbol{x}_b,t)$ has to be subtracted from the lattice velocities \cite{wen_galilean_2014}.
By summing up all the local contributions, i.e.\ all corresponding fluid-solid links in the boundary cells of a specific object, the total hydrodynamic force $\boldsymbol{F}$ and torque $\boldsymbol{T}$ acting on this object can be computed via
\begin{align}
\boldsymbol{F}(t) & = \sum_{\boldsymbol{x}_b}\sum_{q_{\text{f-s}}} \boldsymbol{F}^{q_{\text{f-s}}}(\boldsymbol{x}_b,t), \label{eq:MEM_TotForce}\\
\boldsymbol{T}(t) & = \sum_{\boldsymbol{x}_b}\sum_{q_{\text{f-s}}} (\boldsymbol{x}_b - \boldsymbol{X}_p) \times \boldsymbol{F}^{q_{\text{f-s}}}(\boldsymbol{x}_b,t). \label{eq:MEM_TotTorque}
\end{align}

\subsubsection{PDF reconstruction}
As the momentum exchange method requires an explicit mapping of the object into the fluid domain, the cells regarded as solid cells are not updated in the LBM time step, i.e.\ when executing Eqs.~\eqref{eq:LBM_Collide} and \eqref{eq:LBM_Stream}. 
However, as the particle moves across the computational domain, cells will eventually be converted from solid to fluid and vice versa. 
This requires a consistent reconstruction of the missing PDF information for the cells that returned to the fluid state. 
It has been examined that this refilling can lead to oscillations in the interaction forces and the pressure, e.g.\ in \cite{peng_implementation_2015} where an overview of several refilling schemes is given. 
Following those findings, the normal extrapolation refilling with an explicit enforcing of the velocity constraint is applied here. 
In a first step, the missing PDFs are obtained by extrapolating the PDFs of the neighboring fluid cells in an approximate normal direction $\boldsymbol{c}_{q_n}$ of the particle's surface. 
This direction corresponds to the lattice direction $q_n$ that maximizes $\boldsymbol{n} \cdot \boldsymbol{c}_q$ where $\boldsymbol{n}$ is the local surface normal of the particle. 
Using quadratic extrapolation, the PDFs in a new fluid cell $\boldsymbol{x}_{\text{new}}$ are reconstructed as 
\begin{equation}
f_q(\boldsymbol{x}_{\text{new}},t) =  3f_q(\boldsymbol{x}_{\text{new}}+\boldsymbol{c}_{q_n}\,\Delta t,t) - 3f_q(\boldsymbol{x}_{\text{new}}+2\boldsymbol{c}_{q_n}\,\Delta t,t)+ f_q(\boldsymbol{x}_{\text{new}}+3\boldsymbol{c}_{q_n}\,\Delta t,t). \label{eq:MEM_Reconstruction}
\end{equation}
Secondly, the newly initialized PDFs are transformed into moment space where the local fluid velocity can be set explicitly to match the boundary velocity without affecting the other moments \cite{peng_implementation_2015}. 
This ensures that the no-slip boundary condition is fulfilled and is a key aspect of the refilling. 
If such an extrapolation is not possible, as due to other objects or a boundary nearby, less fluid neighbor cells are included into the extrapolation procedure. 
If no additional fluid cell at all is available in this direction, the refilling is done by setting the PDFs to their equilibrium values $f_q^{\text{eq}}(\bar{\rho}_f, \boldsymbol{v}(\boldsymbol{x}_{\text{new}},t))$ from Eq.~\eqref{eq:LBM_EQ} based on a local average fluid density $\bar{\rho}_f$ and the particle velocity $\boldsymbol{v}$.

\subsubsection{MEM algorithm summary}
An overview of the complete algorithm is displayed in Alg.~\ref{alg:LBM_MEM}. 
Besides the already mentioned parts of the algorithm, it features subcycling loops for the LBM and the rigid body solver part, i.e.\ to execute this part repeatedly within a single global time step. 
For the LBM, it was proposed in \cite{ladd_numerical_1994-1} to use two LBM subcycles to avoid oscillations in the force resulting from the bounce back nature of the boundary treatment.
Since thus two different values for the force are calculated, an averaging step after the LBM subcycling is needed. 
If the bodies are in close contact, as e.g.\ in sediment beds, it might be beneficial to also use subcycling for the rigid body solver to avoid large collision forces because of large overlaps. 
During this type of subcycling, the hydrodynamic forces acting on the particles are kept constant.

\begin{algorithm}
	\begin{algorithmic}
		\STATE Initially map bodies into fluid domain.
		\FOR{each time step $t$}
		\FOR{each LBM subcycle}
		\STATE Perform LBM collision step, Eq.~\eqref{eq:LBM_Collide}, e.g.\ with TRT collision operator Eq.~\eqref{eq:MEM_TRT}.
		\STATE Apply boundary conditions, Eqs.~\eqref{eq:MEM_BB}, \eqref{eq:MEM_CLI} or \eqref{eq:MEM_MR}.
		\STATE Perform LBM stream step, Eq.~\eqref{eq:LBM_Stream}.
		\STATE Calculate hydrodynamic forces on particles, Eq.~\eqref{eq:MEM_Force}.
		\ENDFOR
		\STATE Average forces on particles over LBM subcycles.
		\STATE Obtain total force and torque on particles, Eqs.~\eqref{eq:MEM_TotForce} and \eqref{eq:MEM_TotTorque}.
		\FOR{each rigid body solver subcycle}
		\STATE Perform rigid body solver step (collision detection and resolution, time integration).
		\ENDFOR
		\STATE Update particle mapping and reconstruct PDFs if necessary, Eq.~\eqref{eq:MEM_Reconstruction}.
		\ENDFOR
	\end{algorithmic}
	\caption{LBM algorithm with momentum exchange coupling}
	\label{alg:LBM_MEM}
\end{algorithm}

Summarizing, the momentum exchange coupling algorithm is a convenient and well-established approach for the simulation of particulate flows. 
All parts of the algorithm can be carried out in a strictly local fashion allowing for an efficient implementation for parallel supercomputers. 
The different parts profit directly from new developments in the respective areas like more accurate boundary conditions or more stable collision operators. 
As they are applied in a modular fashion, the user can select different combinations of the components to find the most suitable configuration.
However, in all cases we employ an explicit, binary mapping of the particle into the fluid domain as a key concept. 
This has obvious drawbacks like the necessary PDF reconstruction procedure for newly uncovered cells.
Additionally, the number of solid cells that belong to one particle can vary over time. 
This is one of the reasons why the force acting on a particle as it traverses the computational domain exhibits oscillations that may affect the numerical stability and accuracy of the simulation. 

\subsection{Partially saturated cells method}
\label{sec:PSM}
The partially saturated cells method (PSM), proposed in \cite{noble_lattice-boltzmann_1998}, is also referred to as Noble-Torczynski's method or the immersed moving boundary method.
It differs from the momentum exchange method significantly, as summarized below. 
Instead of an explicit mapping of the object, local solid volume fractions are computed for each computational cell. 
This fill level is then used to modify the LBM collision operator and to evaluate the hydrodynamic forces.

\subsubsection{Modified collision operator}
The PSM collision operator is constructed as a weighted average of the BGK collision operator $\mathcal{C}_q^{\text{BGK}}$, Eq.~\eqref{eq:LBM_SRT}, and a special solid collision operator $\mathcal{C}_q^{\text{solid}}$, i.e. 
\begin{equation}
\mathcal{C}_q^{\text{PSM}}(\boldsymbol{x},t) = (1 - B(\boldsymbol{x},t))\mathcal{C}_q^{\text{BGK}}(\boldsymbol{x},t) + \sum_s B_s(\boldsymbol{x},t) \mathcal{C}_{q,s}^{\text{solid}}(\boldsymbol{x},t). \label{eq:PSM}
\end{equation}
The local weighting factor $B$ is directly related to the cell local solid volume fraction $\varepsilon$. 
Both quantities are summations over the individual contributions from all particles $s$ intersecting with the cell, such that
$B(\boldsymbol{x},t) = \sum_s B_s(\boldsymbol{x},t)$ and $\varepsilon(\boldsymbol{x},t) = \sum_s \varepsilon_s(\boldsymbol{x},t)$.
At least two versions for the calculation of $B_s$ exist \cite{noble_lattice-boltzmann_1998}, given by
\begin{align}
\text{B1: } B_s(\boldsymbol{x},t) &= \varepsilon_s(\boldsymbol{x},t), \label{eq:PSM_B1}\\
\text{B2: } B_s(\boldsymbol{x},t) &= \frac{\varepsilon_s(\boldsymbol{x},t)(\tfrac{\tau}{\Delta t} - \tfrac{1}{2}) }{(1 - \varepsilon_s(\boldsymbol{x},t)) + (\tfrac{\tau}{\Delta t} - \tfrac{1}{2})}, \label{eq:PSM_B2}
\end{align}
respectively. 
Also for $\mathcal{C}_{q,s}^{\text{solid}}$, which is the collision operator only acting when an object $s$ intersects with the cell, different variants have been proposed \cite{noble_lattice-boltzmann_1998,holdych_lattice_2003}:
\begin{align}
\text{M1: } \mathcal{C}_{q,s}^{\text{solid}}(\boldsymbol{x},t) &= \left[f_{\bar{q}}(\boldsymbol{x},t) - f_{\bar{q}}^{\text{eq}}(\rho,\boldsymbol{U})\right] - \left[f_{q}(\boldsymbol{x},t) - f_{q}^{\text{eq}}(\rho,\boldsymbol{v}_s)\right] \label{eq:PSM_M1}\\
\text{M2: } \mathcal{C}_{q,s}^{\text{solid}}(\boldsymbol{x},t) &= \left[f_{q}^{\text{eq}}(\rho,\boldsymbol{v}_s) - f_{q}(\boldsymbol{x},t)\right] + (1- \tfrac{\Delta t}{\tau})\left[f_q(\boldsymbol{x},t) - f_{q}^{\text{eq}}(\rho,\boldsymbol{U})\right] \label{eq:PSM_M2}\\
\text{M3: } \mathcal{C}_{q,s}^{\text{solid}}(\boldsymbol{x},t) &= \left[f_{\bar{q}}(\boldsymbol{x},t) - f_{\bar{q}}^{\text{eq}}(\rho,\boldsymbol{v}_s)\right] - \left[f_{q}(\boldsymbol{x},t) - f_{q}^{\text{eq}}(\rho,\boldsymbol{v}_s)\right] \label{eq:PSM_M3}
\end{align}
Here, the velocity of particle $s$, $\boldsymbol{v}_s = \boldsymbol{v}_s(\boldsymbol{x},t)$ from Eq.~\eqref{eq:bodyVelocity}, is evaluated at the cell center.
The modifications applied in the PSM to the original LBM approach also require to incorporate a weighting in the collision step, now given as
\begin{equation}
\tilde{f}_q(\boldsymbol{x},t) = f_q(\boldsymbol{x},t) + \mathcal{C}_q^{\text{PSM}}(\boldsymbol{x},t) + (1 - B(\boldsymbol{x},t))\mathcal{F}_q(\boldsymbol{x},t),\label{eq:LBM_Collide_PSM}
\end{equation}
and in the calculation of the macroscopic velocity
\begin{equation}
\boldsymbol{u}(\boldsymbol{x},t) =  (1-B(\boldsymbol{x},t))\left(\boldsymbol{U}(\boldsymbol{x},t) + \tfrac{\Delta t}{2\rho_0} \boldsymbol{f}^{\text{ext}}\right) + \sum_s B_s(\boldsymbol{x},t) \boldsymbol{v}_s(\boldsymbol{x},t). \label{eq:MacVel_PSM}
\end{equation} 

Noble and Torczynski suggested to use either (M1) with (B1) or (M2) with (B2) in \cite{noble_lattice-boltzmann_1998}. 
In \cite{holdych_lattice_2003}, (M1) has been modified to become (M3) and \cite{strack_three-dimensional_2007} suggested the usage of (M3) with (B2) for accuracy reasons.

\subsubsection{Solid volume fraction computation}
A crucial and expensive part of the PSM algorithm is the evaluation of the solid volume fraction $\varepsilon_s$ in each cell. 
When the particles move, $\varepsilon_s$ must be recomputed in each time step. 
As summarized in \cite{owen_efficient_2011}, different variants can be used. 
Ideally, an analytical formula exists for the intersection volume of a cubic cell and the particle.
Unfortunately, this is not the case for almost all particle shapes in three dimensions. 
Alternatively, the volume can be approximated by subdividing the cell into smaller sampling points and evaluating whether these points are inside or outside of the object. 
This so called {\em supersampling} method, see e.g. \cite{bartuschat_parallel_2015}, works for any shape for which one can determine whether a point is contained inside.
Since this method is quite efficient and sufficiently accurate when implemented in a recursive fashion, it is used here in the following.
As another approximation, we refer to \cite{galindo-torres_coupled_2013}, where the cell's edges are intersected with the particle and the ratio of covered to total edge length is taken as $\varepsilon_s$. 
As mentioned in \cite{zhou_galilean-invariant_2011}, the partially saturated cells method can also be used if the minimum normalized distance to the particle surface is employed instead of the actual solid volume fraction.

\subsubsection{Fluid-solid interaction force}
\label{sec:PSM_force}
The hydrodynamic force acting on the particle is calculated with the help of $B_s$ and $\mathcal{C}_{q,s}^{\text{solid}}$.
It is applied at the cell center $\boldsymbol{x}_s$ of all cells that intersect with the particle.
By summing up over all lattice directions of all intersecting cells, the total force and torque,
\begin{align}
\boldsymbol{F}(t) &= \tfrac{(\Delta x)^3}{\Delta t} \sum_{\boldsymbol{x}_s} \Big[B_s(\boldsymbol{x}_s,t) \sum_q \left( \mathcal{C}_{q,s}^{\text{solid}}(\boldsymbol{x}_s,t) \boldsymbol{c}_{\bar{q}} \right) \Big], \label{eq:PSM_TotForce}\\
\boldsymbol{T}(t) &= \tfrac{(\Delta x)^3}{\Delta t}\sum_{\boldsymbol{x}_s} \Big[ B_s(\boldsymbol{x}_s,t) (\boldsymbol{x}_s - \boldsymbol{X}_p) \times \sum_q \left( \mathcal{C}_{q,s}^{\text{solid}}(\boldsymbol{x}_s,t) \boldsymbol{c}_{\bar{q}} \right)\Big], \label{eq:PSM_TotTorque}
\end{align}
respectively, can be evaluated.

\subsubsection{PSM algorithm summary}
The partially saturated cells method is summarized in Alg.~\ref{alg:LBM_PSM}. 
The LBM subcycling is omitted here since in our experiments it was not necessary. 
Even though the description of the algorithm might appear shorter than Alg.~\ref{alg:LBM_MEM}, the LBM step here is much more complex than in the momentum exchange case, since it involves the computation of $\mathcal{C}_{q,s}^{\text{solid}}$ and the local force contributions.
Additional complexity is created when several objects intersect with one single cell since this requires to treat the contributions to $B_s$ and $\mathcal{C}_{q,s}^{\text{solid}}$ for each object separately \cite{owen_efficient_2011}.

\begin{algorithm}
	\begin{algorithmic}
		\FOR{each time step $t$}
		\STATE Compute local solid volume fractions in each cell.
		\STATE Perform LBM step, Eqs.~\eqref{eq:LBM_Collide_PSM} and \eqref{eq:LBM_Stream}, with special collision operator from Eq.~\eqref{eq:PSM}.
		\STATE Calculate total force and torque on particles, Eqs.~\eqref{eq:PSM_TotForce} and \eqref{eq:PSM_TotTorque}.
		\FOR{each rigid body solver subcycle}
		\STATE Perform rigid body solver step (collision detection and resolution, time integration).
		\ENDFOR
		\ENDFOR
	\end{algorithmic}
	\caption{LBM algorithm with partially saturated cells coupling}
	\label{alg:LBM_PSM}
\end{algorithm}

The advantage of this coupling method is the smooth variation of the forces acting on the particle when it traverses the domain. 
This originates from the use of the solid volume fraction that varies continuously between cells inside and outside of the particle and thus smoothens the staircase approximation of the momentum exchange method. 
However, as it essentially relies on the BGK collision operator, the simulations suffer from the same inaccuracies and, in particular, exhibit a $\tau$ dependency as shown in Sec.~\ref{sec:DragForce}.
A thorough analysis of the method is still not available since it is characterized by the special solid collision operator and to our knowledge no improvements of its stability and accuracy are yet available in the literature. 
Additionally, the calculation of the solid volume fraction is complex and expensive and may thus consume a major portion of the computing time.

%% file: BenchmarkDragForce.tex
\section{Force on fixed sphere in Stokes flow}
\label{sec:DragForce}
In a short study, we investigate the dependency of the fluid-solid interaction force on a particular choice of the relaxation time $\tau$ and thus the fluid viscosity $\nu$.
For that reason, we evaluate the force on a periodic array of spheres in Stokes flow for which an analytical solution in the form of a series expansion exists \cite{hasimoto_periodic_1959, sangani_slow_1982} .
Such a study has been carried out earlier for different variants of the MEM approach, e.g. in \cite{peng_comparative_2008,bogner_drag_2015,khirevich_coarse-_2015,fattahi_lattice_2016}, but to the best of our knowledge no results are available for the PSM yet.
The simulation setup features a periodic and cubic domain $\Omega$ of size $L \times L \times L$ with a fixed sphere of diameter $D=L/2$ positioned in its center.
This corresponds to the so called \textit{SC} setup.
The flow is driven by an external force density $\boldsymbol{a} = (a,0,0)^\top$ with $a=10^{-5}$ which enters Eq.~\eqref{eq:Forcing} as $\boldsymbol{f}^{\text{ext}}$.
The total force acting on the sphere is the sum of the drag force $\boldsymbol{F}_{d}$, evaluated via Eq.~\eqref{eq:MEM_TotForce} or \eqref{eq:PSM_TotForce}, respectively, and the buoyancy force $\boldsymbol{F}_b = \tfrac{\pi}{6} D^3 \boldsymbol{a}$ due to external forcing \cite{bogner_drag_2015}.
The total dimensionless force in forcing direction is then given as
\begin{equation}
C = \frac{\tfrac{1}{|\boldsymbol{a}|}\boldsymbol{a}^\top(\boldsymbol{F}_d + \boldsymbol{F}_b)}{3 \pi \rho_0 \nu D \bar{u}},
\end{equation}
with the domain average macroscopic velocity 
\begin{equation}
\bar{u} = \frac{\boldsymbol{a}^\top}{|\boldsymbol{a}| L^3} \sum_{\boldsymbol{x}\in\Omega} (\Delta x)^3 \boldsymbol{u}(\boldsymbol{x}).
\end{equation}
The simulation is carried out for $L=32$ and various values of $\nu$ with different coupling methods and is run until $C$ converges.
For the MEM, the TRT collision operator is applied together with either the BB from Eq.~\eqref{eq:MEM_BB}, the CLI from Eq.~\eqref{eq:MEM_CLI}, or the MR scheme from Eq.~\eqref{eq:MEM_MR} for the no-slip boundary condition along the sphere's surface.
We choose $\Lambda_\pm = \tfrac{3}{16}$ which has been shown to yield good flow properties in combination with particles \cite{ginzburg_two-relaxation-time_2008,khirevich_coarse-_2015}.
For the PSM, the commonly used combinations M1B1, M2B2 and M3B2 of Eqs.~\eqref{eq:PSM_B1}--\eqref{eq:PSM_M3} are applied.

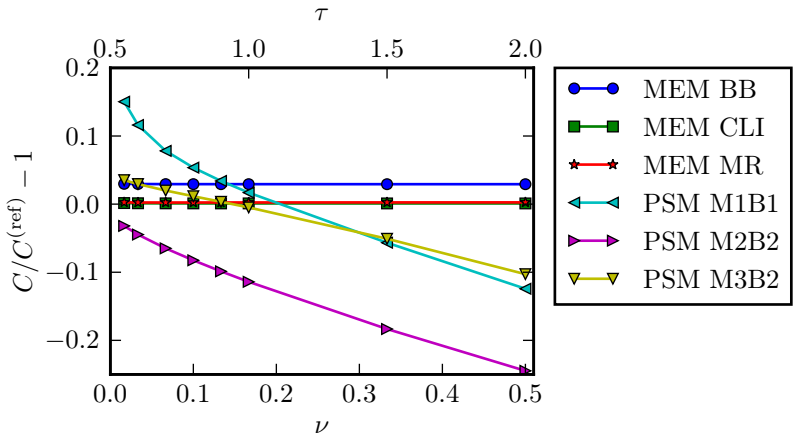
\begin{figure}[t]
	\centering
	\input{plot_tau_DragForce_error.pgf}
	\caption{Relative error of dimensionless force $C$ acting on a periodic array of spheres in Stokes flow for different fluid-solid coupling methods over fluid viscosity $\nu$ and relaxation time $\tau$, respectively.}
	\label{fig:DragForceError}
\end{figure}

The results are shown in Fig.~\ref{fig:DragForceError} as relative error of $C$ to the reference value $C^{\text{(ref)}} = 2.8402$ from \cite{sangani_slow_1982}. 
The MEM variants yield forces that are independent of the particular choice of $\nu$.
The results of CLI and MR are especially accurate whereas BB overestimates the force by less than $5\%$.
In contrast, all PSM variants exhibit a strong dependency on the chosen viscosity and the obtained force gets smaller with increasing $\nu$.
The behavior of M1B1 and M2B2 is similar with a nearly constant offset in between whereas the dependency is less pronounced for M3B2.

This study reveals one general drawback of the PSM which essentially relies on the BGK collision operator and thus cannot yield viscosity independent results.
To the best of our knowledge, no attempts that aim to improve this behavior are reported which is most likely due to the lack of a thorough analysis of the PSM.
As expected, this drawback is not present for the MEM when applying the TRT collision operator with a constant $\Lambda_\pm$ and suitable boundary conditions \cite{ginzburg_two-relaxation-time_2008}.

%% file: plot_tau_DragForce_error.pgf
%% Creator: Matplotlib, PGF backend
%%
%% To include the figure in your LaTeX document, write
%%   \input{<filename>.pgf}
%%
%% Make sure the required packages are loaded in your preamble
%%   \usepackage{pgf}
%%
%% Figures using additional raster images can only be included by \input if
%% they are in the same directory as the main LaTeX file. For loading figures
%% from other directories you can use the `import` package
%%   \usepackage{import}
%% and then include the figures with
%%   \import{<path to file>}{<filename>.pgf}
%%
%% Matplotlib used the following preamble
%%   \usepackage{amsmath}
%%
\begingroup%
\makeatletter%
\begin{pgfpicture}%
\pgfpathrectangle{\pgfpointorigin}{\pgfqpoint{4.264690in}{2.516821in}}%
\pgfusepath{use as bounding box, clip}%
\begin{pgfscope}%
\pgfsetbuttcap%
\pgfsetmiterjoin%
\definecolor{currentfill}{rgb}{1.000000,1.000000,1.000000}%
\pgfsetfillcolor{currentfill}%
\pgfsetlinewidth{0.000000pt}%
\definecolor{currentstroke}{rgb}{1.000000,1.000000,1.000000}%
\pgfsetstrokecolor{currentstroke}%
\pgfsetdash{}{0pt}%
\pgfpathmoveto{\pgfqpoint{0.000000in}{0.000000in}}%
\pgfpathlineto{\pgfqpoint{4.264690in}{0.000000in}}%
\pgfpathlineto{\pgfqpoint{4.264690in}{2.516821in}}%
\pgfpathlineto{\pgfqpoint{0.000000in}{2.516821in}}%
\pgfpathclose%
\pgfusepath{fill}%
\end{pgfscope}%
\begin{pgfscope}%
\pgfsetbuttcap%
\pgfsetmiterjoin%
\definecolor{currentfill}{rgb}{1.000000,1.000000,1.000000}%
\pgfsetfillcolor{currentfill}%
\pgfsetlinewidth{0.000000pt}%
\definecolor{currentstroke}{rgb}{0.000000,0.000000,0.000000}%
\pgfsetstrokecolor{currentstroke}%
\pgfsetstrokeopacity{0.000000}%
\pgfsetdash{}{0pt}%
\pgfpathmoveto{\pgfqpoint{0.636899in}{0.471913in}}%
\pgfpathlineto{\pgfqpoint{2.830149in}{0.471913in}}%
\pgfpathlineto{\pgfqpoint{2.830149in}{2.071913in}}%
\pgfpathlineto{\pgfqpoint{0.636899in}{2.071913in}}%
\pgfpathclose%
\pgfusepath{fill}%
\end{pgfscope}%
\begin{pgfscope}%
\pgfpathrectangle{\pgfqpoint{0.636899in}{0.471913in}}{\pgfqpoint{2.193250in}{1.600000in}} %
\pgfusepath{clip}%
\pgfsetrectcap%
\pgfsetroundjoin%
\pgfsetlinewidth{1.003750pt}%
\definecolor{currentstroke}{rgb}{0.000000,0.000000,1.000000}%
\pgfsetstrokecolor{currentstroke}%
\pgfsetdash{}{0pt}%
\pgfpathmoveto{\pgfqpoint{0.708574in}{1.466777in}}%
\pgfpathlineto{\pgfqpoint{0.780249in}{1.465776in}}%
\pgfpathlineto{\pgfqpoint{0.923599in}{1.465273in}}%
\pgfpathlineto{\pgfqpoint{1.066948in}{1.465118in}}%
\pgfpathlineto{\pgfqpoint{1.210298in}{1.465013in}}%
\pgfpathlineto{\pgfqpoint{1.353648in}{1.464971in}}%
\pgfpathlineto{\pgfqpoint{2.070396in}{1.464880in}}%
\pgfpathlineto{\pgfqpoint{2.787144in}{1.464837in}}%
\pgfusepath{stroke}%
\end{pgfscope}%
\begin{pgfscope}%
\pgfpathrectangle{\pgfqpoint{0.636899in}{0.471913in}}{\pgfqpoint{2.193250in}{1.600000in}} %
\pgfusepath{clip}%
\pgfsetbuttcap%
\pgfsetroundjoin%
\definecolor{currentfill}{rgb}{0.000000,0.000000,1.000000}%
\pgfsetfillcolor{currentfill}%
\pgfsetlinewidth{0.501875pt}%
\definecolor{currentstroke}{rgb}{0.000000,0.000000,0.000000}%
\pgfsetstrokecolor{currentstroke}%
\pgfsetdash{}{0pt}%
\pgfsys@defobject{currentmarker}{\pgfqpoint{-0.027778in}{-0.027778in}}{\pgfqpoint{0.027778in}{0.027778in}}{%
\pgfpathmoveto{\pgfqpoint{0.000000in}{-0.027778in}}%
\pgfpathcurveto{\pgfqpoint{0.007367in}{-0.027778in}}{\pgfqpoint{0.014433in}{-0.024851in}}{\pgfqpoint{0.019642in}{-0.019642in}}%
\pgfpathcurveto{\pgfqpoint{0.024851in}{-0.014433in}}{\pgfqpoint{0.027778in}{-0.007367in}}{\pgfqpoint{0.027778in}{0.000000in}}%
\pgfpathcurveto{\pgfqpoint{0.027778in}{0.007367in}}{\pgfqpoint{0.024851in}{0.014433in}}{\pgfqpoint{0.019642in}{0.019642in}}%
\pgfpathcurveto{\pgfqpoint{0.014433in}{0.024851in}}{\pgfqpoint{0.007367in}{0.027778in}}{\pgfqpoint{0.000000in}{0.027778in}}%
\pgfpathcurveto{\pgfqpoint{-0.007367in}{0.027778in}}{\pgfqpoint{-0.014433in}{0.024851in}}{\pgfqpoint{-0.019642in}{0.019642in}}%
\pgfpathcurveto{\pgfqpoint{-0.024851in}{0.014433in}}{\pgfqpoint{-0.027778in}{0.007367in}}{\pgfqpoint{-0.027778in}{0.000000in}}%
\pgfpathcurveto{\pgfqpoint{-0.027778in}{-0.007367in}}{\pgfqpoint{-0.024851in}{-0.014433in}}{\pgfqpoint{-0.019642in}{-0.019642in}}%
\pgfpathcurveto{\pgfqpoint{-0.014433in}{-0.024851in}}{\pgfqpoint{-0.007367in}{-0.027778in}}{\pgfqpoint{0.000000in}{-0.027778in}}%
\pgfpathclose%
\pgfusepath{stroke,fill}%
}%
\begin{pgfscope}%
\pgfsys@transformshift{0.708574in}{1.466777in}%
\pgfsys@useobject{currentmarker}{}%
\end{pgfscope}%
\begin{pgfscope}%
\pgfsys@transformshift{0.780249in}{1.465776in}%
\pgfsys@useobject{currentmarker}{}%
\end{pgfscope}%
\begin{pgfscope}%
\pgfsys@transformshift{0.923599in}{1.465273in}%
\pgfsys@useobject{currentmarker}{}%
\end{pgfscope}%
\begin{pgfscope}%
\pgfsys@transformshift{1.066948in}{1.465118in}%
\pgfsys@useobject{currentmarker}{}%
\end{pgfscope}%
\begin{pgfscope}%
\pgfsys@transformshift{1.210298in}{1.465013in}%
\pgfsys@useobject{currentmarker}{}%
\end{pgfscope}%
\begin{pgfscope}%
\pgfsys@transformshift{1.353648in}{1.464971in}%
\pgfsys@useobject{currentmarker}{}%
\end{pgfscope}%
\begin{pgfscope}%
\pgfsys@transformshift{2.070396in}{1.464880in}%
\pgfsys@useobject{currentmarker}{}%
\end{pgfscope}%
\begin{pgfscope}%
\pgfsys@transformshift{2.787144in}{1.464837in}%
\pgfsys@useobject{currentmarker}{}%
\end{pgfscope}%
\end{pgfscope}%
\begin{pgfscope}%
\pgfpathrectangle{\pgfqpoint{0.636899in}{0.471913in}}{\pgfqpoint{2.193250in}{1.600000in}} %
\pgfusepath{clip}%
\pgfsetrectcap%
\pgfsetroundjoin%
\pgfsetlinewidth{1.003750pt}%
\definecolor{currentstroke}{rgb}{0.000000,0.500000,0.000000}%
\pgfsetstrokecolor{currentstroke}%
\pgfsetdash{}{0pt}%
\pgfpathmoveto{\pgfqpoint{0.708574in}{1.366376in}}%
\pgfpathlineto{\pgfqpoint{0.780249in}{1.365374in}}%
\pgfpathlineto{\pgfqpoint{0.923599in}{1.364864in}}%
\pgfpathlineto{\pgfqpoint{1.066948in}{1.364693in}}%
\pgfpathlineto{\pgfqpoint{1.210298in}{1.364613in}}%
\pgfpathlineto{\pgfqpoint{1.353648in}{1.364566in}}%
\pgfpathlineto{\pgfqpoint{2.070396in}{1.364467in}}%
\pgfpathlineto{\pgfqpoint{2.787144in}{1.364444in}}%
\pgfusepath{stroke}%
\end{pgfscope}%
\begin{pgfscope}%
\pgfpathrectangle{\pgfqpoint{0.636899in}{0.471913in}}{\pgfqpoint{2.193250in}{1.600000in}} %
\pgfusepath{clip}%
\pgfsetbuttcap%
\pgfsetmiterjoin%
\definecolor{currentfill}{rgb}{0.000000,0.500000,0.000000}%
\pgfsetfillcolor{currentfill}%
\pgfsetlinewidth{0.501875pt}%
\definecolor{currentstroke}{rgb}{0.000000,0.000000,0.000000}%
\pgfsetstrokecolor{currentstroke}%
\pgfsetdash{}{0pt}%
\pgfsys@defobject{currentmarker}{\pgfqpoint{-0.027778in}{-0.027778in}}{\pgfqpoint{0.027778in}{0.027778in}}{%
\pgfpathmoveto{\pgfqpoint{-0.027778in}{-0.027778in}}%
\pgfpathlineto{\pgfqpoint{0.027778in}{-0.027778in}}%
\pgfpathlineto{\pgfqpoint{0.027778in}{0.027778in}}%
\pgfpathlineto{\pgfqpoint{-0.027778in}{0.027778in}}%
\pgfpathclose%
\pgfusepath{stroke,fill}%
}%
\begin{pgfscope}%
\pgfsys@transformshift{0.708574in}{1.366376in}%
\pgfsys@useobject{currentmarker}{}%
\end{pgfscope}%
\begin{pgfscope}%
\pgfsys@transformshift{0.780249in}{1.365374in}%
\pgfsys@useobject{currentmarker}{}%
\end{pgfscope}%
\begin{pgfscope}%
\pgfsys@transformshift{0.923599in}{1.364864in}%
\pgfsys@useobject{currentmarker}{}%
\end{pgfscope}%
\begin{pgfscope}%
\pgfsys@transformshift{1.066948in}{1.364693in}%
\pgfsys@useobject{currentmarker}{}%
\end{pgfscope}%
\begin{pgfscope}%
\pgfsys@transformshift{1.210298in}{1.364613in}%
\pgfsys@useobject{currentmarker}{}%
\end{pgfscope}%
\begin{pgfscope}%
\pgfsys@transformshift{1.353648in}{1.364566in}%
\pgfsys@useobject{currentmarker}{}%
\end{pgfscope}%
\begin{pgfscope}%
\pgfsys@transformshift{2.070396in}{1.364467in}%
\pgfsys@useobject{currentmarker}{}%
\end{pgfscope}%
\begin{pgfscope}%
\pgfsys@transformshift{2.787144in}{1.364444in}%
\pgfsys@useobject{currentmarker}{}%
\end{pgfscope}%
\end{pgfscope}%
\begin{pgfscope}%
\pgfpathrectangle{\pgfqpoint{0.636899in}{0.471913in}}{\pgfqpoint{2.193250in}{1.600000in}} %
\pgfusepath{clip}%
\pgfsetrectcap%
\pgfsetroundjoin%
\pgfsetlinewidth{1.003750pt}%
\definecolor{currentstroke}{rgb}{1.000000,0.000000,0.000000}%
\pgfsetstrokecolor{currentstroke}%
\pgfsetdash{}{0pt}%
\pgfpathmoveto{\pgfqpoint{0.708574in}{1.370172in}}%
\pgfpathlineto{\pgfqpoint{0.780249in}{1.369409in}}%
\pgfpathlineto{\pgfqpoint{0.923599in}{1.369249in}}%
\pgfpathlineto{\pgfqpoint{1.066948in}{1.369324in}}%
\pgfpathlineto{\pgfqpoint{1.210298in}{1.369425in}}%
\pgfpathlineto{\pgfqpoint{1.353648in}{1.369519in}}%
\pgfpathlineto{\pgfqpoint{2.070396in}{1.369815in}}%
\pgfpathlineto{\pgfqpoint{2.787144in}{1.369977in}}%
\pgfusepath{stroke}%
\end{pgfscope}%
\begin{pgfscope}%
\pgfpathrectangle{\pgfqpoint{0.636899in}{0.471913in}}{\pgfqpoint{2.193250in}{1.600000in}} %
\pgfusepath{clip}%
\pgfsetbuttcap%
\pgfsetbeveljoin%
\definecolor{currentfill}{rgb}{1.000000,0.000000,0.000000}%
\pgfsetfillcolor{currentfill}%
\pgfsetlinewidth{0.501875pt}%
\definecolor{currentstroke}{rgb}{0.000000,0.000000,0.000000}%
\pgfsetstrokecolor{currentstroke}%
\pgfsetdash{}{0pt}%
\pgfsys@defobject{currentmarker}{\pgfqpoint{-0.026418in}{-0.022473in}}{\pgfqpoint{0.026418in}{0.027778in}}{%
\pgfpathmoveto{\pgfqpoint{0.000000in}{0.027778in}}%
\pgfpathlineto{\pgfqpoint{-0.006236in}{0.008584in}}%
\pgfpathlineto{\pgfqpoint{-0.026418in}{0.008584in}}%
\pgfpathlineto{\pgfqpoint{-0.010091in}{-0.003279in}}%
\pgfpathlineto{\pgfqpoint{-0.016327in}{-0.022473in}}%
\pgfpathlineto{\pgfqpoint{-0.000000in}{-0.010610in}}%
\pgfpathlineto{\pgfqpoint{0.016327in}{-0.022473in}}%
\pgfpathlineto{\pgfqpoint{0.010091in}{-0.003279in}}%
\pgfpathlineto{\pgfqpoint{0.026418in}{0.008584in}}%
\pgfpathlineto{\pgfqpoint{0.006236in}{0.008584in}}%
\pgfpathclose%
\pgfusepath{stroke,fill}%
}%
\begin{pgfscope}%
\pgfsys@transformshift{0.708574in}{1.370172in}%
\pgfsys@useobject{currentmarker}{}%
\end{pgfscope}%
\begin{pgfscope}%
\pgfsys@transformshift{0.780249in}{1.369409in}%
\pgfsys@useobject{currentmarker}{}%
\end{pgfscope}%
\begin{pgfscope}%
\pgfsys@transformshift{0.923599in}{1.369249in}%
\pgfsys@useobject{currentmarker}{}%
\end{pgfscope}%
\begin{pgfscope}%
\pgfsys@transformshift{1.066948in}{1.369324in}%
\pgfsys@useobject{currentmarker}{}%
\end{pgfscope}%
\begin{pgfscope}%
\pgfsys@transformshift{1.210298in}{1.369425in}%
\pgfsys@useobject{currentmarker}{}%
\end{pgfscope}%
\begin{pgfscope}%
\pgfsys@transformshift{1.353648in}{1.369519in}%
\pgfsys@useobject{currentmarker}{}%
\end{pgfscope}%
\begin{pgfscope}%
\pgfsys@transformshift{2.070396in}{1.369815in}%
\pgfsys@useobject{currentmarker}{}%
\end{pgfscope}%
\begin{pgfscope}%
\pgfsys@transformshift{2.787144in}{1.369977in}%
\pgfsys@useobject{currentmarker}{}%
\end{pgfscope}%
\end{pgfscope}%
\begin{pgfscope}%
\pgfpathrectangle{\pgfqpoint{0.636899in}{0.471913in}}{\pgfqpoint{2.193250in}{1.600000in}} %
\pgfusepath{clip}%
\pgfsetrectcap%
\pgfsetroundjoin%
\pgfsetlinewidth{1.003750pt}%
\definecolor{currentstroke}{rgb}{0.000000,0.750000,0.750000}%
\pgfsetstrokecolor{currentstroke}%
\pgfsetdash{}{0pt}%
\pgfpathmoveto{\pgfqpoint{0.708574in}{1.895405in}}%
\pgfpathlineto{\pgfqpoint{0.780249in}{1.773762in}}%
\pgfpathlineto{\pgfqpoint{0.923599in}{1.638803in}}%
\pgfpathlineto{\pgfqpoint{1.066948in}{1.551246in}}%
\pgfpathlineto{\pgfqpoint{1.210298in}{1.481743in}}%
\pgfpathlineto{\pgfqpoint{1.353648in}{1.420880in}}%
\pgfpathlineto{\pgfqpoint{2.070396in}{1.159289in}}%
\pgfpathlineto{\pgfqpoint{2.787144in}{0.919088in}}%
\pgfusepath{stroke}%
\end{pgfscope}%
\begin{pgfscope}%
\pgfpathrectangle{\pgfqpoint{0.636899in}{0.471913in}}{\pgfqpoint{2.193250in}{1.600000in}} %
\pgfusepath{clip}%
\pgfsetbuttcap%
\pgfsetmiterjoin%
\definecolor{currentfill}{rgb}{0.000000,0.750000,0.750000}%
\pgfsetfillcolor{currentfill}%
\pgfsetlinewidth{0.501875pt}%
\definecolor{currentstroke}{rgb}{0.000000,0.000000,0.000000}%
\pgfsetstrokecolor{currentstroke}%
\pgfsetdash{}{0pt}%
\pgfsys@defobject{currentmarker}{\pgfqpoint{-0.027778in}{-0.027778in}}{\pgfqpoint{0.027778in}{0.027778in}}{%
\pgfpathmoveto{\pgfqpoint{-0.027778in}{0.000000in}}%
\pgfpathlineto{\pgfqpoint{0.027778in}{-0.027778in}}%
\pgfpathlineto{\pgfqpoint{0.027778in}{0.027778in}}%
\pgfpathclose%
\pgfusepath{stroke,fill}%
}%
\begin{pgfscope}%
\pgfsys@transformshift{0.708574in}{1.895405in}%
\pgfsys@useobject{currentmarker}{}%
\end{pgfscope}%
\begin{pgfscope}%
\pgfsys@transformshift{0.780249in}{1.773762in}%
\pgfsys@useobject{currentmarker}{}%
\end{pgfscope}%
\begin{pgfscope}%
\pgfsys@transformshift{0.923599in}{1.638803in}%
\pgfsys@useobject{currentmarker}{}%
\end{pgfscope}%
\begin{pgfscope}%
\pgfsys@transformshift{1.066948in}{1.551246in}%
\pgfsys@useobject{currentmarker}{}%
\end{pgfscope}%
\begin{pgfscope}%
\pgfsys@transformshift{1.210298in}{1.481743in}%
\pgfsys@useobject{currentmarker}{}%
\end{pgfscope}%
\begin{pgfscope}%
\pgfsys@transformshift{1.353648in}{1.420880in}%
\pgfsys@useobject{currentmarker}{}%
\end{pgfscope}%
\begin{pgfscope}%
\pgfsys@transformshift{2.070396in}{1.159289in}%
\pgfsys@useobject{currentmarker}{}%
\end{pgfscope}%
\begin{pgfscope}%
\pgfsys@transformshift{2.787144in}{0.919088in}%
\pgfsys@useobject{currentmarker}{}%
\end{pgfscope}%
\end{pgfscope}%
\begin{pgfscope}%
\pgfpathrectangle{\pgfqpoint{0.636899in}{0.471913in}}{\pgfqpoint{2.193250in}{1.600000in}} %
\pgfusepath{clip}%
\pgfsetrectcap%
\pgfsetroundjoin%
\pgfsetlinewidth{1.003750pt}%
\definecolor{currentstroke}{rgb}{0.750000,0.000000,0.750000}%
\pgfsetstrokecolor{currentstroke}%
\pgfsetdash{}{0pt}%
\pgfpathmoveto{\pgfqpoint{0.708574in}{1.247087in}}%
\pgfpathlineto{\pgfqpoint{0.780249in}{1.201929in}}%
\pgfpathlineto{\pgfqpoint{0.923599in}{1.129924in}}%
\pgfpathlineto{\pgfqpoint{1.066948in}{1.067569in}}%
\pgfpathlineto{\pgfqpoint{1.210298in}{1.009929in}}%
\pgfpathlineto{\pgfqpoint{1.353648in}{0.955304in}}%
\pgfpathlineto{\pgfqpoint{2.070396in}{0.708908in}}%
\pgfpathlineto{\pgfqpoint{2.787144in}{0.490626in}}%
\pgfusepath{stroke}%
\end{pgfscope}%
\begin{pgfscope}%
\pgfpathrectangle{\pgfqpoint{0.636899in}{0.471913in}}{\pgfqpoint{2.193250in}{1.600000in}} %
\pgfusepath{clip}%
\pgfsetbuttcap%
\pgfsetmiterjoin%
\definecolor{currentfill}{rgb}{0.750000,0.000000,0.750000}%
\pgfsetfillcolor{currentfill}%
\pgfsetlinewidth{0.501875pt}%
\definecolor{currentstroke}{rgb}{0.000000,0.000000,0.000000}%
\pgfsetstrokecolor{currentstroke}%
\pgfsetdash{}{0pt}%
\pgfsys@defobject{currentmarker}{\pgfqpoint{-0.027778in}{-0.027778in}}{\pgfqpoint{0.027778in}{0.027778in}}{%
\pgfpathmoveto{\pgfqpoint{0.027778in}{-0.000000in}}%
\pgfpathlineto{\pgfqpoint{-0.027778in}{0.027778in}}%
\pgfpathlineto{\pgfqpoint{-0.027778in}{-0.027778in}}%
\pgfpathclose%
\pgfusepath{stroke,fill}%
}%
\begin{pgfscope}%
\pgfsys@transformshift{0.708574in}{1.247087in}%
\pgfsys@useobject{currentmarker}{}%
\end{pgfscope}%
\begin{pgfscope}%
\pgfsys@transformshift{0.780249in}{1.201929in}%
\pgfsys@useobject{currentmarker}{}%
\end{pgfscope}%
\begin{pgfscope}%
\pgfsys@transformshift{0.923599in}{1.129924in}%
\pgfsys@useobject{currentmarker}{}%
\end{pgfscope}%
\begin{pgfscope}%
\pgfsys@transformshift{1.066948in}{1.067569in}%
\pgfsys@useobject{currentmarker}{}%
\end{pgfscope}%
\begin{pgfscope}%
\pgfsys@transformshift{1.210298in}{1.009929in}%
\pgfsys@useobject{currentmarker}{}%
\end{pgfscope}%
\begin{pgfscope}%
\pgfsys@transformshift{1.353648in}{0.955304in}%
\pgfsys@useobject{currentmarker}{}%
\end{pgfscope}%
\begin{pgfscope}%
\pgfsys@transformshift{2.070396in}{0.708908in}%
\pgfsys@useobject{currentmarker}{}%
\end{pgfscope}%
\begin{pgfscope}%
\pgfsys@transformshift{2.787144in}{0.490626in}%
\pgfsys@useobject{currentmarker}{}%
\end{pgfscope}%
\end{pgfscope}%
\begin{pgfscope}%
\pgfpathrectangle{\pgfqpoint{0.636899in}{0.471913in}}{\pgfqpoint{2.193250in}{1.600000in}} %
\pgfusepath{clip}%
\pgfsetrectcap%
\pgfsetroundjoin%
\pgfsetlinewidth{1.003750pt}%
\definecolor{currentstroke}{rgb}{0.750000,0.750000,0.000000}%
\pgfsetstrokecolor{currentstroke}%
\pgfsetdash{}{0pt}%
\pgfpathmoveto{\pgfqpoint{0.708574in}{1.486000in}}%
\pgfpathlineto{\pgfqpoint{0.780249in}{1.465546in}}%
\pgfpathlineto{\pgfqpoint{0.923599in}{1.431739in}}%
\pgfpathlineto{\pgfqpoint{1.066948in}{1.401961in}}%
\pgfpathlineto{\pgfqpoint{1.210298in}{1.373006in}}%
\pgfpathlineto{\pgfqpoint{1.353648in}{1.343549in}}%
\pgfpathlineto{\pgfqpoint{2.070396in}{1.179994in}}%
\pgfpathlineto{\pgfqpoint{2.787144in}{0.995153in}}%
\pgfusepath{stroke}%
\end{pgfscope}%
\begin{pgfscope}%
\pgfpathrectangle{\pgfqpoint{0.636899in}{0.471913in}}{\pgfqpoint{2.193250in}{1.600000in}} %
\pgfusepath{clip}%
\pgfsetbuttcap%
\pgfsetmiterjoin%
\definecolor{currentfill}{rgb}{0.750000,0.750000,0.000000}%
\pgfsetfillcolor{currentfill}%
\pgfsetlinewidth{0.501875pt}%
\definecolor{currentstroke}{rgb}{0.000000,0.000000,0.000000}%
\pgfsetstrokecolor{currentstroke}%
\pgfsetdash{}{0pt}%
\pgfsys@defobject{currentmarker}{\pgfqpoint{-0.027778in}{-0.027778in}}{\pgfqpoint{0.027778in}{0.027778in}}{%
\pgfpathmoveto{\pgfqpoint{-0.000000in}{-0.027778in}}%
\pgfpathlineto{\pgfqpoint{0.027778in}{0.027778in}}%
\pgfpathlineto{\pgfqpoint{-0.027778in}{0.027778in}}%
\pgfpathclose%
\pgfusepath{stroke,fill}%
}%
\begin{pgfscope}%
\pgfsys@transformshift{0.708574in}{1.486000in}%
\pgfsys@useobject{currentmarker}{}%
\end{pgfscope}%
\begin{pgfscope}%
\pgfsys@transformshift{0.780249in}{1.465546in}%
\pgfsys@useobject{currentmarker}{}%
\end{pgfscope}%
\begin{pgfscope}%
\pgfsys@transformshift{0.923599in}{1.431739in}%
\pgfsys@useobject{currentmarker}{}%
\end{pgfscope}%
\begin{pgfscope}%
\pgfsys@transformshift{1.066948in}{1.401961in}%
\pgfsys@useobject{currentmarker}{}%
\end{pgfscope}%
\begin{pgfscope}%
\pgfsys@transformshift{1.210298in}{1.373006in}%
\pgfsys@useobject{currentmarker}{}%
\end{pgfscope}%
\begin{pgfscope}%
\pgfsys@transformshift{1.353648in}{1.343549in}%
\pgfsys@useobject{currentmarker}{}%
\end{pgfscope}%
\begin{pgfscope}%
\pgfsys@transformshift{2.070396in}{1.179994in}%
\pgfsys@useobject{currentmarker}{}%
\end{pgfscope}%
\begin{pgfscope}%
\pgfsys@transformshift{2.787144in}{0.995153in}%
\pgfsys@useobject{currentmarker}{}%
\end{pgfscope}%
\end{pgfscope}%
\begin{pgfscope}%
\pgfsetrectcap%
\pgfsetmiterjoin%
\pgfsetlinewidth{1.003750pt}%
\definecolor{currentstroke}{rgb}{0.000000,0.000000,0.000000}%
\pgfsetstrokecolor{currentstroke}%
\pgfsetdash{}{0pt}%
\pgfpathmoveto{\pgfqpoint{0.636899in}{2.071913in}}%
\pgfpathlineto{\pgfqpoint{2.830149in}{2.071913in}}%
\pgfusepath{stroke}%
\end{pgfscope}%
\begin{pgfscope}%
\pgfsetrectcap%
\pgfsetmiterjoin%
\pgfsetlinewidth{1.003750pt}%
\definecolor{currentstroke}{rgb}{0.000000,0.000000,0.000000}%
\pgfsetstrokecolor{currentstroke}%
\pgfsetdash{}{0pt}%
\pgfpathmoveto{\pgfqpoint{2.830149in}{0.471913in}}%
\pgfpathlineto{\pgfqpoint{2.830149in}{2.071913in}}%
\pgfusepath{stroke}%
\end{pgfscope}%
\begin{pgfscope}%
\pgfsetrectcap%
\pgfsetmiterjoin%
\pgfsetlinewidth{1.003750pt}%
\definecolor{currentstroke}{rgb}{0.000000,0.000000,0.000000}%
\pgfsetstrokecolor{currentstroke}%
\pgfsetdash{}{0pt}%
\pgfpathmoveto{\pgfqpoint{0.636899in}{0.471913in}}%
\pgfpathlineto{\pgfqpoint{2.830149in}{0.471913in}}%
\pgfusepath{stroke}%
\end{pgfscope}%
\begin{pgfscope}%
\pgfsetrectcap%
\pgfsetmiterjoin%
\pgfsetlinewidth{1.003750pt}%
\definecolor{currentstroke}{rgb}{0.000000,0.000000,0.000000}%
\pgfsetstrokecolor{currentstroke}%
\pgfsetdash{}{0pt}%
\pgfpathmoveto{\pgfqpoint{0.636899in}{0.471913in}}%
\pgfpathlineto{\pgfqpoint{0.636899in}{2.071913in}}%
\pgfusepath{stroke}%
\end{pgfscope}%
\begin{pgfscope}%
\pgfsetbuttcap%
\pgfsetroundjoin%
\definecolor{currentfill}{rgb}{0.000000,0.000000,0.000000}%
\pgfsetfillcolor{currentfill}%
\pgfsetlinewidth{0.501875pt}%
\definecolor{currentstroke}{rgb}{0.000000,0.000000,0.000000}%
\pgfsetstrokecolor{currentstroke}%
\pgfsetdash{}{0pt}%
\pgfsys@defobject{currentmarker}{\pgfqpoint{0.000000in}{0.000000in}}{\pgfqpoint{0.000000in}{0.055556in}}{%
\pgfpathmoveto{\pgfqpoint{0.000000in}{0.000000in}}%
\pgfpathlineto{\pgfqpoint{0.000000in}{0.055556in}}%
\pgfusepath{stroke,fill}%
}%
\begin{pgfscope}%
\pgfsys@transformshift{0.636899in}{0.471913in}%
\pgfsys@useobject{currentmarker}{}%
\end{pgfscope}%
\end{pgfscope}%
\begin{pgfscope}%
\pgftext[x=0.636899in,y=0.416358in,,top]{\rmfamily\fontsize{10.300000}{12.360000}\selectfont 0.0}%
\end{pgfscope}%
\begin{pgfscope}%
\pgfsetbuttcap%
\pgfsetroundjoin%
\definecolor{currentfill}{rgb}{0.000000,0.000000,0.000000}%
\pgfsetfillcolor{currentfill}%
\pgfsetlinewidth{0.501875pt}%
\definecolor{currentstroke}{rgb}{0.000000,0.000000,0.000000}%
\pgfsetstrokecolor{currentstroke}%
\pgfsetdash{}{0pt}%
\pgfsys@defobject{currentmarker}{\pgfqpoint{0.000000in}{0.000000in}}{\pgfqpoint{0.000000in}{0.055556in}}{%
\pgfpathmoveto{\pgfqpoint{0.000000in}{0.000000in}}%
\pgfpathlineto{\pgfqpoint{0.000000in}{0.055556in}}%
\pgfusepath{stroke,fill}%
}%
\begin{pgfscope}%
\pgfsys@transformshift{1.066948in}{0.471913in}%
\pgfsys@useobject{currentmarker}{}%
\end{pgfscope}%
\end{pgfscope}%
\begin{pgfscope}%
\pgftext[x=1.066948in,y=0.416358in,,top]{\rmfamily\fontsize{10.300000}{12.360000}\selectfont 0.1}%
\end{pgfscope}%
\begin{pgfscope}%
\pgfsetbuttcap%
\pgfsetroundjoin%
\definecolor{currentfill}{rgb}{0.000000,0.000000,0.000000}%
\pgfsetfillcolor{currentfill}%
\pgfsetlinewidth{0.501875pt}%
\definecolor{currentstroke}{rgb}{0.000000,0.000000,0.000000}%
\pgfsetstrokecolor{currentstroke}%
\pgfsetdash{}{0pt}%
\pgfsys@defobject{currentmarker}{\pgfqpoint{0.000000in}{0.000000in}}{\pgfqpoint{0.000000in}{0.055556in}}{%
\pgfpathmoveto{\pgfqpoint{0.000000in}{0.000000in}}%
\pgfpathlineto{\pgfqpoint{0.000000in}{0.055556in}}%
\pgfusepath{stroke,fill}%
}%
\begin{pgfscope}%
\pgfsys@transformshift{1.496997in}{0.471913in}%
\pgfsys@useobject{currentmarker}{}%
\end{pgfscope}%
\end{pgfscope}%
\begin{pgfscope}%
\pgftext[x=1.496997in,y=0.416358in,,top]{\rmfamily\fontsize{10.300000}{12.360000}\selectfont 0.2}%
\end{pgfscope}%
\begin{pgfscope}%
\pgfsetbuttcap%
\pgfsetroundjoin%
\definecolor{currentfill}{rgb}{0.000000,0.000000,0.000000}%
\pgfsetfillcolor{currentfill}%
\pgfsetlinewidth{0.501875pt}%
\definecolor{currentstroke}{rgb}{0.000000,0.000000,0.000000}%
\pgfsetstrokecolor{currentstroke}%
\pgfsetdash{}{0pt}%
\pgfsys@defobject{currentmarker}{\pgfqpoint{0.000000in}{0.000000in}}{\pgfqpoint{0.000000in}{0.055556in}}{%
\pgfpathmoveto{\pgfqpoint{0.000000in}{0.000000in}}%
\pgfpathlineto{\pgfqpoint{0.000000in}{0.055556in}}%
\pgfusepath{stroke,fill}%
}%
\begin{pgfscope}%
\pgfsys@transformshift{1.927046in}{0.471913in}%
\pgfsys@useobject{currentmarker}{}%
\end{pgfscope}%
\end{pgfscope}%
\begin{pgfscope}%
\pgftext[x=1.927046in,y=0.416358in,,top]{\rmfamily\fontsize{10.300000}{12.360000}\selectfont 0.3}%
\end{pgfscope}%
\begin{pgfscope}%
\pgfsetbuttcap%
\pgfsetroundjoin%
\definecolor{currentfill}{rgb}{0.000000,0.000000,0.000000}%
\pgfsetfillcolor{currentfill}%
\pgfsetlinewidth{0.501875pt}%
\definecolor{currentstroke}{rgb}{0.000000,0.000000,0.000000}%
\pgfsetstrokecolor{currentstroke}%
\pgfsetdash{}{0pt}%
\pgfsys@defobject{currentmarker}{\pgfqpoint{0.000000in}{0.000000in}}{\pgfqpoint{0.000000in}{0.055556in}}{%
\pgfpathmoveto{\pgfqpoint{0.000000in}{0.000000in}}%
\pgfpathlineto{\pgfqpoint{0.000000in}{0.055556in}}%
\pgfusepath{stroke,fill}%
}%
\begin{pgfscope}%
\pgfsys@transformshift{2.357095in}{0.471913in}%
\pgfsys@useobject{currentmarker}{}%
\end{pgfscope}%
\end{pgfscope}%
\begin{pgfscope}%
\pgftext[x=2.357095in,y=0.416358in,,top]{\rmfamily\fontsize{10.300000}{12.360000}\selectfont 0.4}%
\end{pgfscope}%
\begin{pgfscope}%
\pgfsetbuttcap%
\pgfsetroundjoin%
\definecolor{currentfill}{rgb}{0.000000,0.000000,0.000000}%
\pgfsetfillcolor{currentfill}%
\pgfsetlinewidth{0.501875pt}%
\definecolor{currentstroke}{rgb}{0.000000,0.000000,0.000000}%
\pgfsetstrokecolor{currentstroke}%
\pgfsetdash{}{0pt}%
\pgfsys@defobject{currentmarker}{\pgfqpoint{0.000000in}{0.000000in}}{\pgfqpoint{0.000000in}{0.055556in}}{%
\pgfpathmoveto{\pgfqpoint{0.000000in}{0.000000in}}%
\pgfpathlineto{\pgfqpoint{0.000000in}{0.055556in}}%
\pgfusepath{stroke,fill}%
}%
\begin{pgfscope}%
\pgfsys@transformshift{2.787144in}{0.471913in}%
\pgfsys@useobject{currentmarker}{}%
\end{pgfscope}%
\end{pgfscope}%
\begin{pgfscope}%
\pgftext[x=2.787144in,y=0.416358in,,top]{\rmfamily\fontsize{10.300000}{12.360000}\selectfont 0.5}%
\end{pgfscope}%
\begin{pgfscope}%
\pgftext[x=1.733524in,y=0.223457in,,top]{\rmfamily\fontsize{10.300000}{12.360000}\selectfont \(\displaystyle \nu\)}%
\end{pgfscope}%
\begin{pgfscope}%
\pgfsetbuttcap%
\pgfsetroundjoin%
\definecolor{currentfill}{rgb}{0.000000,0.000000,0.000000}%
\pgfsetfillcolor{currentfill}%
\pgfsetlinewidth{0.501875pt}%
\definecolor{currentstroke}{rgb}{0.000000,0.000000,0.000000}%
\pgfsetstrokecolor{currentstroke}%
\pgfsetdash{}{0pt}%
\pgfsys@defobject{currentmarker}{\pgfqpoint{0.000000in}{0.000000in}}{\pgfqpoint{0.055556in}{0.000000in}}{%
\pgfpathmoveto{\pgfqpoint{0.000000in}{0.000000in}}%
\pgfpathlineto{\pgfqpoint{0.055556in}{0.000000in}}%
\pgfusepath{stroke,fill}%
}%
\begin{pgfscope}%
\pgfsys@transformshift{0.636899in}{0.649691in}%
\pgfsys@useobject{currentmarker}{}%
\end{pgfscope}%
\end{pgfscope}%
\begin{pgfscope}%
\pgfsetbuttcap%
\pgfsetroundjoin%
\definecolor{currentfill}{rgb}{0.000000,0.000000,0.000000}%
\pgfsetfillcolor{currentfill}%
\pgfsetlinewidth{0.501875pt}%
\definecolor{currentstroke}{rgb}{0.000000,0.000000,0.000000}%
\pgfsetstrokecolor{currentstroke}%
\pgfsetdash{}{0pt}%
\pgfsys@defobject{currentmarker}{\pgfqpoint{-0.055556in}{0.000000in}}{\pgfqpoint{0.000000in}{0.000000in}}{%
\pgfpathmoveto{\pgfqpoint{0.000000in}{0.000000in}}%
\pgfpathlineto{\pgfqpoint{-0.055556in}{0.000000in}}%
\pgfusepath{stroke,fill}%
}%
\begin{pgfscope}%
\pgfsys@transformshift{2.830149in}{0.649691in}%
\pgfsys@useobject{currentmarker}{}%
\end{pgfscope}%
\end{pgfscope}%
\begin{pgfscope}%
\pgftext[x=0.581344in,y=0.649691in,right,]{\rmfamily\fontsize{10.300000}{12.360000}\selectfont −0.2}%
\end{pgfscope}%
\begin{pgfscope}%
\pgfsetbuttcap%
\pgfsetroundjoin%
\definecolor{currentfill}{rgb}{0.000000,0.000000,0.000000}%
\pgfsetfillcolor{currentfill}%
\pgfsetlinewidth{0.501875pt}%
\definecolor{currentstroke}{rgb}{0.000000,0.000000,0.000000}%
\pgfsetstrokecolor{currentstroke}%
\pgfsetdash{}{0pt}%
\pgfsys@defobject{currentmarker}{\pgfqpoint{0.000000in}{0.000000in}}{\pgfqpoint{0.055556in}{0.000000in}}{%
\pgfpathmoveto{\pgfqpoint{0.000000in}{0.000000in}}%
\pgfpathlineto{\pgfqpoint{0.055556in}{0.000000in}}%
\pgfusepath{stroke,fill}%
}%
\begin{pgfscope}%
\pgfsys@transformshift{0.636899in}{1.005247in}%
\pgfsys@useobject{currentmarker}{}%
\end{pgfscope}%
\end{pgfscope}%
\begin{pgfscope}%
\pgfsetbuttcap%
\pgfsetroundjoin%
\definecolor{currentfill}{rgb}{0.000000,0.000000,0.000000}%
\pgfsetfillcolor{currentfill}%
\pgfsetlinewidth{0.501875pt}%
\definecolor{currentstroke}{rgb}{0.000000,0.000000,0.000000}%
\pgfsetstrokecolor{currentstroke}%
\pgfsetdash{}{0pt}%
\pgfsys@defobject{currentmarker}{\pgfqpoint{-0.055556in}{0.000000in}}{\pgfqpoint{0.000000in}{0.000000in}}{%
\pgfpathmoveto{\pgfqpoint{0.000000in}{0.000000in}}%
\pgfpathlineto{\pgfqpoint{-0.055556in}{0.000000in}}%
\pgfusepath{stroke,fill}%
}%
\begin{pgfscope}%
\pgfsys@transformshift{2.830149in}{1.005247in}%
\pgfsys@useobject{currentmarker}{}%
\end{pgfscope}%
\end{pgfscope}%
\begin{pgfscope}%
\pgftext[x=0.581344in,y=1.005247in,right,]{\rmfamily\fontsize{10.300000}{12.360000}\selectfont −0.1}%
\end{pgfscope}%
\begin{pgfscope}%
\pgfsetbuttcap%
\pgfsetroundjoin%
\definecolor{currentfill}{rgb}{0.000000,0.000000,0.000000}%
\pgfsetfillcolor{currentfill}%
\pgfsetlinewidth{0.501875pt}%
\definecolor{currentstroke}{rgb}{0.000000,0.000000,0.000000}%
\pgfsetstrokecolor{currentstroke}%
\pgfsetdash{}{0pt}%
\pgfsys@defobject{currentmarker}{\pgfqpoint{0.000000in}{0.000000in}}{\pgfqpoint{0.055556in}{0.000000in}}{%
\pgfpathmoveto{\pgfqpoint{0.000000in}{0.000000in}}%
\pgfpathlineto{\pgfqpoint{0.055556in}{0.000000in}}%
\pgfusepath{stroke,fill}%
}%
\begin{pgfscope}%
\pgfsys@transformshift{0.636899in}{1.360802in}%
\pgfsys@useobject{currentmarker}{}%
\end{pgfscope}%
\end{pgfscope}%
\begin{pgfscope}%
\pgfsetbuttcap%
\pgfsetroundjoin%
\definecolor{currentfill}{rgb}{0.000000,0.000000,0.000000}%
\pgfsetfillcolor{currentfill}%
\pgfsetlinewidth{0.501875pt}%
\definecolor{currentstroke}{rgb}{0.000000,0.000000,0.000000}%
\pgfsetstrokecolor{currentstroke}%
\pgfsetdash{}{0pt}%
\pgfsys@defobject{currentmarker}{\pgfqpoint{-0.055556in}{0.000000in}}{\pgfqpoint{0.000000in}{0.000000in}}{%
\pgfpathmoveto{\pgfqpoint{0.000000in}{0.000000in}}%
\pgfpathlineto{\pgfqpoint{-0.055556in}{0.000000in}}%
\pgfusepath{stroke,fill}%
}%
\begin{pgfscope}%
\pgfsys@transformshift{2.830149in}{1.360802in}%
\pgfsys@useobject{currentmarker}{}%
\end{pgfscope}%
\end{pgfscope}%
\begin{pgfscope}%
\pgftext[x=0.581344in,y=1.360802in,right,]{\rmfamily\fontsize{10.300000}{12.360000}\selectfont 0.0}%
\end{pgfscope}%
\begin{pgfscope}%
\pgfsetbuttcap%
\pgfsetroundjoin%
\definecolor{currentfill}{rgb}{0.000000,0.000000,0.000000}%
\pgfsetfillcolor{currentfill}%
\pgfsetlinewidth{0.501875pt}%
\definecolor{currentstroke}{rgb}{0.000000,0.000000,0.000000}%
\pgfsetstrokecolor{currentstroke}%
\pgfsetdash{}{0pt}%
\pgfsys@defobject{currentmarker}{\pgfqpoint{0.000000in}{0.000000in}}{\pgfqpoint{0.055556in}{0.000000in}}{%
\pgfpathmoveto{\pgfqpoint{0.000000in}{0.000000in}}%
\pgfpathlineto{\pgfqpoint{0.055556in}{0.000000in}}%
\pgfusepath{stroke,fill}%
}%
\begin{pgfscope}%
\pgfsys@transformshift{0.636899in}{1.716358in}%
\pgfsys@useobject{currentmarker}{}%
\end{pgfscope}%
\end{pgfscope}%
\begin{pgfscope}%
\pgfsetbuttcap%
\pgfsetroundjoin%
\definecolor{currentfill}{rgb}{0.000000,0.000000,0.000000}%
\pgfsetfillcolor{currentfill}%
\pgfsetlinewidth{0.501875pt}%
\definecolor{currentstroke}{rgb}{0.000000,0.000000,0.000000}%
\pgfsetstrokecolor{currentstroke}%
\pgfsetdash{}{0pt}%
\pgfsys@defobject{currentmarker}{\pgfqpoint{-0.055556in}{0.000000in}}{\pgfqpoint{0.000000in}{0.000000in}}{%
\pgfpathmoveto{\pgfqpoint{0.000000in}{0.000000in}}%
\pgfpathlineto{\pgfqpoint{-0.055556in}{0.000000in}}%
\pgfusepath{stroke,fill}%
}%
\begin{pgfscope}%
\pgfsys@transformshift{2.830149in}{1.716358in}%
\pgfsys@useobject{currentmarker}{}%
\end{pgfscope}%
\end{pgfscope}%
\begin{pgfscope}%
\pgftext[x=0.581344in,y=1.716358in,right,]{\rmfamily\fontsize{10.300000}{12.360000}\selectfont 0.1}%
\end{pgfscope}%
\begin{pgfscope}%
\pgfsetbuttcap%
\pgfsetroundjoin%
\definecolor{currentfill}{rgb}{0.000000,0.000000,0.000000}%
\pgfsetfillcolor{currentfill}%
\pgfsetlinewidth{0.501875pt}%
\definecolor{currentstroke}{rgb}{0.000000,0.000000,0.000000}%
\pgfsetstrokecolor{currentstroke}%
\pgfsetdash{}{0pt}%
\pgfsys@defobject{currentmarker}{\pgfqpoint{0.000000in}{0.000000in}}{\pgfqpoint{0.055556in}{0.000000in}}{%
\pgfpathmoveto{\pgfqpoint{0.000000in}{0.000000in}}%
\pgfpathlineto{\pgfqpoint{0.055556in}{0.000000in}}%
\pgfusepath{stroke,fill}%
}%
\begin{pgfscope}%
\pgfsys@transformshift{0.636899in}{2.071913in}%
\pgfsys@useobject{currentmarker}{}%
\end{pgfscope}%
\end{pgfscope}%
\begin{pgfscope}%
\pgfsetbuttcap%
\pgfsetroundjoin%
\definecolor{currentfill}{rgb}{0.000000,0.000000,0.000000}%
\pgfsetfillcolor{currentfill}%
\pgfsetlinewidth{0.501875pt}%
\definecolor{currentstroke}{rgb}{0.000000,0.000000,0.000000}%
\pgfsetstrokecolor{currentstroke}%
\pgfsetdash{}{0pt}%
\pgfsys@defobject{currentmarker}{\pgfqpoint{-0.055556in}{0.000000in}}{\pgfqpoint{0.000000in}{0.000000in}}{%
\pgfpathmoveto{\pgfqpoint{0.000000in}{0.000000in}}%
\pgfpathlineto{\pgfqpoint{-0.055556in}{0.000000in}}%
\pgfusepath{stroke,fill}%
}%
\begin{pgfscope}%
\pgfsys@transformshift{2.830149in}{2.071913in}%
\pgfsys@useobject{currentmarker}{}%
\end{pgfscope}%
\end{pgfscope}%
\begin{pgfscope}%
\pgftext[x=0.581344in,y=2.071913in,right,]{\rmfamily\fontsize{10.300000}{12.360000}\selectfont 0.2}%
\end{pgfscope}%
\begin{pgfscope}%
\pgftext[x=0.264985in,y=1.271913in,,bottom,rotate=90.000000]{\rmfamily\fontsize{10.300000}{12.360000}\selectfont \(\displaystyle C/C^{\text{(ref)}}-1\)}%
\end{pgfscope}%
\begin{pgfscope}%
\pgfsetbuttcap%
\pgfsetmiterjoin%
\definecolor{currentfill}{rgb}{1.000000,1.000000,1.000000}%
\pgfsetfillcolor{currentfill}%
\pgfsetlinewidth{1.003750pt}%
\definecolor{currentstroke}{rgb}{0.000000,0.000000,0.000000}%
\pgfsetstrokecolor{currentstroke}%
\pgfsetdash{}{0pt}%
\pgfpathmoveto{\pgfqpoint{2.939812in}{0.836960in}}%
\pgfpathlineto{\pgfqpoint{4.164690in}{0.836960in}}%
\pgfpathlineto{\pgfqpoint{4.164690in}{2.071913in}}%
\pgfpathlineto{\pgfqpoint{2.939812in}{2.071913in}}%
\pgfpathclose%
\pgfusepath{stroke,fill}%
\end{pgfscope}%
\begin{pgfscope}%
\pgfsetrectcap%
\pgfsetroundjoin%
\pgfsetlinewidth{1.003750pt}%
\definecolor{currentstroke}{rgb}{0.000000,0.000000,1.000000}%
\pgfsetstrokecolor{currentstroke}%
\pgfsetdash{}{0pt}%
\pgfpathmoveto{\pgfqpoint{3.039951in}{1.964622in}}%
\pgfpathlineto{\pgfqpoint{3.240228in}{1.964622in}}%
\pgfusepath{stroke}%
\end{pgfscope}%
\begin{pgfscope}%
\pgfsetbuttcap%
\pgfsetroundjoin%
\definecolor{currentfill}{rgb}{0.000000,0.000000,1.000000}%
\pgfsetfillcolor{currentfill}%
\pgfsetlinewidth{0.501875pt}%
\definecolor{currentstroke}{rgb}{0.000000,0.000000,0.000000}%
\pgfsetstrokecolor{currentstroke}%
\pgfsetdash{}{0pt}%
\pgfsys@defobject{currentmarker}{\pgfqpoint{-0.027778in}{-0.027778in}}{\pgfqpoint{0.027778in}{0.027778in}}{%
\pgfpathmoveto{\pgfqpoint{0.000000in}{-0.027778in}}%
\pgfpathcurveto{\pgfqpoint{0.007367in}{-0.027778in}}{\pgfqpoint{0.014433in}{-0.024851in}}{\pgfqpoint{0.019642in}{-0.019642in}}%
\pgfpathcurveto{\pgfqpoint{0.024851in}{-0.014433in}}{\pgfqpoint{0.027778in}{-0.007367in}}{\pgfqpoint{0.027778in}{0.000000in}}%
\pgfpathcurveto{\pgfqpoint{0.027778in}{0.007367in}}{\pgfqpoint{0.024851in}{0.014433in}}{\pgfqpoint{0.019642in}{0.019642in}}%
\pgfpathcurveto{\pgfqpoint{0.014433in}{0.024851in}}{\pgfqpoint{0.007367in}{0.027778in}}{\pgfqpoint{0.000000in}{0.027778in}}%
\pgfpathcurveto{\pgfqpoint{-0.007367in}{0.027778in}}{\pgfqpoint{-0.014433in}{0.024851in}}{\pgfqpoint{-0.019642in}{0.019642in}}%
\pgfpathcurveto{\pgfqpoint{-0.024851in}{0.014433in}}{\pgfqpoint{-0.027778in}{0.007367in}}{\pgfqpoint{-0.027778in}{0.000000in}}%
\pgfpathcurveto{\pgfqpoint{-0.027778in}{-0.007367in}}{\pgfqpoint{-0.024851in}{-0.014433in}}{\pgfqpoint{-0.019642in}{-0.019642in}}%
\pgfpathcurveto{\pgfqpoint{-0.014433in}{-0.024851in}}{\pgfqpoint{-0.007367in}{-0.027778in}}{\pgfqpoint{0.000000in}{-0.027778in}}%
\pgfpathclose%
\pgfusepath{stroke,fill}%
}%
\begin{pgfscope}%
\pgfsys@transformshift{3.039951in}{1.964622in}%
\pgfsys@useobject{currentmarker}{}%
\end{pgfscope}%
\begin{pgfscope}%
\pgfsys@transformshift{3.240228in}{1.964622in}%
\pgfsys@useobject{currentmarker}{}%
\end{pgfscope}%
\end{pgfscope}%
\begin{pgfscope}%
\pgftext[x=3.397589in,y=1.914552in,left,base]{\rmfamily\fontsize{10.300000}{12.360000}\selectfont MEM BB}%
\end{pgfscope}%
\begin{pgfscope}%
\pgfsetrectcap%
\pgfsetroundjoin%
\pgfsetlinewidth{1.003750pt}%
\definecolor{currentstroke}{rgb}{0.000000,0.500000,0.000000}%
\pgfsetstrokecolor{currentstroke}%
\pgfsetdash{}{0pt}%
\pgfpathmoveto{\pgfqpoint{3.039951in}{1.765949in}}%
\pgfpathlineto{\pgfqpoint{3.240228in}{1.765949in}}%
\pgfusepath{stroke}%
\end{pgfscope}%
\begin{pgfscope}%
\pgfsetbuttcap%
\pgfsetmiterjoin%
\definecolor{currentfill}{rgb}{0.000000,0.500000,0.000000}%
\pgfsetfillcolor{currentfill}%
\pgfsetlinewidth{0.501875pt}%
\definecolor{currentstroke}{rgb}{0.000000,0.000000,0.000000}%
\pgfsetstrokecolor{currentstroke}%
\pgfsetdash{}{0pt}%
\pgfsys@defobject{currentmarker}{\pgfqpoint{-0.027778in}{-0.027778in}}{\pgfqpoint{0.027778in}{0.027778in}}{%
\pgfpathmoveto{\pgfqpoint{-0.027778in}{-0.027778in}}%
\pgfpathlineto{\pgfqpoint{0.027778in}{-0.027778in}}%
\pgfpathlineto{\pgfqpoint{0.027778in}{0.027778in}}%
\pgfpathlineto{\pgfqpoint{-0.027778in}{0.027778in}}%
\pgfpathclose%
\pgfusepath{stroke,fill}%
}%
\begin{pgfscope}%
\pgfsys@transformshift{3.039951in}{1.765949in}%
\pgfsys@useobject{currentmarker}{}%
\end{pgfscope}%
\begin{pgfscope}%
\pgfsys@transformshift{3.240228in}{1.765949in}%
\pgfsys@useobject{currentmarker}{}%
\end{pgfscope}%
\end{pgfscope}%
\begin{pgfscope}%
\pgftext[x=3.397589in,y=1.715879in,left,base]{\rmfamily\fontsize{10.300000}{12.360000}\selectfont MEM CLI}%
\end{pgfscope}%
\begin{pgfscope}%
\pgfsetrectcap%
\pgfsetroundjoin%
\pgfsetlinewidth{1.003750pt}%
\definecolor{currentstroke}{rgb}{1.000000,0.000000,0.000000}%
\pgfsetstrokecolor{currentstroke}%
\pgfsetdash{}{0pt}%
\pgfpathmoveto{\pgfqpoint{3.039951in}{1.567276in}}%
\pgfpathlineto{\pgfqpoint{3.240228in}{1.567276in}}%
\pgfusepath{stroke}%
\end{pgfscope}%
\begin{pgfscope}%
\pgfsetbuttcap%
\pgfsetbeveljoin%
\definecolor{currentfill}{rgb}{1.000000,0.000000,0.000000}%
\pgfsetfillcolor{currentfill}%
\pgfsetlinewidth{0.501875pt}%
\definecolor{currentstroke}{rgb}{0.000000,0.000000,0.000000}%
\pgfsetstrokecolor{currentstroke}%
\pgfsetdash{}{0pt}%
\pgfsys@defobject{currentmarker}{\pgfqpoint{-0.026418in}{-0.022473in}}{\pgfqpoint{0.026418in}{0.027778in}}{%
\pgfpathmoveto{\pgfqpoint{0.000000in}{0.027778in}}%
\pgfpathlineto{\pgfqpoint{-0.006236in}{0.008584in}}%
\pgfpathlineto{\pgfqpoint{-0.026418in}{0.008584in}}%
\pgfpathlineto{\pgfqpoint{-0.010091in}{-0.003279in}}%
\pgfpathlineto{\pgfqpoint{-0.016327in}{-0.022473in}}%
\pgfpathlineto{\pgfqpoint{-0.000000in}{-0.010610in}}%
\pgfpathlineto{\pgfqpoint{0.016327in}{-0.022473in}}%
\pgfpathlineto{\pgfqpoint{0.010091in}{-0.003279in}}%
\pgfpathlineto{\pgfqpoint{0.026418in}{0.008584in}}%
\pgfpathlineto{\pgfqpoint{0.006236in}{0.008584in}}%
\pgfpathclose%
\pgfusepath{stroke,fill}%
}%
\begin{pgfscope}%
\pgfsys@transformshift{3.039951in}{1.567276in}%
\pgfsys@useobject{currentmarker}{}%
\end{pgfscope}%
\begin{pgfscope}%
\pgfsys@transformshift{3.240228in}{1.567276in}%
\pgfsys@useobject{currentmarker}{}%
\end{pgfscope}%
\end{pgfscope}%
\begin{pgfscope}%
\pgftext[x=3.397589in,y=1.517207in,left,base]{\rmfamily\fontsize{10.300000}{12.360000}\selectfont MEM MR}%
\end{pgfscope}%
\begin{pgfscope}%
\pgfsetrectcap%
\pgfsetroundjoin%
\pgfsetlinewidth{1.003750pt}%
\definecolor{currentstroke}{rgb}{0.000000,0.750000,0.750000}%
\pgfsetstrokecolor{currentstroke}%
\pgfsetdash{}{0pt}%
\pgfpathmoveto{\pgfqpoint{3.039951in}{1.368603in}}%
\pgfpathlineto{\pgfqpoint{3.240228in}{1.368603in}}%
\pgfusepath{stroke}%
\end{pgfscope}%
\begin{pgfscope}%
\pgfsetbuttcap%
\pgfsetmiterjoin%
\definecolor{currentfill}{rgb}{0.000000,0.750000,0.750000}%
\pgfsetfillcolor{currentfill}%
\pgfsetlinewidth{0.501875pt}%
\definecolor{currentstroke}{rgb}{0.000000,0.000000,0.000000}%
\pgfsetstrokecolor{currentstroke}%
\pgfsetdash{}{0pt}%
\pgfsys@defobject{currentmarker}{\pgfqpoint{-0.027778in}{-0.027778in}}{\pgfqpoint{0.027778in}{0.027778in}}{%
\pgfpathmoveto{\pgfqpoint{-0.027778in}{0.000000in}}%
\pgfpathlineto{\pgfqpoint{0.027778in}{-0.027778in}}%
\pgfpathlineto{\pgfqpoint{0.027778in}{0.027778in}}%
\pgfpathclose%
\pgfusepath{stroke,fill}%
}%
\begin{pgfscope}%
\pgfsys@transformshift{3.039951in}{1.368603in}%
\pgfsys@useobject{currentmarker}{}%
\end{pgfscope}%
\begin{pgfscope}%
\pgfsys@transformshift{3.240228in}{1.368603in}%
\pgfsys@useobject{currentmarker}{}%
\end{pgfscope}%
\end{pgfscope}%
\begin{pgfscope}%
\pgftext[x=3.397589in,y=1.318534in,left,base]{\rmfamily\fontsize{10.300000}{12.360000}\selectfont PSM M1B1}%
\end{pgfscope}%
\begin{pgfscope}%
\pgfsetrectcap%
\pgfsetroundjoin%
\pgfsetlinewidth{1.003750pt}%
\definecolor{currentstroke}{rgb}{0.750000,0.000000,0.750000}%
\pgfsetstrokecolor{currentstroke}%
\pgfsetdash{}{0pt}%
\pgfpathmoveto{\pgfqpoint{3.039951in}{1.169931in}}%
\pgfpathlineto{\pgfqpoint{3.240228in}{1.169931in}}%
\pgfusepath{stroke}%
\end{pgfscope}%
\begin{pgfscope}%
\pgfsetbuttcap%
\pgfsetmiterjoin%
\definecolor{currentfill}{rgb}{0.750000,0.000000,0.750000}%
\pgfsetfillcolor{currentfill}%
\pgfsetlinewidth{0.501875pt}%
\definecolor{currentstroke}{rgb}{0.000000,0.000000,0.000000}%
\pgfsetstrokecolor{currentstroke}%
\pgfsetdash{}{0pt}%
\pgfsys@defobject{currentmarker}{\pgfqpoint{-0.027778in}{-0.027778in}}{\pgfqpoint{0.027778in}{0.027778in}}{%
\pgfpathmoveto{\pgfqpoint{0.027778in}{-0.000000in}}%
\pgfpathlineto{\pgfqpoint{-0.027778in}{0.027778in}}%
\pgfpathlineto{\pgfqpoint{-0.027778in}{-0.027778in}}%
\pgfpathclose%
\pgfusepath{stroke,fill}%
}%
\begin{pgfscope}%
\pgfsys@transformshift{3.039951in}{1.169931in}%
\pgfsys@useobject{currentmarker}{}%
\end{pgfscope}%
\begin{pgfscope}%
\pgfsys@transformshift{3.240228in}{1.169931in}%
\pgfsys@useobject{currentmarker}{}%
\end{pgfscope}%
\end{pgfscope}%
\begin{pgfscope}%
\pgftext[x=3.397589in,y=1.119861in,left,base]{\rmfamily\fontsize{10.300000}{12.360000}\selectfont PSM M2B2}%
\end{pgfscope}%
\begin{pgfscope}%
\pgfsetrectcap%
\pgfsetroundjoin%
\pgfsetlinewidth{1.003750pt}%
\definecolor{currentstroke}{rgb}{0.750000,0.750000,0.000000}%
\pgfsetstrokecolor{currentstroke}%
\pgfsetdash{}{0pt}%
\pgfpathmoveto{\pgfqpoint{3.039951in}{0.971258in}}%
\pgfpathlineto{\pgfqpoint{3.240228in}{0.971258in}}%
\pgfusepath{stroke}%
\end{pgfscope}%
\begin{pgfscope}%
\pgfsetbuttcap%
\pgfsetmiterjoin%
\definecolor{currentfill}{rgb}{0.750000,0.750000,0.000000}%
\pgfsetfillcolor{currentfill}%
\pgfsetlinewidth{0.501875pt}%
\definecolor{currentstroke}{rgb}{0.000000,0.000000,0.000000}%
\pgfsetstrokecolor{currentstroke}%
\pgfsetdash{}{0pt}%
\pgfsys@defobject{currentmarker}{\pgfqpoint{-0.027778in}{-0.027778in}}{\pgfqpoint{0.027778in}{0.027778in}}{%
\pgfpathmoveto{\pgfqpoint{-0.000000in}{-0.027778in}}%
\pgfpathlineto{\pgfqpoint{0.027778in}{0.027778in}}%
\pgfpathlineto{\pgfqpoint{-0.027778in}{0.027778in}}%
\pgfpathclose%
\pgfusepath{stroke,fill}%
}%
\begin{pgfscope}%
\pgfsys@transformshift{3.039951in}{0.971258in}%
\pgfsys@useobject{currentmarker}{}%
\end{pgfscope}%
\begin{pgfscope}%
\pgfsys@transformshift{3.240228in}{0.971258in}%
\pgfsys@useobject{currentmarker}{}%
\end{pgfscope}%
\end{pgfscope}%
\begin{pgfscope}%
\pgftext[x=3.397589in,y=0.921188in,left,base]{\rmfamily\fontsize{10.300000}{12.360000}\selectfont PSM M3B2}%
\end{pgfscope}%
\begin{pgfscope}%
\pgfsetrectcap%
\pgfsetmiterjoin%
\pgfsetlinewidth{1.003750pt}%
\definecolor{currentstroke}{rgb}{0.000000,0.000000,0.000000}%
\pgfsetstrokecolor{currentstroke}%
\pgfsetdash{}{0pt}%
\pgfpathmoveto{\pgfqpoint{0.636899in}{2.071913in}}%
\pgfpathlineto{\pgfqpoint{2.830149in}{2.071913in}}%
\pgfusepath{stroke}%
\end{pgfscope}%
\begin{pgfscope}%
\pgfsetrectcap%
\pgfsetmiterjoin%
\pgfsetlinewidth{1.003750pt}%
\definecolor{currentstroke}{rgb}{0.000000,0.000000,0.000000}%
\pgfsetstrokecolor{currentstroke}%
\pgfsetdash{}{0pt}%
\pgfpathmoveto{\pgfqpoint{2.830149in}{0.471913in}}%
\pgfpathlineto{\pgfqpoint{2.830149in}{2.071913in}}%
\pgfusepath{stroke}%
\end{pgfscope}%
\begin{pgfscope}%
\pgfsetrectcap%
\pgfsetmiterjoin%
\pgfsetlinewidth{1.003750pt}%
\definecolor{currentstroke}{rgb}{0.000000,0.000000,0.000000}%
\pgfsetstrokecolor{currentstroke}%
\pgfsetdash{}{0pt}%
\pgfpathmoveto{\pgfqpoint{0.636899in}{0.471913in}}%
\pgfpathlineto{\pgfqpoint{2.830149in}{0.471913in}}%
\pgfusepath{stroke}%
\end{pgfscope}%
\begin{pgfscope}%
\pgfsetrectcap%
\pgfsetmiterjoin%
\pgfsetlinewidth{1.003750pt}%
\definecolor{currentstroke}{rgb}{0.000000,0.000000,0.000000}%
\pgfsetstrokecolor{currentstroke}%
\pgfsetdash{}{0pt}%
\pgfpathmoveto{\pgfqpoint{0.636899in}{0.471913in}}%
\pgfpathlineto{\pgfqpoint{0.636899in}{2.071913in}}%
\pgfusepath{stroke}%
\end{pgfscope}%
\begin{pgfscope}%
\pgfsetbuttcap%
\pgfsetroundjoin%
\definecolor{currentfill}{rgb}{0.000000,0.000000,0.000000}%
\pgfsetfillcolor{currentfill}%
\pgfsetlinewidth{0.501875pt}%
\definecolor{currentstroke}{rgb}{0.000000,0.000000,0.000000}%
\pgfsetstrokecolor{currentstroke}%
\pgfsetdash{}{0pt}%
\pgfsys@defobject{currentmarker}{\pgfqpoint{0.000000in}{-0.055556in}}{\pgfqpoint{0.000000in}{0.000000in}}{%
\pgfpathmoveto{\pgfqpoint{0.000000in}{0.000000in}}%
\pgfpathlineto{\pgfqpoint{0.000000in}{-0.055556in}}%
\pgfusepath{stroke,fill}%
}%
\begin{pgfscope}%
\pgfsys@transformshift{0.636899in}{2.071913in}%
\pgfsys@useobject{currentmarker}{}%
\end{pgfscope}%
\end{pgfscope}%
\begin{pgfscope}%
\pgftext[x=0.636899in,y=2.127469in,,bottom]{\rmfamily\fontsize{10.300000}{12.360000}\selectfont 0.5}%
\end{pgfscope}%
\begin{pgfscope}%
\pgfsetbuttcap%
\pgfsetroundjoin%
\definecolor{currentfill}{rgb}{0.000000,0.000000,0.000000}%
\pgfsetfillcolor{currentfill}%
\pgfsetlinewidth{0.501875pt}%
\definecolor{currentstroke}{rgb}{0.000000,0.000000,0.000000}%
\pgfsetstrokecolor{currentstroke}%
\pgfsetdash{}{0pt}%
\pgfsys@defobject{currentmarker}{\pgfqpoint{0.000000in}{-0.055556in}}{\pgfqpoint{0.000000in}{0.000000in}}{%
\pgfpathmoveto{\pgfqpoint{0.000000in}{0.000000in}}%
\pgfpathlineto{\pgfqpoint{0.000000in}{-0.055556in}}%
\pgfusepath{stroke,fill}%
}%
\begin{pgfscope}%
\pgfsys@transformshift{1.353648in}{2.071913in}%
\pgfsys@useobject{currentmarker}{}%
\end{pgfscope}%
\end{pgfscope}%
\begin{pgfscope}%
\pgftext[x=1.353648in,y=2.127469in,,bottom]{\rmfamily\fontsize{10.300000}{12.360000}\selectfont 1.0}%
\end{pgfscope}%
\begin{pgfscope}%
\pgfsetbuttcap%
\pgfsetroundjoin%
\definecolor{currentfill}{rgb}{0.000000,0.000000,0.000000}%
\pgfsetfillcolor{currentfill}%
\pgfsetlinewidth{0.501875pt}%
\definecolor{currentstroke}{rgb}{0.000000,0.000000,0.000000}%
\pgfsetstrokecolor{currentstroke}%
\pgfsetdash{}{0pt}%
\pgfsys@defobject{currentmarker}{\pgfqpoint{0.000000in}{-0.055556in}}{\pgfqpoint{0.000000in}{0.000000in}}{%
\pgfpathmoveto{\pgfqpoint{0.000000in}{0.000000in}}%
\pgfpathlineto{\pgfqpoint{0.000000in}{-0.055556in}}%
\pgfusepath{stroke,fill}%
}%
\begin{pgfscope}%
\pgfsys@transformshift{2.070396in}{2.071913in}%
\pgfsys@useobject{currentmarker}{}%
\end{pgfscope}%
\end{pgfscope}%
\begin{pgfscope}%
\pgftext[x=2.070396in,y=2.127469in,,bottom]{\rmfamily\fontsize{10.300000}{12.360000}\selectfont 1.5}%
\end{pgfscope}%
\begin{pgfscope}%
\pgfsetbuttcap%
\pgfsetroundjoin%
\definecolor{currentfill}{rgb}{0.000000,0.000000,0.000000}%
\pgfsetfillcolor{currentfill}%
\pgfsetlinewidth{0.501875pt}%
\definecolor{currentstroke}{rgb}{0.000000,0.000000,0.000000}%
\pgfsetstrokecolor{currentstroke}%
\pgfsetdash{}{0pt}%
\pgfsys@defobject{currentmarker}{\pgfqpoint{0.000000in}{-0.055556in}}{\pgfqpoint{0.000000in}{0.000000in}}{%
\pgfpathmoveto{\pgfqpoint{0.000000in}{0.000000in}}%
\pgfpathlineto{\pgfqpoint{0.000000in}{-0.055556in}}%
\pgfusepath{stroke,fill}%
}%
\begin{pgfscope}%
\pgfsys@transformshift{2.787144in}{2.071913in}%
\pgfsys@useobject{currentmarker}{}%
\end{pgfscope}%
\end{pgfscope}%
\begin{pgfscope}%
\pgftext[x=2.787144in,y=2.127469in,,bottom]{\rmfamily\fontsize{10.300000}{12.360000}\selectfont 2.0}%
\end{pgfscope}%
\begin{pgfscope}%
\pgftext[x=1.733524in,y=2.320370in,,base]{\rmfamily\fontsize{10.300000}{12.360000}\selectfont \(\displaystyle \tau\)}%
\end{pgfscope}%
\end{pgfpicture}%
\makeatother%
\endgroup%

%% file: BenchmarkMSHS.tex
\section{Motion of a single heavy sphere in ambient fluid}
\label{chap:benchmark}

\subsection{Description}
\label{sec:description}
In this benchmark scenario, a sphere that is heavier than the surrounding fluid and has a diameter $D$ is exposed to an external flow field. 
The simulation domain is rectangular and of size $5.34D \times 5.34D \times 16D$ and set up as periodic in the horizontal directions $x$ and $y$.
A constant inflow from below with the velocity $\boldsymbol{u}_\infty$ is imposed and an outflow condition is applied at the top of the domain.
The sphere is initially placed at the position $(2.67D, 2.67D, 5.34D)$. 
In order to keep the sphere positioned inside the simulation domain, the net force due to buoyancy and gravity is chosen to balance the acting drag force $F_{\text{drag}}$ in vertical direction.
This setup corresponds to the one used in \cite{uhlmann_motion_2014} for the simulation with the immersed boundary method from \cite{uhlmann_immersed_2005}. 
As a reference, the results from the spectral element method will be used which are also reported in \cite{uhlmann_motion_2014}.

More specifically, the ratio between solid and fluid density, $\rho_p / \rho_f$, is fixed to $1.5$ for all simulations.
This density ratio allows the definition of an important characteristic quantity, the Galileo number
\begin{equation}
Ga = \frac{\sqrt{ |\frac{\rho_p}{\rho_f}-1| g D^3}}{\nu}, \label{eq:Galileo}
\end{equation}
with the magnitude of the gravitational acceleration $g = |\boldsymbol{g}|$. % and the kinematic viscosity $\nu$.
The Galileo number characterizes different regimes of the sphere's motion which will be explored and investigated with this test case.
For increasing $Ga$, the setup undergoes transitions from the steady axisymmetric regime to the steady oblique regime and to the oscillating oblique regime until it arrives at the chaotic regime.

However, the Galileo number cannot be set directly in this test case since it depends on the gravitational acceleration which is chosen with respect to the acting drag force as mentioned earlier. 
This force is mainly determined by the inflow velocity and the fluid viscosity, but is also influenced by the applied fluid-solid coupling algorithm.
Therefore, the execution of the test case consists of two parts: 
At the beginning, the simulation is started with the sphere held fixed at its initial position.
The viscosity is initially set to $0.01$ in lattice units and independent of the numerical resolution.
The inflow velocity as well as the initial fluid velocity are determined as $\boldsymbol{u}_\infty = \left( 0,0, Re_{||,\text{ref}}\nu / D\right)^\top$ where $Re_{||,\text{ref}}$ is the particle Reynolds number reported in Tabs. 2-5 of \cite{uhlmann_motion_2014} for the reference simulations.
The simulation is then run until the drag force on the sphere is sufficiently converged.
Then, the gravitational acceleration is determined via 
\begin{equation}
	\boldsymbol{g} = \left(0,0, -\frac{F_{\text{drag}}}{\tfrac{\pi}{6} D^3|\tfrac{\rho_p}{\rho_f} -1|}\right)^\top,
\end{equation}
and the Galileo number can be computed via Eq.~\eqref{eq:Galileo}.
If its value is close enough to the desired one, the second part of the simulation can be started. 
In all presented simulations, the relative difference between the actual and the targeted Galileo number is below $10^{-4}$. 
Else, the viscosity is adjusted via $\nu \leftarrow \tfrac{Ga}{Ga_{\text{target}}}\nu$ and the simulation is continued, repeating the evaluation procedure with the updated viscosity.
In our experiments, this routine typically converged after a few iterations and thus avoids tedious manual experimentation. 

In the second part of the simulation, the sphere is released and allowed to move freely while maintaining the flow field from the first part. 
Its position and velocity are updated in each simulation step based on the acting forces via an explicit Euler integration.
Unless stated otherwise, all simulations are carried out for $250$ dimensionless time steps, with $t_{\text{ref}} = D / u_{\text{ref}}$ and $u_{\text{ref}} = \sqrt{|\rho_p/\rho_f -1| g D}$.

As inflow condition, the common velocity bounce-back scheme like in Eq.~\eqref{eq:MEM_BB} with the inflow velocity $\boldsymbol{u}_\infty$ is applied. 
The outflow is modeled following \cite{ginzburg_generic_2005} with $\rho_{\text{out}}=1$ in lattice units to yield
\begin{equation}
f_{\bar{q}}(\boldsymbol{x},t+1) = - \tilde{f}_q(\boldsymbol{x},t) + 2 f^{\text{eq},+}_q(\rho_{\text{out}}, \boldsymbol{u}(\boldsymbol{x},t)).
\end{equation}
To obtain the forces on the sphere, two LBM subcycles are executed and the calculated forces are averaged over these two cycles. 

For the actual test case, the four different regimes are simulated by choosing a corresponding Galileo number.
Here we consider a range of the Galileo numbers from $144$ up to $250$ which corresponds to particle Reynolds numbers of approximately $185$ to $365$.
The quantities of interest for each regime are evaluated, e.g. the magnitudes of the vertical and horizontal relative sphere velocity, $u_{pV}$ and $u_{pH}$, and of the vertical and horizontal rotational velocities, $\omega_{pV}$ and $\omega_{pH}$.
The exact definition of all quantities and coordinate systems is given in \cite{uhlmann_motion_2014} and the reported values here are made dimensionless by the same scaling.
Then, the relative errors of these quantities with respect to the spectral element results are evaluated analogously to \cite{uhlmann_motion_2014}.
In particular, the errors in velocity components are always evaluated relative to the reference value of $u_{pV}$.

For the MEM from Sec.~\ref{sec:MEM}, the TRT collision operator with $\Lambda_\pm = \tfrac{3}{16}$ with extrapolation reconstruction technique of Eq.~\eqref{eq:MEM_Reconstruction} is used in all cases. 
The variants only differ in the no-slip boundary condition along the sphere's surface which are the BB from Eq.~\eqref{eq:MEM_BB}, the CLI from Eq.~\eqref{eq:MEM_CLI} and the MR scheme from Eq.~\eqref{eq:MEM_MR}.
For the PSM from Sec.~\ref{sec:PSM}, the combinations M1B1, M2B2 and M3B2 of Eqs.~\eqref{eq:PSM_B1}--\eqref{eq:PSM_M3} are applied.
The respective labels are chosen to reflect the test regime (A-D), the coupling method, and the cells per diameter $D/\Delta x$, where $\Delta x$ is the cell size. 
Values between $18$ and $48$ are typically used for $D/\Delta x$, which then also determines the overall resolution of the simulation.

\subsection{Steady axisymmetric regime}
\label{sec:regime1}

\begin{figure}[t]
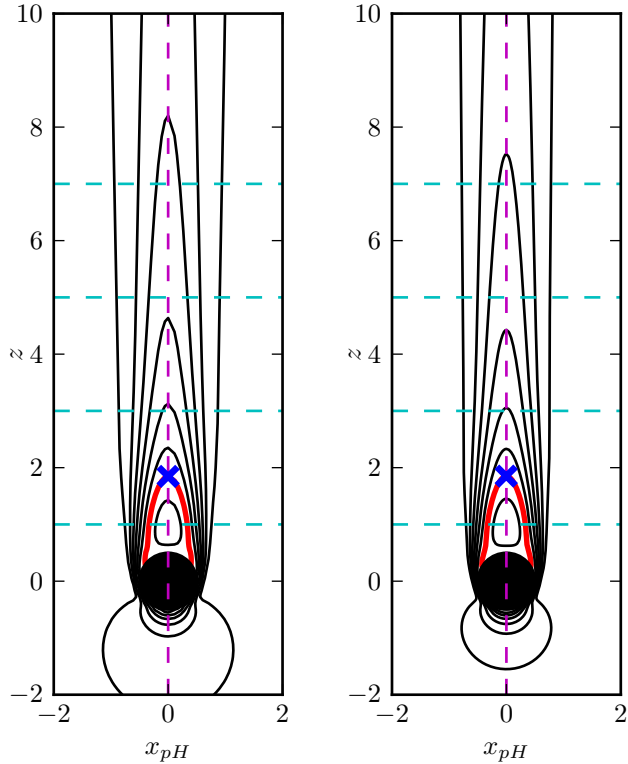

	\centering
	\input{plot_144_CLI_18_plot_contour_xpH_Z_urpar.pgf}
	\input{plot_144_CLI_48_plot_contour_xpH_Z_urpar.pgf}
	\captionof{figure}{Contour lines of the projected relative velocity $u_{r\parallel}$ for case A-CLI-18 (left) and A-CLI-48 (right) for $Ga=144$. Contours are at (-0.2:0.2:1.2) where the red line outlines the recirculation area with $u_{r\parallel}=0$. The blue cross marks the location taken for the calculation of the recirculation length $L_r$.}
	\label{fig:G144_contour}
\end{figure}

\begin{figure}[t]
	\centering
	\input{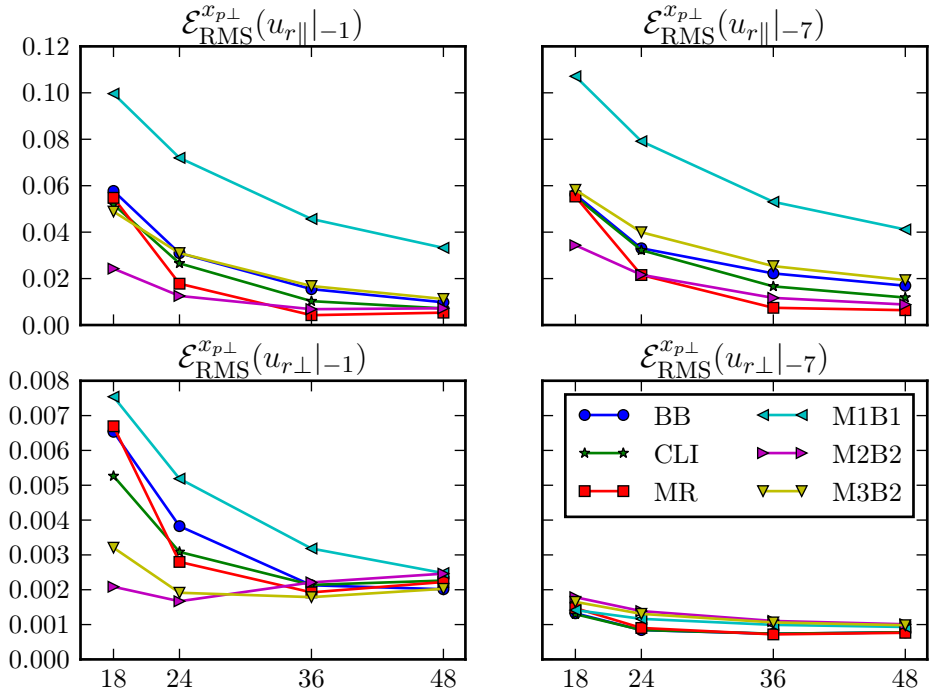}
	\caption{Root mean square errors of different velocity components along $x_{p\perp}$ at positions $x_{p\parallel}\in\{-1,-7\}$ with respect to $D/\Delta x$ for case A ($Ga=144$).}
	\label{fig:G144_err_vel}
\end{figure}
\begin{figure}[t]
	\centering
	\input{plot_144_D_rms_errors_cp.pgf}
	\caption{Root mean square errors of the pressure coefficient $c_p$ with respect to $D/\Delta x$ for case A ($G=144$). The values are evaluated along a circle close to the sphere's surface with angle $\theta_1$. Line styles as in Fig.~\ref{fig:G144_err_vel}.}
	\label{fig:G144_err_cp}
\end{figure}
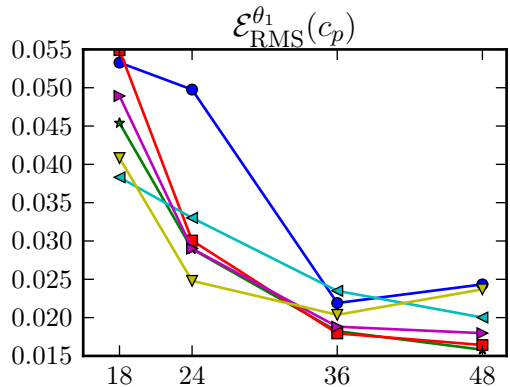

The first regime with a Galileo number of $144$ features a steady and axisymmetric flow around the sphere. 
The simulation has been carried out for four different resolutions $D/\Delta x \in \{18,24,36,48\}$.
The contour plots of the vertical relative flow velocity for $D/\Delta = 18$ and $48$ are shown in Fig.~\ref{fig:G144_contour} for the CLI variant.
The results for the vertical sphere velocity and the recirculation length are given in Tab.~\ref{tab:G144}, together with their errors with respect to the reference values from the spectral element results in \cite{uhlmann_motion_2014}.
Generally, the values for $u_{pV}$ converge against the desired reference value in all cases.
The differences in the MEM variants are overall rather small but especially at higher resolutions CLI and MR are more accurate.
Regarding the PSM variants, clearly M1B1 performs worse than M2B2 and M3B2.
In contrast to that, all LBM coupling methods converge to a recirculation length that is smaller than in the reference case. 

In addition to those two values, reference results for the flow velocity profiles along $x_{p\parallel}\in\{-1,-3,-5,-7\}$ and the pressure coefficient $c_p$ along a circle around the sphere are reported in \cite{uhlmann_motion_2014}.
To evaluate the data in a compact form, they are compressed into single root mean square (RMS) values of the absolute deviation from these values. 
For a quantity $\phi$, the RMS error is here defined as 
\begin{equation}
\mathcal{E}_{\text{RMS}}^{x_{\text{coord}}}(\phi) = \sqrt{ \frac{1}{N} \sum_{i=1}^{N} \left(\phi^{\text{(ref)}}(x_{\text{coord}}(i)) - \phi(x_{\text{coord}}(i))\right)^2}. \label{eq:RMSerror}
\end{equation}
The $N$ evaluation points along the axis $x_{\text{coord}}$ are prescribed by the available reference data.
The RMS errors for $u_{r\parallel}$ and $u_{r\perp}$ along the $x_{p\perp}$ axis and at the positions $x_{p\parallel} = -1$ and $-7$, denoted by $|_{-1}$ and $|_{-7}$ respectively, are given in the graphs of Fig.~\ref{fig:G144_err_vel}.
Additionally, the RMS errors of $c_p$ along a circle near the sphere's surface and evaluated at different angles $\theta_1$ are shown in Fig.~\ref{fig:G144_err_cp}.
The differences between the MEM variants are again rather small. 
The MR scheme shows overall slightly better results than CLI, particularly for the higher resolution cases of $u_{r\parallel}$.
BB is less accurate than CLI and MR in almost all cases, especially for $c_p$. 
M2B2 is the most accurate PSM variant for $u_{r\parallel}$ and overall similar to M3B2 for $u_{r\perp}$ and $c_p$. 
The errors of M1B1 are up to four times larger for $u_{r\parallel}$ and for $u_{r\perp}$ at $x_{p\parallel} = -1$ in comparisons to the other PSM combinations. 
 
Summarizing, all MEM variants are able to capture the characteristics of this regime very well. 
The outcomes of M2B2 and M3B2 are comparable to the ones from the MEM variants but M1B1 clearly falls behind the other two variants.

\subsection{Steady oblique regime}
\label{sec:regime2}

\begin{figure}[t]
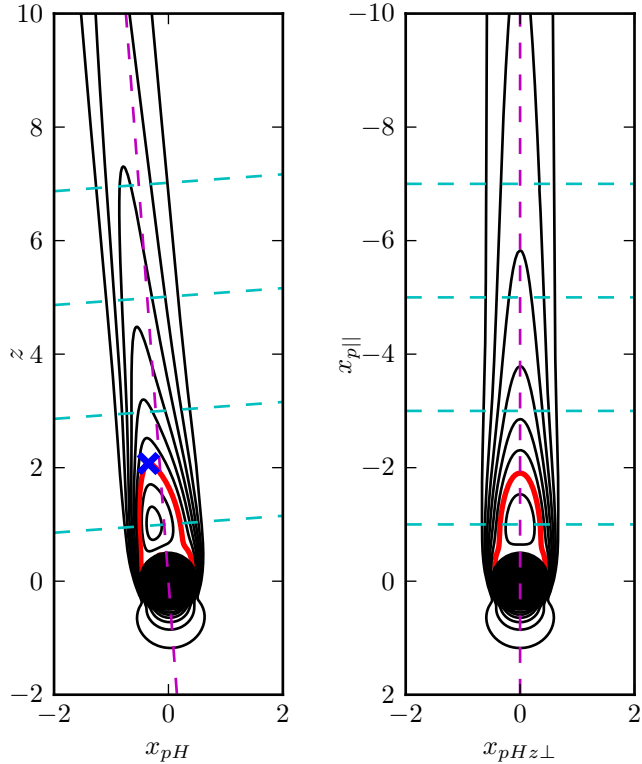

	\centering
	\input{plot_178_CLI_48_plot_contour_xpH_Z_urpar.pgf}
	\input{plot_178_CLI_48_plot_contour_xpHzperp_xppar_urpar.pgf}
	\caption{Contour lines of the projected relative velocity $u_{r\parallel}$ for case B-CLI-48 ($Ga=178.46$). Contours are at (-0.4:0.2:1.2) where the red line outlines the recirculation area with $u_{r\parallel}=0$. The blue cross in the left plot marks the location taken for the calculation of the recirculation length $L_r$.}
	\label{fig:G178_contour}
\end{figure}

\begin{figure}[t]
	\centering
	\input{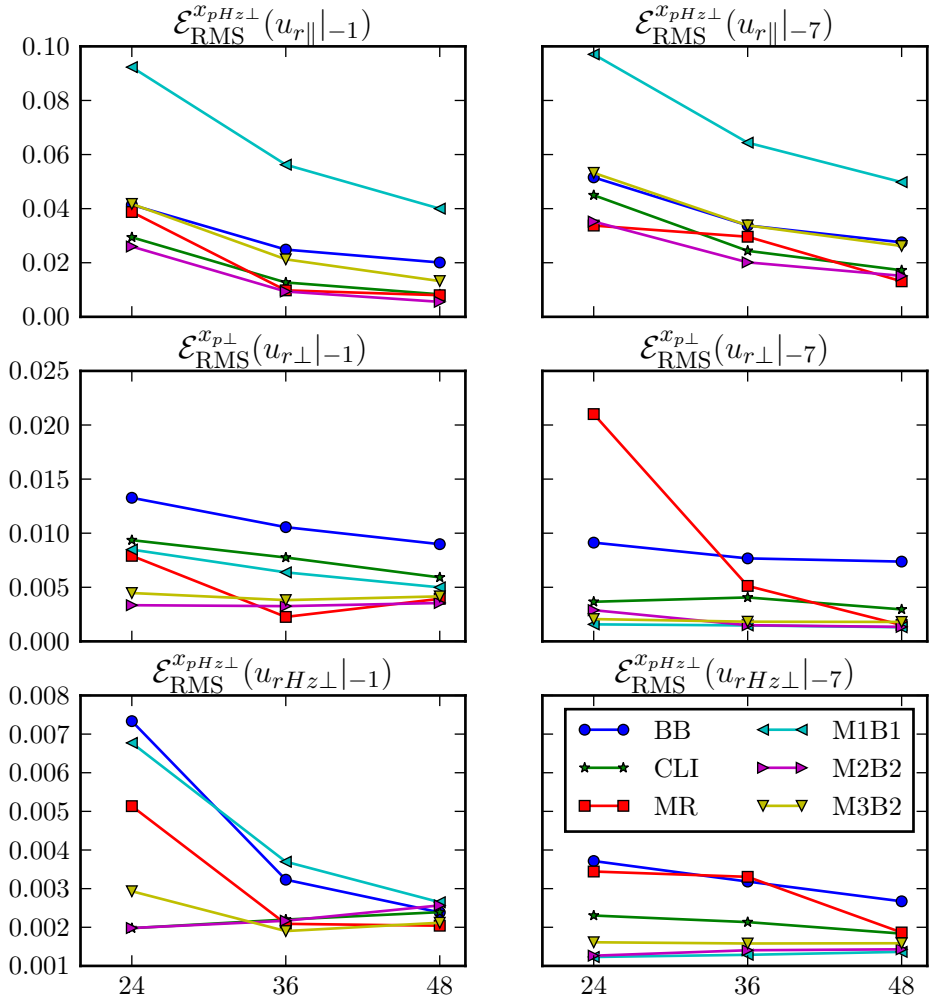}
	\caption{Root mean square errors of different velocity components along $x_{pHz\perp}$ and $x_{p\perp}$ at positions $x_{p\parallel}\in\{-1,-7\}$ with respect to $D/\Delta x$ for case B ($Ga=178.46$).}
	\label{fig:G178_err_vel}
\end{figure}

\begin{figure}[t]
	\centering
	\input{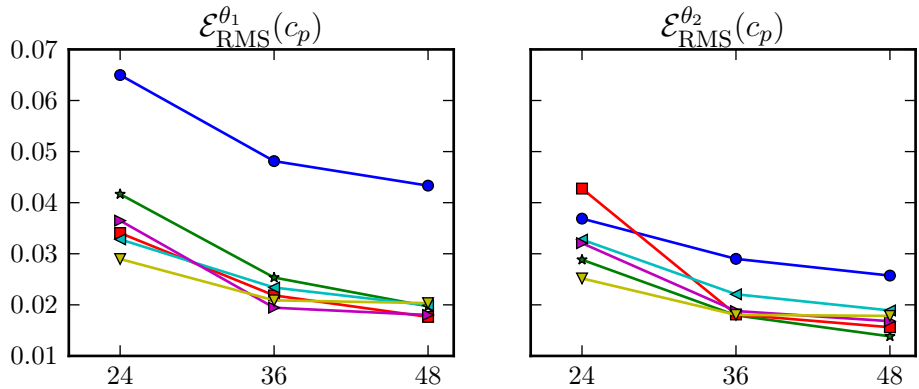}
	\caption{Root mean square errors of the pressure coefficient $c_p$ with respect to $D/\Delta x$ for case B ($Ga=178.46$). The values are evaluated along two circles close to the sphere's surface with angles $\theta_1$ and $\theta_2$. Line styles as in Fig.~\ref{fig:G178_err_vel}.}
	\label{fig:G178_err_cp}
\end{figure}

When the Galileo number is increased to a nominal value of $178.46$, which corresponds to a particle Reynolds number of approximately $243$, the sphere eventually starts to move in the horizontal plane with a constant horizontal velocity.
To trigger this movement stemming from an instability, the stabilizing numerical symmetry of this system is broken by slightly perturbing the sphere's initial horizontal position.
Note, however, that this does not change the physical setting due to the periodic boundaries in horizontal directions.
In Fig.~\ref{fig:G178_contour}, the contour lines of the projected flow velocities are shown for two different planes.
As a result of the horizontal movement, the horizontal and rotational sphere velocities, $u_{pH}$ and $\omega_{pH}$, can additionally be used for a comparison. 
The values and errors of these four quantities are given in Tab.~\ref{tab:G178}.
Following the reasoning given in \cite{uhlmann_motion_2014}, the relative errors for quantities triggered by instabilities are calculated with respect to the reference value of $u_{pV}$ in all cases. 
For our setup, the simulations become unstable for the lowest resolution of $D/\Delta x = 18$ which are therefore excluded from the table.
Regarding the vertical sphere velocity, the MEM variants with errors below $2\%$ for the finest resolution perform systematically better than the PSM variants.
The opposite is the case for $u_{pH}$ where the PSM variants exhibit very small errors almost independent of the resolution. 
The results obtained by CLI and MR are still acceptable but BB shows significant deviations from the reference value.
The horizontal rotational velocity is captured reasonably well by all methods with errors below $1\%$ for resolutions of $36$ and $48$ except for BB.
Like for case A, all methods tend to have a shorter recirculation length.
It has to be noted, however, that the results from MR show significant oscillations over the simulation time and cannot be regarded as steady state results. 
This explains the fluctuations in the results for the different resolutions.

Due to the steady nature of the flow, profiles of various velocity components can again be compared and will be expressed by RMS errors with the definition from Eq.~\eqref{eq:RMSerror}.
Graphs of some representative errors over the resolution are shown in Fig.~\ref{fig:G178_err_vel}.
Once more, it can be seen that BB for $\mathcal{E}_{\text{RMS}}^{x_{p\perp}}(u_{r\perp})$ and M1B1 for $\mathcal{E}_{\text{RMS}}^{x_{pHz\perp}}(u_{r\parallel})$ fall behind the accuracy of the other methods. 
M2B2 and M3B2 yield the overall best results followed by CLI.
The large errors in MR for the lowest resolution can again be explained by the unsteady results there. 
When analyzing the RMS errors of $c_p$ along $\theta_1$ and $\theta_2$, as done in Fig.~\ref{fig:G178_err_cp}, BB performs significantly worse than the rest. 
The results of all other methods are comparable with slightly more accurate results for PSM at low resolutions.

To summarize the findings for this regime, again M1B1 and now also BB show significantly worse results compared to the other variants.
The MEM variants seem to be better suited for predicting the vertical sphere velocity whereas the PSM variants work very well for the horizontal sphere velocity.
The flow properties are best represented by M2B2 followed by CLI and M3B2.
The higher order MR scheme seems to destabilize the system as the results show oscillations which are expected to occur only at larger Galileo numbers like the one used for the next regime.

\subsection{Oscillating oblique regime}
\label{sec:regime3}

\begin{figure}[t]
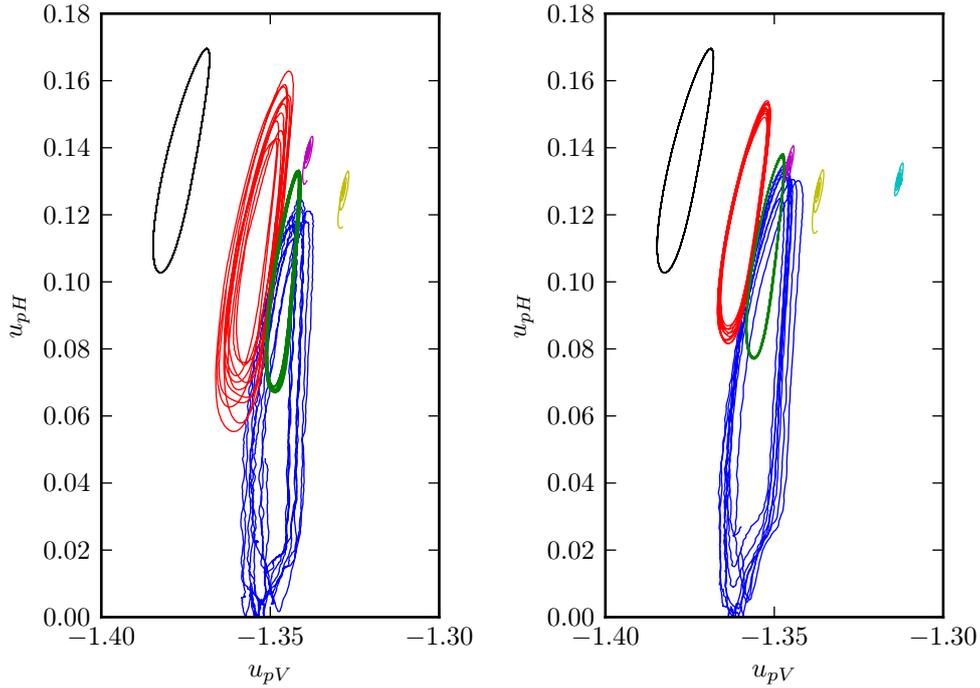

	\centering
	\begin{subfigure}{0.45\textwidth}
		\centering
		\input{plot_190_36_plot_upv_uph.pgf}
	\end{subfigure}
	\begin{subfigure}{0.45\textwidth}
		\centering
		\input{plot_190_48_plot_upv_uph.pgf}
	\end{subfigure}
	\caption{Phase-space plot of the vertical and horizontal sphere velocity components for case C ($Ga=190$) with $D/\Delta x=36$ (left) and $48$ (right). {\color{blue} \textbf{---}} BB, {\color{OliveGreen} \textbf{---}} CLI, {\color{red}\textbf{---}} MR, {\color{Turquoise} \textbf{---}} M1B1, {\color{Mulberry}\textbf{---}} M2B2, {\color{GreenYellow}\textbf{---}} M3B2, {\color{black}\textbf{---}} reference.}
	\label{fig:G190_upv_uph}
\end{figure}

For a Galileo number of $190$ ($Re_{\parallel} \approx 263$), the sphere is expected to exhibit an oscillating motion with a constant frequency $f$.
As explained in \cite{uhlmann_motion_2014}, this behavior is triggered by a secondary bifurcation and is only present in a rather narrow range of Galileo numbers.
Numerical inaccuracies can lead to a shift of the bifurcation point and the respective method will then fail to capture this motion at $Ga=190$.
In Fig.~\ref{fig:G190_upv_uph}, a phase-space diagram of the results for the different coupling methods together with the reference data is shown for the two resolutions $D/\Delta x=36$ and $48$.
The expected time-periodic behavior is a closed curve around a fixed midpoint.
Even for the finer resolution, only CLI and MR are able to capture this oscillating motion accurately. 
Oscillations can also be found for BB but the amplitude in $u_{pH}$ is too large and the value for $u_{pV}$ around which the curve oscillates is slightly changing in time. 
On the other hand, all PSM variants yield exponentially decaying oscillations and thus fail to capture this instability.
It is worth to note that CLI is also able to reproduce the time-periodic oscillations with a resolution of $D/\Delta x = 36$, whereas MR shows strong deviations from a closed curve.
This motion can be analyzed in more detail by calculating the time average and fluctuation values of the different sphere velocities. 
These values are given in Tab.~\ref{tab:G190}, where $\overline{\phi}$ denotes the average and $\phi'$ the fluctuation part of a quantity $\phi$.
Their exact definitions can be found in \cite{uhlmann_motion_2014}. 
Table~\ref{tab:G190} also shows the frequency of the oscillation calculated with the help of a discrete Fourier transformation.
It can be seen that the average of $u_{pV}$ is captured well by the MEM variants with errors well below $2\%$ for the fine resolution. 
In contrast to that, the errors of the PSM variants are larger and always above $2\%$.
They, however, predict the average of $u_{pH}$ significantly better than the MEM counterparts with errors below $1\%$ and approximately independent of the resolution.
The MEM variants tend to underestimate this quantity.
Regarding $\overline{\omega}_{pH}$, all methods yield comparable results except for BB being off by a factor of three and M3B2 which is particularly close to the reference results.
Having a look at the fluctuation parts, all PSM variants show large deviations as expected from their inability to capture the oscillating motion already discussed in the phase-space diagram in Fig.~\ref{fig:G190_upv_uph}.
BB shows too large fluctuations, whereas CLI and MR offer overall good predictions, especially for the finer resolution. 
Comparing the obtained oscillation frequencies $f$, CLI and MR can be considered again acceptable whereas BB underestimates the frequency considerably.
All PSM variants predict the frequency more accurately than the MEM variants.

Summarizing, the PSM variants cannot capture the expected oscillating motion in this regime. 
However, they perform well in predicting the frequency and the temporal averages of the horizontal quantities, with M2B2 and M3B2 being the best choice.
The MEM variants show two distinct behaviors.
BB is too inaccurate and, except for $\overline{u}_{pV}$, all its values deviate significantly from the reference values.
CLI and MR, on the other hand, are the methods that overall represent the expected trajectories best and additionally yield acceptable results for the rotational velocity and the frequency. 

\subsection{Chaotic regime}
\label{sec:regime4}

\begin{figure}[t]
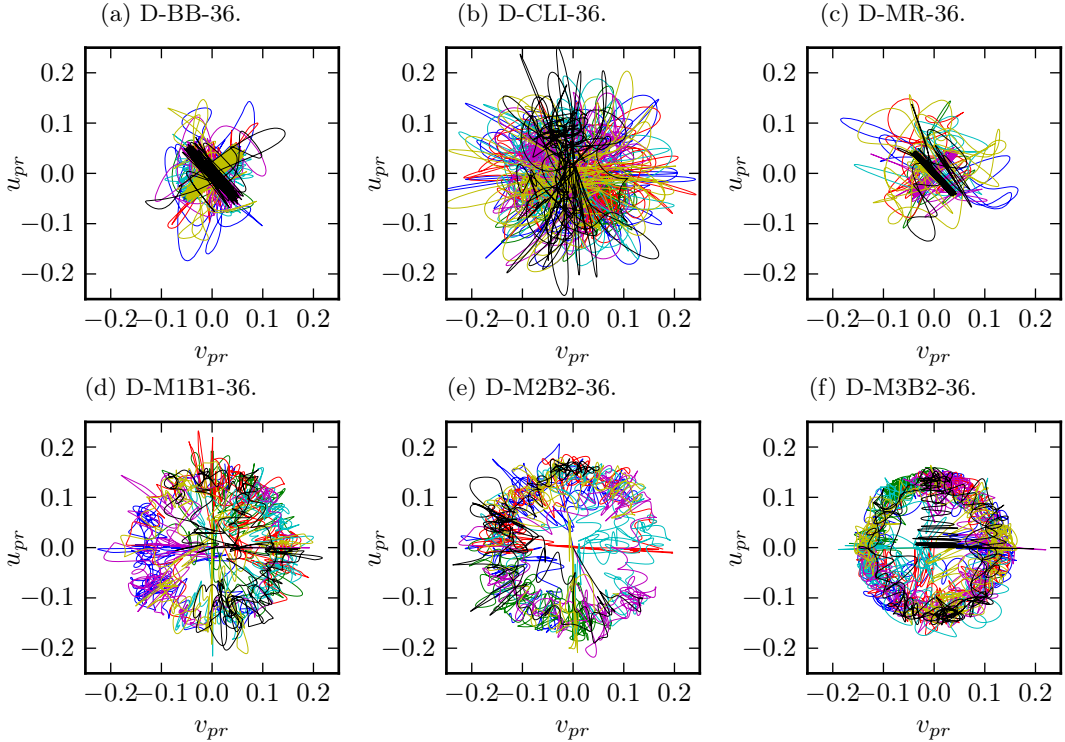

	\centering
	\begin{subfigure}{0.32\textwidth}
		\centering
		\caption{D-BB-36.}
		\input{plot_250_BB_vpr_upr.pgf}	
	\end{subfigure}
	\begin{subfigure}{0.32\textwidth}
		\centering
		\caption{D-CLI-36.}
		\input{plot_250_CLI_vpr_upr.pgf}
	\end{subfigure}
	\begin{subfigure}{0.32\textwidth}
		\centering
		\caption{D-MR-36.}
		\input{plot_250_MR_vpr_upr.pgf}
	\end{subfigure}
	\begin{subfigure}{0.32\textwidth}
		\centering
		\caption{D-M1B1-36.}
		\input{plot_250_M1B1_vpr_upr.pgf}
	\end{subfigure}
	\begin{subfigure}{0.32\textwidth}
		\centering
		\caption{D-M2B2-36.}
		\input{plot_250_M2B2_vpr_upr.pgf}
	\end{subfigure}
	\begin{subfigure}{0.32\textwidth}
		\centering
		\caption{D-M3B2-36.}
		\input{plot_250_M3B2_vpr_upr.pgf}
	\end{subfigure}	
	\caption{Phase-space plot of the two horizontal sphere velocity components for case D ($Ga=250$) with $D/\Delta x=36$. The colors depict the results of the seven sampling runs.}
	\label{fig:G250_vpr_upr}
\end{figure}

The sphere sets into a chaotic motion when the Galileo number is increased to $250$, corresponding to a particle Reynolds number of around $365$.
In order to obtain reliable data for a comparison with the reference values, the simulation for this regime is run seven times.
Each time, the initial sphere position is slightly perturbed at random in the horizontal plane resulting in identical physical conditions for each run.
All samples combined amount to a total dimensionless simulation time of $3570$ units.
These seven samples are then used to calculate average and fluctuation quantities as defined in \cite{uhlmann_motion_2014}.
Fig.~\ref{fig:G250_vpr_upr} shows the phase-space diagrams of the horizontal velocity components of the seven samples.
The BB variant, like the MR, shows only a few chaotic trajectories at the beginning of each sample which quickly settle into a rather time-periodic motion.
In contrast to that, the motion obtained by CLI can be considered chaotic.
All PSM variants exhibit an essentially circular structure in the phase-space with distinct chaotic oscillations inside this structure.
More insight can be gained from the sample averages and the average amplitudes of the instantaneous fluctuations, given in Tab.~\ref{tab:G250}.
The only non-zero sample average, $\langle u_{pV} \rangle$, is captured reasonably well by all methods with CLI and MR showing the best results.
The fluctuation of $u_{pV}$ is best reproduced by CLI, closely followed by the PSM variants.
For $u_{pr}$ fluctuations, M3B2 gives the best result followed by M1B1 and CLI.
Regarding the fluctuation values for the rotational velocities, CLI is again the closest to the reference value for $\omega_{pV}$. 
Remarkably, M3B2's value is off by a factor of $10$ whereas the other two PSM variants perform quite well.
The last quantity, the fluctuation of $\omega_{px}$, seems to be especially difficult to capture since only M3B2 comes close to the reference value. 
BB and MR show large deviations for almost all fluctuation quantities due to the essentially non-chaotic movement. 

\begin{figure}[t]
	\centering
	\begin{subfigure}{0.48\textwidth}
		\centering
		\input{G250_CLI_plot_pdf_translational.pgf}
	\end{subfigure}
	\begin{subfigure}{0.48\textwidth}
		\centering
		\input{G250_M3B2_plot_pdf_translational.pgf}
	\end{subfigure}
	\caption{Normalized probability density function of translational sphere velocity for D-CLI-36 (left) and D-M3B2-36 (right) with $Ga=250$. {\color{black} \textbf{---}} vertical component $u_{pV}$, {\color{red} \textbf{---}} horizontal component $u_{pr}$, {\color{blue} \textbf{---}} Gaussian reference curve. The reference data from \cite{uhlmann_motion_2014} is shown as dashed lines.}
	\label{fig:G250_pdf_trans}
\end{figure}

\begin{figure}[t]
	\centering
	\begin{subfigure}{0.48\textwidth}
		\centering
		\input{G250_CLI_plot_pdf_rotational.pgf}
	\end{subfigure}
	\begin{subfigure}{0.48\textwidth}
		\centering
		\input{G250_M3B2_plot_pdf_rotational.pgf}
	\end{subfigure}
	\caption{Normalized probability density function of rotational sphere velocity for D-CLI-36 (left) and D-M3B2-36 (right) with $Ga=250$. {\color{black} \textbf{---}} vertical component $\omega_{pV}$, {\color{red} \textbf{---}} horizontal component $\omega_{px}$, {\color{blue} \textbf{---}} Gaussian reference curve. The reference data from \cite{uhlmann_motion_2014} is shown as dashed lines.}
	\label{fig:G250_pdf_rot}
\end{figure}

\begin{figure}[t]
	\centering
	\begin{subfigure}{0.48\textwidth}
		\centering
		\input{G250_CLI_plot_autocorr_translational.pgf}
	\end{subfigure}
	\begin{subfigure}{0.48\textwidth}
		\centering
		\input{G250_M3B2_plot_autocorr_translational.pgf}
	\end{subfigure}
	\caption{Temporal auto-correlation of translational sphere velocity for D-CLI-36 (left) and D-M3B2-36 (right) with $Ga=250$. {\color{black} \textbf{---}} vertical component $u_{pV}$, {\color{red} \textbf{---}} horizontal component $u_{pr}$. The reference data from \cite{uhlmann_motion_2014} is shown as dashed lines.}
	\label{fig:G250_autocorr_trans}
\end{figure}
\begin{figure}[t]
	\centering
	\begin{subfigure}{0.48\textwidth}
		\centering
		\input{G250_CLI_plot_autocorr_rotational.pgf}
	\end{subfigure}
	\begin{subfigure}{0.48\textwidth}
		\centering
		\input{G250_M3B2_plot_autocorr_rotational.pgf}
	\end{subfigure}
	\caption{Temporal auto-correlation of rotational sphere velocity for D-CLI-36 (left) and D-M3B2-36 (right) with $Ga=250$. {\color{black} \textbf{---}} vertical component $\omega_{pV}$, {\color{red} \textbf{---}} horizontal component $\omega_{px}$. The reference data from \cite{uhlmann_motion_2014} is shown as dashed lines.}
	\label{fig:G250_autocorr_rot}
\end{figure}

Apart from these statistical quantities, also the probability density functions (pdfs) of the translational and rotational sphere velocities, Figs. \ref{fig:G250_pdf_trans} and \ref{fig:G250_pdf_rot}, as well as their temporal auto-correlations, Figs. \ref{fig:G250_autocorr_trans} and \ref{fig:G250_autocorr_rot}, can be compared to the reference values.
Only the cases D-CLI-36 and D-M3B2-36 are shown as BB and MR could not capture the chaotic motion and the results of the PSM variants do not differ much.

Analyzing the graphs in Fig.~\ref{fig:G250_pdf_trans} first, CLI shows a clearly visible peak in the vertical component which creates a deviation from the assumed Gaussian shape. 
The plateau-like shape of $u_{pr}$ is present with a less steep descent around the value for the standard deviation $\sigma$.
The PSM methods, on the other hand, show this plateau very distinctly and feature a sharp drop like the reference data.
The vertical component is more pronounced towards negative fluctuations.
Regarding the pdfs of the rotational velocities, the CLI results capture the Gaussian distribution quite well and also show the expected mild peak in the middle for the vertical component.
The PSM methods feature a slightly narrower distribution for both, the vertical and horizontal component, and also exhibit the mild peak in $\omega_{pV}$.

Finally, the auto-correlations for both coupling techniques are compared.
CLI captures the decay and dominant frequencies of the vertical translational sphere velocity very well, but shows deviations for the horizontal component with a too fast decay.
As opposed to this, the M3B2 results reproduce the horizontal component accurately but contain a higher frequency in the vertical component.
In Fig.~\ref{fig:G250_autocorr_rot}, CLI shows the same frequency in the vertical rotational sphere velocity as the reference case with less fluctuations. 
Again, the horizontal component's auto-correlation deviates from the reference in a way that the decay is slower and the frequency lower.
The PSM method exhibits a very slow decay in both components but the frequency of the small visible fluctuations matches the reference ones.

Putting the findings for the chaotic regime together, CLI seems to be a better choice to capture the characteristics of vertical velocity components more accurately. 
The PSM variants have their strength in the prediction of the horizontal components where again M2B2 and M3B2 perform better than M1B1.

\subsection{Discussion}
\label{sec:discussion}
Based on all results from the different tested regimes, substantial differences between the applied fluid-solid coupling methods can be seen.
The momentum exchange method consistently yields good approximations of the sphere's streamwise velocity component whereas the partially saturated cells method seems to have its strength in predicting the spanwise component.
Apart from those general differences of the two approaches, also their variants show large variations in some cases. 

The very common bounce back scheme shows good agreement with the reference data for the small Reynolds number regime but then clearly falls behind the other methods for larger Reynolds numbers. 
It cannot reproduce the desired oscillating and chaotic motion of cases C and D even at high resolutions.
When applying the central linear interpolation scheme, which makes use of the exact surface position for achieving better accuracy, the results improve drastically.
The CLI scheme can reproduce the expected characteristics of all regimes with good accuracy and yields acceptable results already at lower resolutions.
This comes at the cost of a slightly increased computational effort compared to BB but the results clearly demonstrate that this extra cost is worth the effort.
A further improvement by the multi-reflection scheme can only be observed at lower Reynolds numbers which is, however, not significant. 
The MR scheme seems to destabilize the system as the instability that triggers the oscillatory motion is already present at lower Galileo numbers.
This behavior could potentially be improved by applying even finer resolutions.
But here, this higher order scheme does not yet pay off at the coarser resolutions that are relevant for current computational practice.
Another noticeable drawback of MR is the correction term, see Eq.~\eqref{eq:MEM_MR}, that requires old PDF values which have to be stored additionally.
This leads to a noticeable slow down of the method.
As CLI seems to work better and more robust than BB or MR for this benchmark scenario, its usage is recommended when applying the momentum exchange coupling for particulate flows.

Comparing the variants of the partially saturated cells method, the M1B1 version clearly falls behind the other two schemes in terms of accuracy.
As it also does not offer any additional benefits, this version is not considered favorable for coupled simulations.
On the other hand, M2B2 and M3B2 both are viable choices that yield reasonable accuracy for the tested regimes.
This could mean that it is advisable to apply Eq.~\eqref{eq:PSM_B2} instead of Eq.~\eqref{eq:PSM_B1} in these methods
This is an interesting finding also for possible future developments and improvements of the PSM formulation. 
All PSM methods have in common that they seemingly introduce more dissipation into the system.
They shift the threshold for instabilities to occur towards higher Galileo numbers and they produce damped oscillations and less chaotic motion.
Such a behavior is also reported for the immersed boundary method applied in \cite{uhlmann_motion_2014} in the context of finite difference computations which could hint towards a common feature of the boundary treatment through immersed interface methods.
In fact, when comparing the PSM collision operator, Eq.~\eqref{eq:PSM}, to the BGK collision operator, Eq.~\eqref{eq:LBM_SRT}, one could argue that the PSM method is a BGK model with modified relaxation time $\tau^{\text{PSM}} = \tfrac{\tau}{(1-B)}$ and an additional external forcing operator given by the second term of Eq.~\eqref{eq:PSM}, i.e. $\sum_s B_s \mathcal{C}_{q,s}^{\text{solid}}$. 
As a consequence, $\tau^{\text{PSM}}$ increases in the vicinity of a particle and thus the apparent fluid viscosity, Eq.~\eqref{eq:Viscosity}, locally increases as well.

It remains to note that all applied LBM coupling methods converge to values for the recirculation length that are smaller than the reference ones. 
The exact reason for this behavior is unclear since all other investigated quantities converge to the reference values.
Issues with the current implementation are therefore unlikely.
Possible influences originating from the applied inflow and outflow boundary conditions can also be excluded as the results remain unchanged when extending the domain by $16D$ in positive and negative $z$ direction and thus increasing the vertical domain size threefold.
We note, however, that evaluating the recirculation length requires extensive interpolations of the velocity values and here differences in the implementation (also for the reference data of \cite{uhlmann_motion_2014}) can easily result in substantially different values.
In the present work, the $x_{pH}-z-$plane is first triangulated in the vicinity of the assumed recirculation point and then the velocity values are interpolated in a cubic trilinear fashion to obtain the contour lines using VTK 6.3.0. functionalities.

%% file: plot_144_D_rms_errors_cp.pgf
%% Creator: Matplotlib, PGF backend
%%
%% To include the figure in your LaTeX document, write
%%   \input{<filename>.pgf}
%%
%% Make sure the required packages are loaded in your preamble
%%   \usepackage{pgf}
%%
%% Figures using additional raster images can only be included by \input if
%% they are in the same directory as the main LaTeX file. For loading figures
%% from other directories you can use the `import` package
%%   \usepackage{import}
%% and then include the figures with
%%   \import{<path to file>}{<filename>.pgf}
%%
%% Matplotlib used the following preamble
%%   \usepackage{amsmath}
%%
\begingroup%
\makeatletter%
\begin{pgfpicture}%
\pgfpathrectangle{\pgfpointorigin}{\pgfqpoint{2.765165in}{2.206558in}}%
\pgfusepath{use as bounding box, clip}%
\begin{pgfscope}%
\pgfsetbuttcap%
\pgfsetmiterjoin%
\definecolor{currentfill}{rgb}{1.000000,1.000000,1.000000}%
\pgfsetfillcolor{currentfill}%
\pgfsetlinewidth{0.000000pt}%
\definecolor{currentstroke}{rgb}{1.000000,1.000000,1.000000}%
\pgfsetstrokecolor{currentstroke}%
\pgfsetdash{}{0pt}%
\pgfpathmoveto{\pgfqpoint{0.000000in}{0.000000in}}%
\pgfpathlineto{\pgfqpoint{2.765165in}{0.000000in}}%
\pgfpathlineto{\pgfqpoint{2.765165in}{2.206558in}}%
\pgfpathlineto{\pgfqpoint{0.000000in}{2.206558in}}%
\pgfpathclose%
\pgfusepath{fill}%
\end{pgfscope}%
\begin{pgfscope}%
\pgfsetbuttcap%
\pgfsetmiterjoin%
\definecolor{currentfill}{rgb}{1.000000,1.000000,1.000000}%
\pgfsetfillcolor{currentfill}%
\pgfsetlinewidth{0.000000pt}%
\definecolor{currentstroke}{rgb}{0.000000,0.000000,0.000000}%
\pgfsetstrokecolor{currentstroke}%
\pgfsetstrokeopacity{0.000000}%
\pgfsetdash{}{0pt}%
\pgfpathmoveto{\pgfqpoint{0.471915in}{0.279012in}}%
\pgfpathlineto{\pgfqpoint{2.665165in}{0.279012in}}%
\pgfpathlineto{\pgfqpoint{2.665165in}{1.879012in}}%
\pgfpathlineto{\pgfqpoint{0.471915in}{1.879012in}}%
\pgfpathclose%
\pgfusepath{fill}%
\end{pgfscope}%
\begin{pgfscope}%
\pgfpathrectangle{\pgfqpoint{0.471915in}{0.279012in}}{\pgfqpoint{2.193250in}{1.600000in}} %
\pgfusepath{clip}%
\pgfsetrectcap%
\pgfsetroundjoin%
\pgfsetlinewidth{1.003750pt}%
\definecolor{currentstroke}{rgb}{0.000000,0.000000,1.000000}%
\pgfsetstrokecolor{currentstroke}%
\pgfsetdash{}{0pt}%
\pgfpathmoveto{\pgfqpoint{0.659907in}{1.810318in}}%
\pgfpathlineto{\pgfqpoint{1.035893in}{1.669339in}}%
\pgfpathlineto{\pgfqpoint{1.787865in}{0.555235in}}%
\pgfpathlineto{\pgfqpoint{2.539836in}{0.652432in}}%
\pgfusepath{stroke}%
\end{pgfscope}%
\begin{pgfscope}%
\pgfpathrectangle{\pgfqpoint{0.471915in}{0.279012in}}{\pgfqpoint{2.193250in}{1.600000in}} %
\pgfusepath{clip}%
\pgfsetbuttcap%
\pgfsetroundjoin%
\definecolor{currentfill}{rgb}{0.000000,0.000000,1.000000}%
\pgfsetfillcolor{currentfill}%
\pgfsetlinewidth{0.501875pt}%
\definecolor{currentstroke}{rgb}{0.000000,0.000000,0.000000}%
\pgfsetstrokecolor{currentstroke}%
\pgfsetdash{}{0pt}%
\pgfsys@defobject{currentmarker}{\pgfqpoint{-0.027778in}{-0.027778in}}{\pgfqpoint{0.027778in}{0.027778in}}{%
\pgfpathmoveto{\pgfqpoint{0.000000in}{-0.027778in}}%
\pgfpathcurveto{\pgfqpoint{0.007367in}{-0.027778in}}{\pgfqpoint{0.014433in}{-0.024851in}}{\pgfqpoint{0.019642in}{-0.019642in}}%
\pgfpathcurveto{\pgfqpoint{0.024851in}{-0.014433in}}{\pgfqpoint{0.027778in}{-0.007367in}}{\pgfqpoint{0.027778in}{0.000000in}}%
\pgfpathcurveto{\pgfqpoint{0.027778in}{0.007367in}}{\pgfqpoint{0.024851in}{0.014433in}}{\pgfqpoint{0.019642in}{0.019642in}}%
\pgfpathcurveto{\pgfqpoint{0.014433in}{0.024851in}}{\pgfqpoint{0.007367in}{0.027778in}}{\pgfqpoint{0.000000in}{0.027778in}}%
\pgfpathcurveto{\pgfqpoint{-0.007367in}{0.027778in}}{\pgfqpoint{-0.014433in}{0.024851in}}{\pgfqpoint{-0.019642in}{0.019642in}}%
\pgfpathcurveto{\pgfqpoint{-0.024851in}{0.014433in}}{\pgfqpoint{-0.027778in}{0.007367in}}{\pgfqpoint{-0.027778in}{0.000000in}}%
\pgfpathcurveto{\pgfqpoint{-0.027778in}{-0.007367in}}{\pgfqpoint{-0.024851in}{-0.014433in}}{\pgfqpoint{-0.019642in}{-0.019642in}}%
\pgfpathcurveto{\pgfqpoint{-0.014433in}{-0.024851in}}{\pgfqpoint{-0.007367in}{-0.027778in}}{\pgfqpoint{0.000000in}{-0.027778in}}%
\pgfpathclose%
\pgfusepath{stroke,fill}%
}%
\begin{pgfscope}%
\pgfsys@transformshift{0.659907in}{1.810318in}%
\pgfsys@useobject{currentmarker}{}%
\end{pgfscope}%
\begin{pgfscope}%
\pgfsys@transformshift{1.035893in}{1.669339in}%
\pgfsys@useobject{currentmarker}{}%
\end{pgfscope}%
\begin{pgfscope}%
\pgfsys@transformshift{1.787865in}{0.555235in}%
\pgfsys@useobject{currentmarker}{}%
\end{pgfscope}%
\begin{pgfscope}%
\pgfsys@transformshift{2.539836in}{0.652432in}%
\pgfsys@useobject{currentmarker}{}%
\end{pgfscope}%
\end{pgfscope}%
\begin{pgfscope}%
\pgfpathrectangle{\pgfqpoint{0.471915in}{0.279012in}}{\pgfqpoint{2.193250in}{1.600000in}} %
\pgfusepath{clip}%
\pgfsetrectcap%
\pgfsetroundjoin%
\pgfsetlinewidth{1.003750pt}%
\definecolor{currentstroke}{rgb}{0.000000,0.500000,0.000000}%
\pgfsetstrokecolor{currentstroke}%
\pgfsetdash{}{0pt}%
\pgfpathmoveto{\pgfqpoint{0.659907in}{1.494718in}}%
\pgfpathlineto{\pgfqpoint{1.035893in}{0.836600in}}%
\pgfpathlineto{\pgfqpoint{1.787865in}{0.408390in}}%
\pgfpathlineto{\pgfqpoint{2.539836in}{0.311441in}}%
\pgfusepath{stroke}%
\end{pgfscope}%
\begin{pgfscope}%
\pgfpathrectangle{\pgfqpoint{0.471915in}{0.279012in}}{\pgfqpoint{2.193250in}{1.600000in}} %
\pgfusepath{clip}%
\pgfsetbuttcap%
\pgfsetbeveljoin%
\definecolor{currentfill}{rgb}{0.000000,0.500000,0.000000}%
\pgfsetfillcolor{currentfill}%
\pgfsetlinewidth{0.501875pt}%
\definecolor{currentstroke}{rgb}{0.000000,0.000000,0.000000}%
\pgfsetstrokecolor{currentstroke}%
\pgfsetdash{}{0pt}%
\pgfsys@defobject{currentmarker}{\pgfqpoint{-0.026418in}{-0.022473in}}{\pgfqpoint{0.026418in}{0.027778in}}{%
\pgfpathmoveto{\pgfqpoint{0.000000in}{0.027778in}}%
\pgfpathlineto{\pgfqpoint{-0.006236in}{0.008584in}}%
\pgfpathlineto{\pgfqpoint{-0.026418in}{0.008584in}}%
\pgfpathlineto{\pgfqpoint{-0.010091in}{-0.003279in}}%
\pgfpathlineto{\pgfqpoint{-0.016327in}{-0.022473in}}%
\pgfpathlineto{\pgfqpoint{-0.000000in}{-0.010610in}}%
\pgfpathlineto{\pgfqpoint{0.016327in}{-0.022473in}}%
\pgfpathlineto{\pgfqpoint{0.010091in}{-0.003279in}}%
\pgfpathlineto{\pgfqpoint{0.026418in}{0.008584in}}%
\pgfpathlineto{\pgfqpoint{0.006236in}{0.008584in}}%
\pgfpathclose%
\pgfusepath{stroke,fill}%
}%
\begin{pgfscope}%
\pgfsys@transformshift{0.659907in}{1.494718in}%
\pgfsys@useobject{currentmarker}{}%
\end{pgfscope}%
\begin{pgfscope}%
\pgfsys@transformshift{1.035893in}{0.836600in}%
\pgfsys@useobject{currentmarker}{}%
\end{pgfscope}%
\begin{pgfscope}%
\pgfsys@transformshift{1.787865in}{0.408390in}%
\pgfsys@useobject{currentmarker}{}%
\end{pgfscope}%
\begin{pgfscope}%
\pgfsys@transformshift{2.539836in}{0.311441in}%
\pgfsys@useobject{currentmarker}{}%
\end{pgfscope}%
\end{pgfscope}%
\begin{pgfscope}%
\pgfpathrectangle{\pgfqpoint{0.471915in}{0.279012in}}{\pgfqpoint{2.193250in}{1.600000in}} %
\pgfusepath{clip}%
\pgfsetrectcap%
\pgfsetroundjoin%
\pgfsetlinewidth{1.003750pt}%
\definecolor{currentstroke}{rgb}{1.000000,0.000000,0.000000}%
\pgfsetstrokecolor{currentstroke}%
\pgfsetdash{}{0pt}%
\pgfpathmoveto{\pgfqpoint{0.659907in}{1.875305in}}%
\pgfpathlineto{\pgfqpoint{1.035893in}{0.881226in}}%
\pgfpathlineto{\pgfqpoint{1.787865in}{0.396757in}}%
\pgfpathlineto{\pgfqpoint{2.539836in}{0.335956in}}%
\pgfusepath{stroke}%
\end{pgfscope}%
\begin{pgfscope}%
\pgfpathrectangle{\pgfqpoint{0.471915in}{0.279012in}}{\pgfqpoint{2.193250in}{1.600000in}} %
\pgfusepath{clip}%
\pgfsetbuttcap%
\pgfsetmiterjoin%
\definecolor{currentfill}{rgb}{1.000000,0.000000,0.000000}%
\pgfsetfillcolor{currentfill}%
\pgfsetlinewidth{0.501875pt}%
\definecolor{currentstroke}{rgb}{0.000000,0.000000,0.000000}%
\pgfsetstrokecolor{currentstroke}%
\pgfsetdash{}{0pt}%
\pgfsys@defobject{currentmarker}{\pgfqpoint{-0.027778in}{-0.027778in}}{\pgfqpoint{0.027778in}{0.027778in}}{%
\pgfpathmoveto{\pgfqpoint{-0.027778in}{-0.027778in}}%
\pgfpathlineto{\pgfqpoint{0.027778in}{-0.027778in}}%
\pgfpathlineto{\pgfqpoint{0.027778in}{0.027778in}}%
\pgfpathlineto{\pgfqpoint{-0.027778in}{0.027778in}}%
\pgfpathclose%
\pgfusepath{stroke,fill}%
}%
\begin{pgfscope}%
\pgfsys@transformshift{0.659907in}{1.875305in}%
\pgfsys@useobject{currentmarker}{}%
\end{pgfscope}%
\begin{pgfscope}%
\pgfsys@transformshift{1.035893in}{0.881226in}%
\pgfsys@useobject{currentmarker}{}%
\end{pgfscope}%
\begin{pgfscope}%
\pgfsys@transformshift{1.787865in}{0.396757in}%
\pgfsys@useobject{currentmarker}{}%
\end{pgfscope}%
\begin{pgfscope}%
\pgfsys@transformshift{2.539836in}{0.335956in}%
\pgfsys@useobject{currentmarker}{}%
\end{pgfscope}%
\end{pgfscope}%
\begin{pgfscope}%
\pgfpathrectangle{\pgfqpoint{0.471915in}{0.279012in}}{\pgfqpoint{2.193250in}{1.600000in}} %
\pgfusepath{clip}%
\pgfsetrectcap%
\pgfsetroundjoin%
\pgfsetlinewidth{1.003750pt}%
\definecolor{currentstroke}{rgb}{0.000000,0.750000,0.750000}%
\pgfsetstrokecolor{currentstroke}%
\pgfsetdash{}{0pt}%
\pgfpathmoveto{\pgfqpoint{0.659907in}{1.211481in}}%
\pgfpathlineto{\pgfqpoint{1.035893in}{0.999729in}}%
\pgfpathlineto{\pgfqpoint{1.787865in}{0.618795in}}%
\pgfpathlineto{\pgfqpoint{2.539836in}{0.479646in}}%
\pgfusepath{stroke}%
\end{pgfscope}%
\begin{pgfscope}%
\pgfpathrectangle{\pgfqpoint{0.471915in}{0.279012in}}{\pgfqpoint{2.193250in}{1.600000in}} %
\pgfusepath{clip}%
\pgfsetbuttcap%
\pgfsetmiterjoin%
\definecolor{currentfill}{rgb}{0.000000,0.750000,0.750000}%
\pgfsetfillcolor{currentfill}%
\pgfsetlinewidth{0.501875pt}%
\definecolor{currentstroke}{rgb}{0.000000,0.000000,0.000000}%
\pgfsetstrokecolor{currentstroke}%
\pgfsetdash{}{0pt}%
\pgfsys@defobject{currentmarker}{\pgfqpoint{-0.027778in}{-0.027778in}}{\pgfqpoint{0.027778in}{0.027778in}}{%
\pgfpathmoveto{\pgfqpoint{-0.027778in}{0.000000in}}%
\pgfpathlineto{\pgfqpoint{0.027778in}{-0.027778in}}%
\pgfpathlineto{\pgfqpoint{0.027778in}{0.027778in}}%
\pgfpathclose%
\pgfusepath{stroke,fill}%
}%
\begin{pgfscope}%
\pgfsys@transformshift{0.659907in}{1.211481in}%
\pgfsys@useobject{currentmarker}{}%
\end{pgfscope}%
\begin{pgfscope}%
\pgfsys@transformshift{1.035893in}{0.999729in}%
\pgfsys@useobject{currentmarker}{}%
\end{pgfscope}%
\begin{pgfscope}%
\pgfsys@transformshift{1.787865in}{0.618795in}%
\pgfsys@useobject{currentmarker}{}%
\end{pgfscope}%
\begin{pgfscope}%
\pgfsys@transformshift{2.539836in}{0.479646in}%
\pgfsys@useobject{currentmarker}{}%
\end{pgfscope}%
\end{pgfscope}%
\begin{pgfscope}%
\pgfpathrectangle{\pgfqpoint{0.471915in}{0.279012in}}{\pgfqpoint{2.193250in}{1.600000in}} %
\pgfusepath{clip}%
\pgfsetrectcap%
\pgfsetroundjoin%
\pgfsetlinewidth{1.003750pt}%
\definecolor{currentstroke}{rgb}{0.750000,0.000000,0.750000}%
\pgfsetstrokecolor{currentstroke}%
\pgfsetdash{}{0pt}%
\pgfpathmoveto{\pgfqpoint{0.659907in}{1.635995in}}%
\pgfpathlineto{\pgfqpoint{1.035893in}{0.839220in}}%
\pgfpathlineto{\pgfqpoint{1.787865in}{0.431635in}}%
\pgfpathlineto{\pgfqpoint{2.539836in}{0.397801in}}%
\pgfusepath{stroke}%
\end{pgfscope}%
\begin{pgfscope}%
\pgfpathrectangle{\pgfqpoint{0.471915in}{0.279012in}}{\pgfqpoint{2.193250in}{1.600000in}} %
\pgfusepath{clip}%
\pgfsetbuttcap%
\pgfsetmiterjoin%
\definecolor{currentfill}{rgb}{0.750000,0.000000,0.750000}%
\pgfsetfillcolor{currentfill}%
\pgfsetlinewidth{0.501875pt}%
\definecolor{currentstroke}{rgb}{0.000000,0.000000,0.000000}%
\pgfsetstrokecolor{currentstroke}%
\pgfsetdash{}{0pt}%
\pgfsys@defobject{currentmarker}{\pgfqpoint{-0.027778in}{-0.027778in}}{\pgfqpoint{0.027778in}{0.027778in}}{%
\pgfpathmoveto{\pgfqpoint{0.027778in}{-0.000000in}}%
\pgfpathlineto{\pgfqpoint{-0.027778in}{0.027778in}}%
\pgfpathlineto{\pgfqpoint{-0.027778in}{-0.027778in}}%
\pgfpathclose%
\pgfusepath{stroke,fill}%
}%
\begin{pgfscope}%
\pgfsys@transformshift{0.659907in}{1.635995in}%
\pgfsys@useobject{currentmarker}{}%
\end{pgfscope}%
\begin{pgfscope}%
\pgfsys@transformshift{1.035893in}{0.839220in}%
\pgfsys@useobject{currentmarker}{}%
\end{pgfscope}%
\begin{pgfscope}%
\pgfsys@transformshift{1.787865in}{0.431635in}%
\pgfsys@useobject{currentmarker}{}%
\end{pgfscope}%
\begin{pgfscope}%
\pgfsys@transformshift{2.539836in}{0.397801in}%
\pgfsys@useobject{currentmarker}{}%
\end{pgfscope}%
\end{pgfscope}%
\begin{pgfscope}%
\pgfpathrectangle{\pgfqpoint{0.471915in}{0.279012in}}{\pgfqpoint{2.193250in}{1.600000in}} %
\pgfusepath{clip}%
\pgfsetrectcap%
\pgfsetroundjoin%
\pgfsetlinewidth{1.003750pt}%
\definecolor{currentstroke}{rgb}{0.750000,0.750000,0.000000}%
\pgfsetstrokecolor{currentstroke}%
\pgfsetdash{}{0pt}%
\pgfpathmoveto{\pgfqpoint{0.659907in}{1.311930in}}%
\pgfpathlineto{\pgfqpoint{1.035893in}{0.670827in}}%
\pgfpathlineto{\pgfqpoint{1.787865in}{0.493627in}}%
\pgfpathlineto{\pgfqpoint{2.539836in}{0.626875in}}%
\pgfusepath{stroke}%
\end{pgfscope}%
\begin{pgfscope}%
\pgfpathrectangle{\pgfqpoint{0.471915in}{0.279012in}}{\pgfqpoint{2.193250in}{1.600000in}} %
\pgfusepath{clip}%
\pgfsetbuttcap%
\pgfsetmiterjoin%
\definecolor{currentfill}{rgb}{0.750000,0.750000,0.000000}%
\pgfsetfillcolor{currentfill}%
\pgfsetlinewidth{0.501875pt}%
\definecolor{currentstroke}{rgb}{0.000000,0.000000,0.000000}%
\pgfsetstrokecolor{currentstroke}%
\pgfsetdash{}{0pt}%
\pgfsys@defobject{currentmarker}{\pgfqpoint{-0.027778in}{-0.027778in}}{\pgfqpoint{0.027778in}{0.027778in}}{%
\pgfpathmoveto{\pgfqpoint{-0.000000in}{-0.027778in}}%
\pgfpathlineto{\pgfqpoint{0.027778in}{0.027778in}}%
\pgfpathlineto{\pgfqpoint{-0.027778in}{0.027778in}}%
\pgfpathclose%
\pgfusepath{stroke,fill}%
}%
\begin{pgfscope}%
\pgfsys@transformshift{0.659907in}{1.311930in}%
\pgfsys@useobject{currentmarker}{}%
\end{pgfscope}%
\begin{pgfscope}%
\pgfsys@transformshift{1.035893in}{0.670827in}%
\pgfsys@useobject{currentmarker}{}%
\end{pgfscope}%
\begin{pgfscope}%
\pgfsys@transformshift{1.787865in}{0.493627in}%
\pgfsys@useobject{currentmarker}{}%
\end{pgfscope}%
\begin{pgfscope}%
\pgfsys@transformshift{2.539836in}{0.626875in}%
\pgfsys@useobject{currentmarker}{}%
\end{pgfscope}%
\end{pgfscope}%
\begin{pgfscope}%
\pgfsetrectcap%
\pgfsetmiterjoin%
\pgfsetlinewidth{1.003750pt}%
\definecolor{currentstroke}{rgb}{0.000000,0.000000,0.000000}%
\pgfsetstrokecolor{currentstroke}%
\pgfsetdash{}{0pt}%
\pgfpathmoveto{\pgfqpoint{0.471915in}{1.879012in}}%
\pgfpathlineto{\pgfqpoint{2.665165in}{1.879012in}}%
\pgfusepath{stroke}%
\end{pgfscope}%
\begin{pgfscope}%
\pgfsetrectcap%
\pgfsetmiterjoin%
\pgfsetlinewidth{1.003750pt}%
\definecolor{currentstroke}{rgb}{0.000000,0.000000,0.000000}%
\pgfsetstrokecolor{currentstroke}%
\pgfsetdash{}{0pt}%
\pgfpathmoveto{\pgfqpoint{2.665165in}{0.279012in}}%
\pgfpathlineto{\pgfqpoint{2.665165in}{1.879012in}}%
\pgfusepath{stroke}%
\end{pgfscope}%
\begin{pgfscope}%
\pgfsetrectcap%
\pgfsetmiterjoin%
\pgfsetlinewidth{1.003750pt}%
\definecolor{currentstroke}{rgb}{0.000000,0.000000,0.000000}%
\pgfsetstrokecolor{currentstroke}%
\pgfsetdash{}{0pt}%
\pgfpathmoveto{\pgfqpoint{0.471915in}{0.279012in}}%
\pgfpathlineto{\pgfqpoint{2.665165in}{0.279012in}}%
\pgfusepath{stroke}%
\end{pgfscope}%
\begin{pgfscope}%
\pgfsetrectcap%
\pgfsetmiterjoin%
\pgfsetlinewidth{1.003750pt}%
\definecolor{currentstroke}{rgb}{0.000000,0.000000,0.000000}%
\pgfsetstrokecolor{currentstroke}%
\pgfsetdash{}{0pt}%
\pgfpathmoveto{\pgfqpoint{0.471915in}{0.279012in}}%
\pgfpathlineto{\pgfqpoint{0.471915in}{1.879012in}}%
\pgfusepath{stroke}%
\end{pgfscope}%
\begin{pgfscope}%
\pgfsetbuttcap%
\pgfsetroundjoin%
\definecolor{currentfill}{rgb}{0.000000,0.000000,0.000000}%
\pgfsetfillcolor{currentfill}%
\pgfsetlinewidth{0.501875pt}%
\definecolor{currentstroke}{rgb}{0.000000,0.000000,0.000000}%
\pgfsetstrokecolor{currentstroke}%
\pgfsetdash{}{0pt}%
\pgfsys@defobject{currentmarker}{\pgfqpoint{0.000000in}{0.000000in}}{\pgfqpoint{0.000000in}{0.055556in}}{%
\pgfpathmoveto{\pgfqpoint{0.000000in}{0.000000in}}%
\pgfpathlineto{\pgfqpoint{0.000000in}{0.055556in}}%
\pgfusepath{stroke,fill}%
}%
\begin{pgfscope}%
\pgfsys@transformshift{0.659907in}{0.279012in}%
\pgfsys@useobject{currentmarker}{}%
\end{pgfscope}%
\end{pgfscope}%
\begin{pgfscope}%
\pgfsetbuttcap%
\pgfsetroundjoin%
\definecolor{currentfill}{rgb}{0.000000,0.000000,0.000000}%
\pgfsetfillcolor{currentfill}%
\pgfsetlinewidth{0.501875pt}%
\definecolor{currentstroke}{rgb}{0.000000,0.000000,0.000000}%
\pgfsetstrokecolor{currentstroke}%
\pgfsetdash{}{0pt}%
\pgfsys@defobject{currentmarker}{\pgfqpoint{0.000000in}{-0.055556in}}{\pgfqpoint{0.000000in}{0.000000in}}{%
\pgfpathmoveto{\pgfqpoint{0.000000in}{0.000000in}}%
\pgfpathlineto{\pgfqpoint{0.000000in}{-0.055556in}}%
\pgfusepath{stroke,fill}%
}%
\begin{pgfscope}%
\pgfsys@transformshift{0.659907in}{1.879012in}%
\pgfsys@useobject{currentmarker}{}%
\end{pgfscope}%
\end{pgfscope}%
\begin{pgfscope}%
\pgftext[x=0.659907in,y=0.223457in,,top]{\rmfamily\fontsize{10.300000}{12.360000}\selectfont 18}%
\end{pgfscope}%
\begin{pgfscope}%
\pgfsetbuttcap%
\pgfsetroundjoin%
\definecolor{currentfill}{rgb}{0.000000,0.000000,0.000000}%
\pgfsetfillcolor{currentfill}%
\pgfsetlinewidth{0.501875pt}%
\definecolor{currentstroke}{rgb}{0.000000,0.000000,0.000000}%
\pgfsetstrokecolor{currentstroke}%
\pgfsetdash{}{0pt}%
\pgfsys@defobject{currentmarker}{\pgfqpoint{0.000000in}{0.000000in}}{\pgfqpoint{0.000000in}{0.055556in}}{%
\pgfpathmoveto{\pgfqpoint{0.000000in}{0.000000in}}%
\pgfpathlineto{\pgfqpoint{0.000000in}{0.055556in}}%
\pgfusepath{stroke,fill}%
}%
\begin{pgfscope}%
\pgfsys@transformshift{1.035893in}{0.279012in}%
\pgfsys@useobject{currentmarker}{}%
\end{pgfscope}%
\end{pgfscope}%
\begin{pgfscope}%
\pgfsetbuttcap%
\pgfsetroundjoin%
\definecolor{currentfill}{rgb}{0.000000,0.000000,0.000000}%
\pgfsetfillcolor{currentfill}%
\pgfsetlinewidth{0.501875pt}%
\definecolor{currentstroke}{rgb}{0.000000,0.000000,0.000000}%
\pgfsetstrokecolor{currentstroke}%
\pgfsetdash{}{0pt}%
\pgfsys@defobject{currentmarker}{\pgfqpoint{0.000000in}{-0.055556in}}{\pgfqpoint{0.000000in}{0.000000in}}{%
\pgfpathmoveto{\pgfqpoint{0.000000in}{0.000000in}}%
\pgfpathlineto{\pgfqpoint{0.000000in}{-0.055556in}}%
\pgfusepath{stroke,fill}%
}%
\begin{pgfscope}%
\pgfsys@transformshift{1.035893in}{1.879012in}%
\pgfsys@useobject{currentmarker}{}%
\end{pgfscope}%
\end{pgfscope}%
\begin{pgfscope}%
\pgftext[x=1.035893in,y=0.223457in,,top]{\rmfamily\fontsize{10.300000}{12.360000}\selectfont 24}%
\end{pgfscope}%
\begin{pgfscope}%
\pgfsetbuttcap%
\pgfsetroundjoin%
\definecolor{currentfill}{rgb}{0.000000,0.000000,0.000000}%
\pgfsetfillcolor{currentfill}%
\pgfsetlinewidth{0.501875pt}%
\definecolor{currentstroke}{rgb}{0.000000,0.000000,0.000000}%
\pgfsetstrokecolor{currentstroke}%
\pgfsetdash{}{0pt}%
\pgfsys@defobject{currentmarker}{\pgfqpoint{0.000000in}{0.000000in}}{\pgfqpoint{0.000000in}{0.055556in}}{%
\pgfpathmoveto{\pgfqpoint{0.000000in}{0.000000in}}%
\pgfpathlineto{\pgfqpoint{0.000000in}{0.055556in}}%
\pgfusepath{stroke,fill}%
}%
\begin{pgfscope}%
\pgfsys@transformshift{1.787865in}{0.279012in}%
\pgfsys@useobject{currentmarker}{}%
\end{pgfscope}%
\end{pgfscope}%
\begin{pgfscope}%
\pgfsetbuttcap%
\pgfsetroundjoin%
\definecolor{currentfill}{rgb}{0.000000,0.000000,0.000000}%
\pgfsetfillcolor{currentfill}%
\pgfsetlinewidth{0.501875pt}%
\definecolor{currentstroke}{rgb}{0.000000,0.000000,0.000000}%
\pgfsetstrokecolor{currentstroke}%
\pgfsetdash{}{0pt}%
\pgfsys@defobject{currentmarker}{\pgfqpoint{0.000000in}{-0.055556in}}{\pgfqpoint{0.000000in}{0.000000in}}{%
\pgfpathmoveto{\pgfqpoint{0.000000in}{0.000000in}}%
\pgfpathlineto{\pgfqpoint{0.000000in}{-0.055556in}}%
\pgfusepath{stroke,fill}%
}%
\begin{pgfscope}%
\pgfsys@transformshift{1.787865in}{1.879012in}%
\pgfsys@useobject{currentmarker}{}%
\end{pgfscope}%
\end{pgfscope}%
\begin{pgfscope}%
\pgftext[x=1.787865in,y=0.223457in,,top]{\rmfamily\fontsize{10.300000}{12.360000}\selectfont 36}%
\end{pgfscope}%
\begin{pgfscope}%
\pgfsetbuttcap%
\pgfsetroundjoin%
\definecolor{currentfill}{rgb}{0.000000,0.000000,0.000000}%
\pgfsetfillcolor{currentfill}%
\pgfsetlinewidth{0.501875pt}%
\definecolor{currentstroke}{rgb}{0.000000,0.000000,0.000000}%
\pgfsetstrokecolor{currentstroke}%
\pgfsetdash{}{0pt}%
\pgfsys@defobject{currentmarker}{\pgfqpoint{0.000000in}{0.000000in}}{\pgfqpoint{0.000000in}{0.055556in}}{%
\pgfpathmoveto{\pgfqpoint{0.000000in}{0.000000in}}%
\pgfpathlineto{\pgfqpoint{0.000000in}{0.055556in}}%
\pgfusepath{stroke,fill}%
}%
\begin{pgfscope}%
\pgfsys@transformshift{2.539836in}{0.279012in}%
\pgfsys@useobject{currentmarker}{}%
\end{pgfscope}%
\end{pgfscope}%
\begin{pgfscope}%
\pgfsetbuttcap%
\pgfsetroundjoin%
\definecolor{currentfill}{rgb}{0.000000,0.000000,0.000000}%
\pgfsetfillcolor{currentfill}%
\pgfsetlinewidth{0.501875pt}%
\definecolor{currentstroke}{rgb}{0.000000,0.000000,0.000000}%
\pgfsetstrokecolor{currentstroke}%
\pgfsetdash{}{0pt}%
\pgfsys@defobject{currentmarker}{\pgfqpoint{0.000000in}{-0.055556in}}{\pgfqpoint{0.000000in}{0.000000in}}{%
\pgfpathmoveto{\pgfqpoint{0.000000in}{0.000000in}}%
\pgfpathlineto{\pgfqpoint{0.000000in}{-0.055556in}}%
\pgfusepath{stroke,fill}%
}%
\begin{pgfscope}%
\pgfsys@transformshift{2.539836in}{1.879012in}%
\pgfsys@useobject{currentmarker}{}%
\end{pgfscope}%
\end{pgfscope}%
\begin{pgfscope}%
\pgftext[x=2.539836in,y=0.223457in,,top]{\rmfamily\fontsize{10.300000}{12.360000}\selectfont 48}%
\end{pgfscope}%
\begin{pgfscope}%
\pgfsetbuttcap%
\pgfsetroundjoin%
\definecolor{currentfill}{rgb}{0.000000,0.000000,0.000000}%
\pgfsetfillcolor{currentfill}%
\pgfsetlinewidth{0.501875pt}%
\definecolor{currentstroke}{rgb}{0.000000,0.000000,0.000000}%
\pgfsetstrokecolor{currentstroke}%
\pgfsetdash{}{0pt}%
\pgfsys@defobject{currentmarker}{\pgfqpoint{0.000000in}{0.000000in}}{\pgfqpoint{0.055556in}{0.000000in}}{%
\pgfpathmoveto{\pgfqpoint{0.000000in}{0.000000in}}%
\pgfpathlineto{\pgfqpoint{0.055556in}{0.000000in}}%
\pgfusepath{stroke,fill}%
}%
\begin{pgfscope}%
\pgfsys@transformshift{0.471915in}{0.279012in}%
\pgfsys@useobject{currentmarker}{}%
\end{pgfscope}%
\end{pgfscope}%
\begin{pgfscope}%
\pgfsetbuttcap%
\pgfsetroundjoin%
\definecolor{currentfill}{rgb}{0.000000,0.000000,0.000000}%
\pgfsetfillcolor{currentfill}%
\pgfsetlinewidth{0.501875pt}%
\definecolor{currentstroke}{rgb}{0.000000,0.000000,0.000000}%
\pgfsetstrokecolor{currentstroke}%
\pgfsetdash{}{0pt}%
\pgfsys@defobject{currentmarker}{\pgfqpoint{-0.055556in}{0.000000in}}{\pgfqpoint{0.000000in}{0.000000in}}{%
\pgfpathmoveto{\pgfqpoint{0.000000in}{0.000000in}}%
\pgfpathlineto{\pgfqpoint{-0.055556in}{0.000000in}}%
\pgfusepath{stroke,fill}%
}%
\begin{pgfscope}%
\pgfsys@transformshift{2.665165in}{0.279012in}%
\pgfsys@useobject{currentmarker}{}%
\end{pgfscope}%
\end{pgfscope}%
\begin{pgfscope}%
\pgftext[x=0.416359in,y=0.279012in,right,]{\rmfamily\fontsize{10.300000}{12.360000}\selectfont 0.015}%
\end{pgfscope}%
\begin{pgfscope}%
\pgfsetbuttcap%
\pgfsetroundjoin%
\definecolor{currentfill}{rgb}{0.000000,0.000000,0.000000}%
\pgfsetfillcolor{currentfill}%
\pgfsetlinewidth{0.501875pt}%
\definecolor{currentstroke}{rgb}{0.000000,0.000000,0.000000}%
\pgfsetstrokecolor{currentstroke}%
\pgfsetdash{}{0pt}%
\pgfsys@defobject{currentmarker}{\pgfqpoint{0.000000in}{0.000000in}}{\pgfqpoint{0.055556in}{0.000000in}}{%
\pgfpathmoveto{\pgfqpoint{0.000000in}{0.000000in}}%
\pgfpathlineto{\pgfqpoint{0.055556in}{0.000000in}}%
\pgfusepath{stroke,fill}%
}%
\begin{pgfscope}%
\pgfsys@transformshift{0.471915in}{0.479012in}%
\pgfsys@useobject{currentmarker}{}%
\end{pgfscope}%
\end{pgfscope}%
\begin{pgfscope}%
\pgfsetbuttcap%
\pgfsetroundjoin%
\definecolor{currentfill}{rgb}{0.000000,0.000000,0.000000}%
\pgfsetfillcolor{currentfill}%
\pgfsetlinewidth{0.501875pt}%
\definecolor{currentstroke}{rgb}{0.000000,0.000000,0.000000}%
\pgfsetstrokecolor{currentstroke}%
\pgfsetdash{}{0pt}%
\pgfsys@defobject{currentmarker}{\pgfqpoint{-0.055556in}{0.000000in}}{\pgfqpoint{0.000000in}{0.000000in}}{%
\pgfpathmoveto{\pgfqpoint{0.000000in}{0.000000in}}%
\pgfpathlineto{\pgfqpoint{-0.055556in}{0.000000in}}%
\pgfusepath{stroke,fill}%
}%
\begin{pgfscope}%
\pgfsys@transformshift{2.665165in}{0.479012in}%
\pgfsys@useobject{currentmarker}{}%
\end{pgfscope}%
\end{pgfscope}%
\begin{pgfscope}%
\pgftext[x=0.416359in,y=0.479012in,right,]{\rmfamily\fontsize{10.300000}{12.360000}\selectfont 0.020}%
\end{pgfscope}%
\begin{pgfscope}%
\pgfsetbuttcap%
\pgfsetroundjoin%
\definecolor{currentfill}{rgb}{0.000000,0.000000,0.000000}%
\pgfsetfillcolor{currentfill}%
\pgfsetlinewidth{0.501875pt}%
\definecolor{currentstroke}{rgb}{0.000000,0.000000,0.000000}%
\pgfsetstrokecolor{currentstroke}%
\pgfsetdash{}{0pt}%
\pgfsys@defobject{currentmarker}{\pgfqpoint{0.000000in}{0.000000in}}{\pgfqpoint{0.055556in}{0.000000in}}{%
\pgfpathmoveto{\pgfqpoint{0.000000in}{0.000000in}}%
\pgfpathlineto{\pgfqpoint{0.055556in}{0.000000in}}%
\pgfusepath{stroke,fill}%
}%
\begin{pgfscope}%
\pgfsys@transformshift{0.471915in}{0.679012in}%
\pgfsys@useobject{currentmarker}{}%
\end{pgfscope}%
\end{pgfscope}%
\begin{pgfscope}%
\pgfsetbuttcap%
\pgfsetroundjoin%
\definecolor{currentfill}{rgb}{0.000000,0.000000,0.000000}%
\pgfsetfillcolor{currentfill}%
\pgfsetlinewidth{0.501875pt}%
\definecolor{currentstroke}{rgb}{0.000000,0.000000,0.000000}%
\pgfsetstrokecolor{currentstroke}%
\pgfsetdash{}{0pt}%
\pgfsys@defobject{currentmarker}{\pgfqpoint{-0.055556in}{0.000000in}}{\pgfqpoint{0.000000in}{0.000000in}}{%
\pgfpathmoveto{\pgfqpoint{0.000000in}{0.000000in}}%
\pgfpathlineto{\pgfqpoint{-0.055556in}{0.000000in}}%
\pgfusepath{stroke,fill}%
}%
\begin{pgfscope}%
\pgfsys@transformshift{2.665165in}{0.679012in}%
\pgfsys@useobject{currentmarker}{}%
\end{pgfscope}%
\end{pgfscope}%
\begin{pgfscope}%
\pgftext[x=0.416359in,y=0.679012in,right,]{\rmfamily\fontsize{10.300000}{12.360000}\selectfont 0.025}%
\end{pgfscope}%
\begin{pgfscope}%
\pgfsetbuttcap%
\pgfsetroundjoin%
\definecolor{currentfill}{rgb}{0.000000,0.000000,0.000000}%
\pgfsetfillcolor{currentfill}%
\pgfsetlinewidth{0.501875pt}%
\definecolor{currentstroke}{rgb}{0.000000,0.000000,0.000000}%
\pgfsetstrokecolor{currentstroke}%
\pgfsetdash{}{0pt}%
\pgfsys@defobject{currentmarker}{\pgfqpoint{0.000000in}{0.000000in}}{\pgfqpoint{0.055556in}{0.000000in}}{%
\pgfpathmoveto{\pgfqpoint{0.000000in}{0.000000in}}%
\pgfpathlineto{\pgfqpoint{0.055556in}{0.000000in}}%
\pgfusepath{stroke,fill}%
}%
\begin{pgfscope}%
\pgfsys@transformshift{0.471915in}{0.879012in}%
\pgfsys@useobject{currentmarker}{}%
\end{pgfscope}%
\end{pgfscope}%
\begin{pgfscope}%
\pgfsetbuttcap%
\pgfsetroundjoin%
\definecolor{currentfill}{rgb}{0.000000,0.000000,0.000000}%
\pgfsetfillcolor{currentfill}%
\pgfsetlinewidth{0.501875pt}%
\definecolor{currentstroke}{rgb}{0.000000,0.000000,0.000000}%
\pgfsetstrokecolor{currentstroke}%
\pgfsetdash{}{0pt}%
\pgfsys@defobject{currentmarker}{\pgfqpoint{-0.055556in}{0.000000in}}{\pgfqpoint{0.000000in}{0.000000in}}{%
\pgfpathmoveto{\pgfqpoint{0.000000in}{0.000000in}}%
\pgfpathlineto{\pgfqpoint{-0.055556in}{0.000000in}}%
\pgfusepath{stroke,fill}%
}%
\begin{pgfscope}%
\pgfsys@transformshift{2.665165in}{0.879012in}%
\pgfsys@useobject{currentmarker}{}%
\end{pgfscope}%
\end{pgfscope}%
\begin{pgfscope}%
\pgftext[x=0.416359in,y=0.879012in,right,]{\rmfamily\fontsize{10.300000}{12.360000}\selectfont 0.030}%
\end{pgfscope}%
\begin{pgfscope}%
\pgfsetbuttcap%
\pgfsetroundjoin%
\definecolor{currentfill}{rgb}{0.000000,0.000000,0.000000}%
\pgfsetfillcolor{currentfill}%
\pgfsetlinewidth{0.501875pt}%
\definecolor{currentstroke}{rgb}{0.000000,0.000000,0.000000}%
\pgfsetstrokecolor{currentstroke}%
\pgfsetdash{}{0pt}%
\pgfsys@defobject{currentmarker}{\pgfqpoint{0.000000in}{0.000000in}}{\pgfqpoint{0.055556in}{0.000000in}}{%
\pgfpathmoveto{\pgfqpoint{0.000000in}{0.000000in}}%
\pgfpathlineto{\pgfqpoint{0.055556in}{0.000000in}}%
\pgfusepath{stroke,fill}%
}%
\begin{pgfscope}%
\pgfsys@transformshift{0.471915in}{1.079012in}%
\pgfsys@useobject{currentmarker}{}%
\end{pgfscope}%
\end{pgfscope}%
\begin{pgfscope}%
\pgfsetbuttcap%
\pgfsetroundjoin%
\definecolor{currentfill}{rgb}{0.000000,0.000000,0.000000}%
\pgfsetfillcolor{currentfill}%
\pgfsetlinewidth{0.501875pt}%
\definecolor{currentstroke}{rgb}{0.000000,0.000000,0.000000}%
\pgfsetstrokecolor{currentstroke}%
\pgfsetdash{}{0pt}%
\pgfsys@defobject{currentmarker}{\pgfqpoint{-0.055556in}{0.000000in}}{\pgfqpoint{0.000000in}{0.000000in}}{%
\pgfpathmoveto{\pgfqpoint{0.000000in}{0.000000in}}%
\pgfpathlineto{\pgfqpoint{-0.055556in}{0.000000in}}%
\pgfusepath{stroke,fill}%
}%
\begin{pgfscope}%
\pgfsys@transformshift{2.665165in}{1.079012in}%
\pgfsys@useobject{currentmarker}{}%
\end{pgfscope}%
\end{pgfscope}%
\begin{pgfscope}%
\pgftext[x=0.416359in,y=1.079012in,right,]{\rmfamily\fontsize{10.300000}{12.360000}\selectfont 0.035}%
\end{pgfscope}%
\begin{pgfscope}%
\pgfsetbuttcap%
\pgfsetroundjoin%
\definecolor{currentfill}{rgb}{0.000000,0.000000,0.000000}%
\pgfsetfillcolor{currentfill}%
\pgfsetlinewidth{0.501875pt}%
\definecolor{currentstroke}{rgb}{0.000000,0.000000,0.000000}%
\pgfsetstrokecolor{currentstroke}%
\pgfsetdash{}{0pt}%
\pgfsys@defobject{currentmarker}{\pgfqpoint{0.000000in}{0.000000in}}{\pgfqpoint{0.055556in}{0.000000in}}{%
\pgfpathmoveto{\pgfqpoint{0.000000in}{0.000000in}}%
\pgfpathlineto{\pgfqpoint{0.055556in}{0.000000in}}%
\pgfusepath{stroke,fill}%
}%
\begin{pgfscope}%
\pgfsys@transformshift{0.471915in}{1.279012in}%
\pgfsys@useobject{currentmarker}{}%
\end{pgfscope}%
\end{pgfscope}%
\begin{pgfscope}%
\pgfsetbuttcap%
\pgfsetroundjoin%
\definecolor{currentfill}{rgb}{0.000000,0.000000,0.000000}%
\pgfsetfillcolor{currentfill}%
\pgfsetlinewidth{0.501875pt}%
\definecolor{currentstroke}{rgb}{0.000000,0.000000,0.000000}%
\pgfsetstrokecolor{currentstroke}%
\pgfsetdash{}{0pt}%
\pgfsys@defobject{currentmarker}{\pgfqpoint{-0.055556in}{0.000000in}}{\pgfqpoint{0.000000in}{0.000000in}}{%
\pgfpathmoveto{\pgfqpoint{0.000000in}{0.000000in}}%
\pgfpathlineto{\pgfqpoint{-0.055556in}{0.000000in}}%
\pgfusepath{stroke,fill}%
}%
\begin{pgfscope}%
\pgfsys@transformshift{2.665165in}{1.279012in}%
\pgfsys@useobject{currentmarker}{}%
\end{pgfscope}%
\end{pgfscope}%
\begin{pgfscope}%
\pgftext[x=0.416359in,y=1.279012in,right,]{\rmfamily\fontsize{10.300000}{12.360000}\selectfont 0.040}%
\end{pgfscope}%
\begin{pgfscope}%
\pgfsetbuttcap%
\pgfsetroundjoin%
\definecolor{currentfill}{rgb}{0.000000,0.000000,0.000000}%
\pgfsetfillcolor{currentfill}%
\pgfsetlinewidth{0.501875pt}%
\definecolor{currentstroke}{rgb}{0.000000,0.000000,0.000000}%
\pgfsetstrokecolor{currentstroke}%
\pgfsetdash{}{0pt}%
\pgfsys@defobject{currentmarker}{\pgfqpoint{0.000000in}{0.000000in}}{\pgfqpoint{0.055556in}{0.000000in}}{%
\pgfpathmoveto{\pgfqpoint{0.000000in}{0.000000in}}%
\pgfpathlineto{\pgfqpoint{0.055556in}{0.000000in}}%
\pgfusepath{stroke,fill}%
}%
\begin{pgfscope}%
\pgfsys@transformshift{0.471915in}{1.479012in}%
\pgfsys@useobject{currentmarker}{}%
\end{pgfscope}%
\end{pgfscope}%
\begin{pgfscope}%
\pgfsetbuttcap%
\pgfsetroundjoin%
\definecolor{currentfill}{rgb}{0.000000,0.000000,0.000000}%
\pgfsetfillcolor{currentfill}%
\pgfsetlinewidth{0.501875pt}%
\definecolor{currentstroke}{rgb}{0.000000,0.000000,0.000000}%
\pgfsetstrokecolor{currentstroke}%
\pgfsetdash{}{0pt}%
\pgfsys@defobject{currentmarker}{\pgfqpoint{-0.055556in}{0.000000in}}{\pgfqpoint{0.000000in}{0.000000in}}{%
\pgfpathmoveto{\pgfqpoint{0.000000in}{0.000000in}}%
\pgfpathlineto{\pgfqpoint{-0.055556in}{0.000000in}}%
\pgfusepath{stroke,fill}%
}%
\begin{pgfscope}%
\pgfsys@transformshift{2.665165in}{1.479012in}%
\pgfsys@useobject{currentmarker}{}%
\end{pgfscope}%
\end{pgfscope}%
\begin{pgfscope}%
\pgftext[x=0.416359in,y=1.479012in,right,]{\rmfamily\fontsize{10.300000}{12.360000}\selectfont 0.045}%
\end{pgfscope}%
\begin{pgfscope}%
\pgfsetbuttcap%
\pgfsetroundjoin%
\definecolor{currentfill}{rgb}{0.000000,0.000000,0.000000}%
\pgfsetfillcolor{currentfill}%
\pgfsetlinewidth{0.501875pt}%
\definecolor{currentstroke}{rgb}{0.000000,0.000000,0.000000}%
\pgfsetstrokecolor{currentstroke}%
\pgfsetdash{}{0pt}%
\pgfsys@defobject{currentmarker}{\pgfqpoint{0.000000in}{0.000000in}}{\pgfqpoint{0.055556in}{0.000000in}}{%
\pgfpathmoveto{\pgfqpoint{0.000000in}{0.000000in}}%
\pgfpathlineto{\pgfqpoint{0.055556in}{0.000000in}}%
\pgfusepath{stroke,fill}%
}%
\begin{pgfscope}%
\pgfsys@transformshift{0.471915in}{1.679012in}%
\pgfsys@useobject{currentmarker}{}%
\end{pgfscope}%
\end{pgfscope}%
\begin{pgfscope}%
\pgfsetbuttcap%
\pgfsetroundjoin%
\definecolor{currentfill}{rgb}{0.000000,0.000000,0.000000}%
\pgfsetfillcolor{currentfill}%
\pgfsetlinewidth{0.501875pt}%
\definecolor{currentstroke}{rgb}{0.000000,0.000000,0.000000}%
\pgfsetstrokecolor{currentstroke}%
\pgfsetdash{}{0pt}%
\pgfsys@defobject{currentmarker}{\pgfqpoint{-0.055556in}{0.000000in}}{\pgfqpoint{0.000000in}{0.000000in}}{%
\pgfpathmoveto{\pgfqpoint{0.000000in}{0.000000in}}%
\pgfpathlineto{\pgfqpoint{-0.055556in}{0.000000in}}%
\pgfusepath{stroke,fill}%
}%
\begin{pgfscope}%
\pgfsys@transformshift{2.665165in}{1.679012in}%
\pgfsys@useobject{currentmarker}{}%
\end{pgfscope}%
\end{pgfscope}%
\begin{pgfscope}%
\pgftext[x=0.416359in,y=1.679012in,right,]{\rmfamily\fontsize{10.300000}{12.360000}\selectfont 0.050}%
\end{pgfscope}%
\begin{pgfscope}%
\pgfsetbuttcap%
\pgfsetroundjoin%
\definecolor{currentfill}{rgb}{0.000000,0.000000,0.000000}%
\pgfsetfillcolor{currentfill}%
\pgfsetlinewidth{0.501875pt}%
\definecolor{currentstroke}{rgb}{0.000000,0.000000,0.000000}%
\pgfsetstrokecolor{currentstroke}%
\pgfsetdash{}{0pt}%
\pgfsys@defobject{currentmarker}{\pgfqpoint{0.000000in}{0.000000in}}{\pgfqpoint{0.055556in}{0.000000in}}{%
\pgfpathmoveto{\pgfqpoint{0.000000in}{0.000000in}}%
\pgfpathlineto{\pgfqpoint{0.055556in}{0.000000in}}%
\pgfusepath{stroke,fill}%
}%
\begin{pgfscope}%
\pgfsys@transformshift{0.471915in}{1.879012in}%
\pgfsys@useobject{currentmarker}{}%
\end{pgfscope}%
\end{pgfscope}%
\begin{pgfscope}%
\pgfsetbuttcap%
\pgfsetroundjoin%
\definecolor{currentfill}{rgb}{0.000000,0.000000,0.000000}%
\pgfsetfillcolor{currentfill}%
\pgfsetlinewidth{0.501875pt}%
\definecolor{currentstroke}{rgb}{0.000000,0.000000,0.000000}%
\pgfsetstrokecolor{currentstroke}%
\pgfsetdash{}{0pt}%
\pgfsys@defobject{currentmarker}{\pgfqpoint{-0.055556in}{0.000000in}}{\pgfqpoint{0.000000in}{0.000000in}}{%
\pgfpathmoveto{\pgfqpoint{0.000000in}{0.000000in}}%
\pgfpathlineto{\pgfqpoint{-0.055556in}{0.000000in}}%
\pgfusepath{stroke,fill}%
}%
\begin{pgfscope}%
\pgfsys@transformshift{2.665165in}{1.879012in}%
\pgfsys@useobject{currentmarker}{}%
\end{pgfscope}%
\end{pgfscope}%
\begin{pgfscope}%
\pgftext[x=0.416359in,y=1.879012in,right,]{\rmfamily\fontsize{10.300000}{12.360000}\selectfont 0.055}%
\end{pgfscope}%
\begin{pgfscope}%
\pgftext[x=1.568540in,y=1.948457in,,base]{\rmfamily\fontsize{12.360000}{14.832000}\selectfont \(\displaystyle \mathcal{E}_{\text{RMS}}^{\theta_1}(c_p)\)}%
\end{pgfscope}%
\end{pgfpicture}%
\makeatother%
\endgroup%

%% file: Conclusion.tex
\section{Conclusion}
\label{chap:conclusion}

In this work, two popular, yet substantially different fluid-particle coupling approaches for the lattice Boltzmann method have been revisited and three variants of each method have been compared in detail.
The applied test case of a single solid sphere settling in an ambient fluid has been found to be a suitable benchmark scenario for the evaluation of coupling methods that fully resolve the submerged particles.
In contrast to other commonly applied test cases in this area, the particle Reynolds number is at the order of $100$ and the sphere is moving throughout the simulation.
We find that the momentum exchange method can provide accurate predictions of the streamwise sphere velocity whereas Noble and Torczynski's partially saturated cells method yields more accurate spanwise velocities.
Even though no unique explicit recommendation for one of the methods can be given, we find that the central linear interpolation boundary scheme \cite{ginzburg_two-relaxation-time_2008} from the family of the momentum exchange methods yields highly accurate and reliable predictions for all of the tested flow regimes.
Compared to this scheme, both the bounce back and the multi-reflection scheme perform worse in the case of higher Reynolds numbers.

The partially saturated cells methods are found to seemingly add numerical diffusion which prevents the prediction of certain flow and coupling characteristics that appear at higher Reynolds numbers.
Even though this affects the quality of the results, it provides a partial reason why this method is sometimes the preferred coupling method for high Reynolds number particulate flows.
The extra dissipation of this method is likely to numerically stabilize the computation and thus to permit stable simulations even at lower resolutions.

To reach errors below 5\% for the velocity components of the sphere, essentially all methods require a resolution of $24$ computational cells per diameter already for Reynolds numbers of $185$ and $243$.
Depending on the applied method, $36$ or even $48$ cells must be used for similarly accurate predictions of the regimes with $Re=263$ and $365$.
These resolutions are significantly higher than those commonly reported and used in current computational practice where often a single particle is resolved with around ten cells only. 
These values usually find their justification in drag force evaluations when using fixed spheres in Stokes flow.
Even though such a low resolution might be enough to capture the relevant length scales in the flow, the choice for a certain resolution should also take into account the accuracy of the fluid-particle coupling at the respective flow regime.
These findings agree with the arguments made by \cite{tenneti_drag_2011} that the boundary layers along the particles must be resolved sufficiently well for flows at higher Reynolds numbers.
The evaluation in this article shows that a genuine direct numerical simulation of particles in inertial flows may require such an increased resolution.
Future work should apply this test case also to other available fluid-particle coupling approaches not covered in this work like the immersed boundary method for LBM to obtain a complete overview of their properties.

These remarks must be seen in the light that one cannot directly relate the accuracy of a single particle system
to that achieved for a system with several thousand particles, as they may be needed to simulate industrially relevant applications.
In such cases, systematic evaluations must consider additional aspects like particle-particle interaction models and lubrication forces since they constitute additional numerical errors.
Moreover, the evaluation of the quantities of interest when simulating such systems will often require statistical averaging.
Because of such effects it is yet unclear how the findings of this article will carry over to simulations with many embedded particles.
Therefore, future work must focus on developing meaningful test and benchmark cases for the direct numerical simulations of particulate flow scenarios with several particles.

\section*{Acknowledgements}
The authors would like to thank Simon Bogner and Ehsan Fattahi for valuable discussions.
They gratefully acknowledge the Gauss Centre for Supercomputing e.V. (\url{www.gauss-centre.eu}) for funding this project by providing computing time on the GCS Supercomputer SuperMUC at Leibniz Supercomputing Centre (LRZ, \url{www.lrz.de}).

%% file: Tables.tex
\begin{table}[t]
	\caption{Results obtained for the LBM with different coupling methods for case A, with nominal Galileo number $Ga=144$. The error is computed with respect to the results of the reference case AL in \cite{uhlmann_motion_2014}.}
	\label{tab:G144}
	\centering
	\begin{tabular}{r l l l l}
		\hline
		& $u_{pV}$ & $\mathcal{E}(u_{pV})$ & $L_r$ & $\mathcal{E}(L_r)$  \\
		\hline
		Reference & $-1.285$ & & $1.383$ & \\ 
		\hline
		A-BB-18 & $-1.2161$ & $0.0536$ & $1.4126$ & $0.0214$ \\
		A-BB-24 & $-1.2386$ & $0.0361$ & $1.3884$ & $0.0039$ \\
		A-BB-36 & $-1.2487$ & $0.0283$ & $1.3701$ & $0.0093$ \\
		A-BB-48 & $-1.2538$ & $0.0243$ & $1.3634$ & $0.0142$ \\  
		\hline
		A-CLI-18 & $-1.2150$ & $0.0545$ & $1.3462$ & $0.0266$ \\
		A-CLI-24 & $-1.2383$ & $0.0363$ & $1.3616$ & $0.0155$ \\
		A-CLI-36 & $-1.2538$ & $0.0243$ & $1.3540$ & $0.0210$ \\
		A-CLI-48 & $-1.2586$ & $0.0205$ & $1.3511$ & $0.0231$ \\
		\hline
		A-MR-18 & $-1.2172$ & $0.0528$ & $1.4919$ & $0.0787$ \\
		A-MR-24 & $-1.2510$ & $0.0265$ & $1.4248$ & $0.0303$ \\
		A-MR-36 & $-1.2643$ & $0.0161$ & $1.3775$ & $0.0040$ \\
		A-MR-48 & $-1.2646$ & $0.0159$ & $1.3610$ & $0.0159$ \\
		\hline
		A-M1B1-18 & $-1.1646$ & $0.0937$ & $1.4685$ & $0.0618$ \\
		A-M1B1-24 & $-1.1924$ & $0.0721$ & $1.4330$ & $0.0361$ \\
		A-M1B1-36 & $-1.2182$ & $0.0520$ & $1.4015$ & $0.0134$ \\
		A-M1B1-48 & $-1.2299$ & $0.0429$ & $1.3873$ & $0.0031$ \\
		\hline
		A-M2B2-18 & $-1.2363$ & $0.0379$ & $1.3820$ & $0.0008$ \\
		A-M2B2-24 & $-1.2490$ & $0.0280$ & $1.3665$ & $0.0119$ \\
		A-M2B2-36 & $-1.2588$ & $0.0204$ & $1.3521$ & $0.0224$ \\
		A-M2B2-48 & $-1.2616$ & $0.0182$ & $1.3482$ & $0.0251$ \\
		\hline
		A-M3B2-18 & $-1.2128$ & $0.0562$ & $1.4128$ & $0.0216$ \\
		A-M3B2-24 & $-1.2309$ & $0.0421$ & $1.3876$ & $0.0033$ \\
		A-M3B2-36 & $-1.2453$ & $0.0309$ & $1.3692$ & $0.0099$ \\
		A-M3B2-48 & $-1.2513$ & $0.0262$ & $1.3612$ & $0.0158$ \\
		\hline
	\end{tabular}
\end{table}

\begin{table}[t]
	\caption{Results obtained for the LBM with different coupling methods for case B, with nominal Galileo number $Ga=178.46$. The error is computed with respect to the results of the reference case BL in \cite{uhlmann_motion_2014}. No steady solution could be obtained for simulations marked with (*).}
	\label{tab:G178}
	\centering
	\begin{tabular}{c l l l l l l l l}
		\hline
		& $u_{pV}$ & $\mathcal{E}(u_{pV})$ & $u_{pH}$ & $\mathcal{E}(u_{pH})$ & $\omega_{pH}$ & $\mathcal{E}(\omega_{pH})$& $L_r$ & $\mathcal{E}(L_r)$ \\
		\hline
		Reference  & $-1.356$ & & $0.1245$ & & $0.0137$ & & $1.629$ & \\ 
		\hline      
		B-BB-24 & $-1.3097$ & $0.0341$ & $0.0494$ & $0.0554$ & $0.0605$ & $0.0345$ & $1.6593$ & $0.0186$ \\
		B-BB-36 & $-1.3265$ & $0.0218$ & $0.0652$ & $0.0437$ & $0.0490$ & $0.0260$ & $1.6324$ & $0.0021$ \\
		B-BB-48 & $-1.3325$ & $0.0173$ & $0.0722$ & $0.0386$ & $0.0367$ & $0.0170$ & $1.6181$ & $0.0067$ \\\hline
		B-CLI-24 & $-1.3056$ & $0.0371$ & $0.0883$ & $0.0267$ & $0.0230$ & $0.0069$ & $1.6023$ & $0.0164$ \\
		B-CLI-36 & $-1.3267$ & $0.0216$ & $0.0934$ & $0.0229$ & $0.0175$ & $0.0028$ & $1.6058$ & $0.0143$ \\
		B-CLI-48 & $-1.3328$ & $0.0171$ & $0.0991$ & $0.0187$ & $0.0104$ & $0.0025$ & $1.6025$ & $0.0163$ \\\hline
		B-MR-24(*) & $-1.3037$ & $0.0385$ & $0.1178$ & $0.0049$ & $0.0092$ & $0.0033$ & $1.6625$ & $0.0206$ \\
		B-MR-36(*) & $-1.3358$ & $0.0149$ & $0.1097$ & $0.0109$ & $0.0038$ & $0.0073$ & $1.6486$ & $0.0120$ \\
		B-MR-48(*) & $-1.3384$ & $0.0130$ & $0.1108$ & $0.0101$ & $0.0013$ & $0.0091$ & $1.6216$ & $0.0046$ \\\hline
		B-M1B1-24 & $-1.2455$ & $0.0815$ & $0.1171$ & $0.0055$ & $0.0332$ & $0.0144$ & $1.7104$ & $0.0500$ \\
		B-M1B1-36 & $-1.2795$ & $0.0564$ & $0.1153$ & $0.0068$ & $0.0220$ & $0.0061$ & $1.6634$ & $0.0211$ \\
		B-M1B1-48 & $-1.2944$ & $0.0454$ & $0.1159$ & $0.0063$ & $0.0186$ & $0.0036$ & $1.6432$ & $0.0087$ \\\hline
		B-M2B2-24 & $-1.3040$ & $0.0384$ & $0.1329$ & $0.0062$ & $0.0437$ & $0.0221$ & $1.6225$ & $0.0040$ \\
		B-M2B2-36 & $-1.3218$ & $0.0252$ & $0.1238$ & $0.0005$ & $0.0255$ & $0.0087$ & $1.6047$ & $0.0149$ \\
		B-M2B2-48 & $-1.3278$ & $0.0208$ & $0.1209$ & $0.0027$ & $0.0198$ & $0.0045$ & $1.5972$ & $0.0195$ \\\hline
		B-M3B2-24 & $-1.2904$ & $0.0484$ & $0.1150$ & $0.0070$ & $0.0223$ & $0.0064$ & $1.6497$ & $0.0127$ \\
		B-M3B2-36 & $-1.3112$ & $0.0331$ & $0.1128$ & $0.0086$ & $0.0135$ & $0.0001$ & $1.6233$ & $0.0035$ \\
		B-M3B2-48 & $-1.3192$ & $0.0271$ & $0.1124$ & $0.0089$ & $0.0108$ & $0.0021$ & $1.6120$ & $0.0104$ \\
		\hline
	\end{tabular}
\end{table}

\begin{table}[t]
	\caption{Results obtained for the LBM with different coupling methods for case C, with nominal Galileo number $Ga=190$. The error is computed with respect to the results of the reference case CL in \cite{uhlmann_motion_2014}.}
	\label{tab:G190}
	\centering
	\begin{tabular}{l r l l l l l l}
		\hline
		& $\overline{u}_{pV}$ \hspace{0.7cm} & $\overline{u}_{pH}$ & $\overline{\omega}_{pH}$ & $u'_{pV}$ & $u'_{pH}$ & $\omega'_{pH}$ & $f$      \\
		\hline
		Reference                   & $-1.376$ \hspace{0.4ex}            & $0.136$             & $0.012$                  & $0.008$   & $0.033$   & $0.008$        & $0.071$  \\ \hline
		C-BB-36 & $-1.3479$ & $0.0624$ & $0.0452$ & $0.0108$ & $0.0624$ & $0.0439$ & $0.0415$ \\
		$\mathcal{E}^{(\text{C-BB-36})}$ & $0.0206$ & $0.0536$ & $0.0241$ & $0.0018$ & $0.0210$ & $0.0258$ & $0.4152$ \\
		C-BB-48 & $-1.3540$ & $0.0678$ & $0.0351$ & $0.0126$ & $0.0678$ & $0.0350$ & $0.0356$ \\
		$\mathcal{E}^{(\text{C-BB-48})}$ & $0.0162$ & $0.0496$ & $0.0167$ & $0.0030$ & $0.0250$ & $0.0193$ & $0.4984$ \\ \hline
		C-CLI-36 & $-1.3461$ & $0.1001$ & $0.0284$ & $0.0053$ & $0.0333$ & $0.0120$ & $0.0637$ \\
		$\mathcal{E}^{(\text{C-CLI-36})}$ & $0.0220$ & $0.0262$ & $0.0118$ & $0.0022$ & $0.0001$ & $0.0026$ & $0.1030$ \\
		C-CLI-48 & $-1.3527$ & $0.1076$ & $0.0195$ & $0.0058$ & $0.0307$ & $0.0087$ & $0.0668$ \\
		$\mathcal{E}^{(\text{C-CLI-48})}$ & $0.0172$ & $0.0207$ & $0.0054$ & $0.0019$ & $0.0020$ & $0.0003$ & $0.0599$ \\ \hline
		C-MR-36 & $-1.3547$ & $0.1091$ & $0.0143$ & $0.0115$ & $0.0538$ & $0.0143$ & $0.0636$ \\
		$\mathcal{E}^{(\text{C-MR-36})}$ & $0.0158$ & $0.0197$ & $0.0016$ & $0.0023$ & $0.0148$ & $0.0043$ & $0.1042$ \\
		C-MR-48 & $-1.3590$ & $0.1178$ & $0.0080$ & $0.0078$ & $0.0362$ & $0.0080$ & $0.0667$ \\ 
		$\mathcal{E}^{(\text{C-MR-48})}$ & $0.0126$ & $0.0134$ & $0.0030$ & $0.0004$ & $0.0021$ & $0.0003$ & $0.0607$ \\ \hline
		C-M1B1-36 & $-1.2978$ & $0.1295$ & $0.0223$ & $0.0015$ & $0.0065$ & $0.0019$ & $0.0680$ \\
		$\mathcal{E}^{(\text{C-M1B1-36})}$ & $0.0571$ & $0.0049$ & $0.0074$ & $0.0050$ & $0.0195$ & $0.0047$ & $0.0421$ \\
		C-M1B1-48 & $-1.3131$ & $0.1306$ & $0.0173$ & $0.0013$ & $0.0049$ & $0.0013$ & $0.0669$ \\
		$\mathcal{E}^{(\text{C-M1B1-48})}$ & $0.0459$ & $0.0041$ & $0.0038$ & $0.0052$ & $0.0207$ & $0.0051$ & $0.0576$ \\\hline
		C-M2B2-36 & $-1.3389$ & $0.1363$ & $0.0250$ & $0.0015$ & $0.0072$ & $0.0018$ & $0.0690$ \\
		$\mathcal{E}^{(\text{C-M2B2-36})}$ & $0.0272$ & $0.0001$ & $0.0094$ & $0.0050$ & $0.0191$ & $0.0047$ & $0.0287$ \\
		C-M2B2-48 & $-1.3455$ & $0.1355$ & $0.0185$ & $0.0013$ & $0.0052$ & $0.0013$ & $0.0712$ \\
		$\mathcal{E}^{(\text{C-M2B2-48})}$ & $0.0224$ & $0.0005$ & $0.0046$ & $0.0051$ & $0.0205$ & $0.0052$ & $0.0024$ \\\hline
		C-M3B2-36 & $-1.3283$ & $0.1245$ & $0.0117$ & $0.0016$ & $0.0086$ & $0.0021$ & $0.0733$ \\
		$\mathcal{E}^{(\text{C-M3B2-36})}$ & $0.0349$ & $0.0084$ & $0.0003$ & $0.0049$ & $0.0181$ & $0.0046$ & $0.0328$ \\
		C-M3B2-48 & $-1.3371$ & $0.1240$ & $0.0084$ & $0.0017$ & $0.0095$ & $0.0021$ & $0.0729$ \\
		$\mathcal{E}^{(\text{C-M3B2-48})}$ & $0.0285$ & $0.0089$ & $0.0027$ & $0.0048$ & $0.0174$ & $0.0045$ & $0.0273$ \\
		\hline
	\end{tabular}
\end{table}

\begin{table}[t]
	\caption{Results obtained for the LBM with different coupling methods for case D, with nominal Galileo number $Ga=250$. The error is computed with respect to the results of the reference case DL in \cite{uhlmann_motion_2014}.}
	\label{tab:G250}
	\centering
	\begin{tabular}{l r l l l l}
		\hline
		& $\langle u_{pV} \rangle $ & $\langle u_{pV}''u_{pV}'' \rangle^{1/2}$ & $\langle u_{pr}''u_{pr}'' \rangle^{1/2}$ & $\langle \omega_{pV}''\omega_{pV}'' \rangle^{1/2}$ & $\langle \omega_{px}''\omega_{px}'' \rangle^{1/2}$ \\
		\hline
		Reference                   & $-1.4604$                 & $0.0087$                                 & $0.0854$                                 & $0.0013$                                           & $0.0067$                                           \\ \hline
		D-BB-36                     & $-1.4111$ & $0.0049$ & $0.0342$ & $0.0005$ & $0.0200$ \\
		$\mathcal{E}^{(\text{D-BB-36})}$   & $0.0338$ & $0.0026$ & $0.0351$ & $0.0005$ & $0.0091$ \\ \hline
		D-CLI-36                    & $-1.4114$ & $0.0075$ & $0.0701$ & $0.0016$ & $0.0358$ \\
		$\mathcal{E}^{(\text{D-CLI-36})}$  & $0.0336$ & $0.0008$ & $0.0105$ & $0.0002$ & $0.0199$ \\ \hline
		%MR results without correction
		%D-MR-36                     & $-1.4241$ & $0.0068$ & $0.0649$ & $0.0019$ & $0.0377$ \\
		%$\mathcal{E}^{(\text{D-MR-36})}$   & $0.0249$ & $0.0013$ & $0.0140$ & $0.0004$ & $0.0212$ \\ \hline
		%MR results with correction, post collision values
		%D-MR-36 & $-1.4211$ & $0.0027$ & $0.0340$ & $0.0007$ & $0.0114$ \\
		%$\mathcal{E}^{(\text{D-MR-36})}$ & $0.0269$ & $0.0041$ & $0.0352$ & $0.0004$ & $0.0032$ \\ \hline
		%MR results with correction, pre collision values
		D-MR-36 & $-1.4187$ & $0.0037$ & $0.0348$ & $0.0008$ & $0.0116$ \\
		$\mathcal{E}^{(\text{D-MR-36})}$ & $0.0286$ & $0.0034$ & $0.0347$ & $0.0004$ & $0.0033$ \\ \hline
		D-M1B1-36                   & $-1.3701$ & $0.0109$ & $0.0913$ & $0.0008$ & $0.0160$ \\
		$\mathcal{E}^{(\text{D-M1B1-36})}$ & $0.0618$ & $0.0015$ & $0.0041$ & $0.0003$ & $0.0064$ \\ \hline
		D-M2B2-36                   & $-1.4093$ & $0.0102$ & $0.1045$ & $0.0009$ & $0.0209$ \\
		$\mathcal{E}^{(\text{D-M2B2-36})}$ & $0.0350$ & $0.0011$ & $0.0131$ & $0.0003$ & $0.0097$ \\ \hline
		D-M3B2-36                   & $-1.4010$ & $0.0066$ & $0.0881$ & $0.0158$ & $0.0082$ \\
		$\mathcal{E}^{(\text{D-M3B2-36})}$ & $0.0407$ & $0.0014$ & $0.0019$ & $0.0099$ & $0.0010$ \\
		\hline
	\end{tabular}
\end{table}